\documentclass[aps,showpacs, showkeys,prd]{revtex4}
\usepackage [utf8]{inputenc} 
\inputencoding{latin1}
\usepackage[USenglish]{babel}
\usepackage{graphicx}
\usepackage{subfloat}
\usepackage{amsmath}
\usepackage{latexsym}
\usepackage{amstext}
\usepackage{amssymb}
\usepackage{color}
\usepackage[dvipsnames]{xcolor} 
\usepackage{tikz}
\usepackage{amsfonts}
\usepackage{multirow} 

\usepackage{ulem}

\usepackage{amsmath}
\usepackage{subfig}
\usepackage{float}

\newcommand{\beq}{\begin{equation}}

\newcommand{\eeq}{\end{equation}}

\newcommand{\beqa}{\begin{eqnarray}}

\newcommand{\eeqa}{\end{eqnarray}}

\newcommand{\mbf}{\mathbf}

\begin{document}

\title{Radiation from  a dipole  perpendicular to the interface between two planar semi-infinite  magnetoelectric media }

\author{O. J.  Franca} \email{francamentesantiago@ciencias.unam.mx}  
\affiliation{Institut f\"ur Physik, Universit\"at Kassel, Heinrich-Plett-Stra\ss e 40, 34132 Kassel, Germany}
\affiliation{Instituto de Ciencias Nucleares, Universidad Nacional Aut\'onoma de M\'exico, \\ 04510 M\'exico, Distrito Federal, M\'exico}

\author{ Luis F. Urrutia}
\email{urrutia@nucleares.unam.mx}
\affiliation{Instituto de Ciencias Nucleares, Universidad Nacional Aut\'onoma de M\'exico, \\ 04510 M\'exico, Distrito Federal, M\'exico}

\begin{abstract}
We consider  two semi-infinite magnetoelectric media with constant dielectric permittivity separated by a planar interface, whose electromagnetic response is described by non-dynamical axion electrodynamics and investigate  the radiation of a  point-like electric dipole located perpendicularly to the interface. We start from the exact Green's function for the electromagnetic potential, whose far-field approximation is obtained using a modified steepest descent approximation.  We compute  the angular distribution of the radiation and the total radiated power  finding  different interference patterns,  depending on  the relative position dipole-observer, and  polarization mixing effects   which are all absent  in the standard dipole radiation. They are a manifestation of the magnetoelectric effect induced by axion electrodynamics.  We illustrate our findings with some numerical estimations employing realistic media as well as some hypothetical choices in order to illuminate the effects of the magnetoelectric coupling which is usually very small.
\end{abstract}

\date{\today}

\maketitle

\section{Introduction}

\label{INTRO}

The  radiation produced by an electric dipole near a planar interface has been well studied over the years and has 
remained a relevant  subject of research 
for  physicists and engineers  due its relevance in a wide range of phenomena  like  practical  applications in radio communications \cite{Michalski-Mosig}, THz Zenneck wave propagation \cite{Jeon}, near-field optics \cite{Novotny JOSA}, plasmonics \cite{Maier} and nanophotonics \cite{Gordon}, just to mention a few examples. In 1909, Sommerfeld \cite{Sommerfeld 1} published a theory for a radiating  vertically oriented dipole above a planar and lossy ground which formed the basis for  subsequent investigations  \cite{Hoerschelmann, Sommerfeld 2, Weyl, Strutt,VdP N}.  Probably, by the fact that the early theory of dipole emission near planar interfaces was written in German, although there was an English version  summarized in Sommerfeld's lectures on theoretical physics \cite{Sommerfeld 3}, many aspects of the theory were reinvented and clarified over the years \cite{Banhos,Felsen-Marcuvitz,Brekhovskikh,Books}.

Here  we consider planar interfaces constructed  with linear homogeneous and isotropic  magnetoelectric  (ME)  media giving rise to the so called magnetoelectric effect (MEE), whereby electric (magnetic) fields are able to induce additional (polarization) magnetization (polarization)  in the material. This effect, which  was predicted \cite{DIZIA} and discovered \cite{ASTROV} in antiferromagnets, has been widely studied along the years
in multiferroic materials and it is codified in and additional parameter of the material: the magnetoelectric polarizability (MEP) \cite{FIEBIG}. The recent discovery of three-dimensional topological insulators (TIs) has boosted the interest in this topic by providing  new materials where this effect is predicted to occur \cite{Qi PRB,Wang,Morimoto,Essin,Mogi,Nomura}. Generally speaking, TIs  belong to a novel state of matter {in which the characterization of their quantum states does not fit into the standard paradigm of condensed matter physics whereby the phases of the material 
are classified  according to order parameters arising from the spontaneous symmetry breaking of the corresponding Hamiltonian according to the phenomenological Ginzburg-Landau theory. A distinguished example of this classification are normal superconductors, where gauge invariance is spontaneously broken. Instead, these states are classified according to topological invariants that arise in the Hilbert space generated by the corresponding Hamiltonians in the reciprocal space of the crystal lattice. They are}
protected by time-reversal-symmetry (TRS) and  admit an insulating bulk together with  { conducting surface (edge)  states.} The imposition of this symmetry yields 
two classes of materials: standard insulators labeled by a zero MEP and TIs characterized by a MEP  equal to $\pi$. {These new materials host a number of exceptional electromagnetic properties. Among them we have:  (i) they can carry currents along the edge channels without dissipation, (ii) their  MEP is quantized, (iii) the conducting edge states can be interpreted as quasi-particles being  massless Dirac fermions. This by itself is an important feature that makes contact with high energy physics and which  provides the opportunity to investigate the existence of unseen particles like  Majorana fermions, for example. (iv) they are predicted to exhibit the quantized photogalvanic effect in which light can induce a quantized current. For an extensive review of the properties of TIs see  for example \cite{Hasan,Qi Review,Franz}. All these new features provide additional motivation to reconsider the problem of radiation in magnetolectric media.}

The effective  theoretical framework to deal with magnetoelectric media  {is motivated by }  axion electrodynamics \cite{SIKIVIE}, which consists of adding the term 
{ $\mathcal{L}_{a}=g_{a \gamma \gamma}\,a(t, {\mathbf x})  F_{\mu\nu} {\tilde F}^{\mu\nu} $  to the  Maxwell Lagrangian density $\cal L_{\rm em}$,  plus a kinetic and a mass terms for the pseudoscalar field $a(t, {\mathbf x})$. The so called axion field $a(t, {\mathbf x})$ was introduced in Ref. \cite{PECCEI} to propose a solution for the strong CP problem  in strong interactions \cite{WEINBERG,Wilczek}. In the original formulation \cite{SIKIVIE}, the coupling constant $g_{a \gamma \gamma}$ arised from  a specific grand unification model  of the strong and electroweak interactions. Also, $F_{\mu\nu}$ is the electromagnetic tensor and ${\tilde F}^{\mu\nu}=\frac{1}{2}\epsilon^{\mu\nu\alpha\beta} F_{\alpha \beta}$ is the  dual tensor. The well known relation $F_{\mu\nu} {\tilde F}^{\mu\nu} =-4\, {\mathbf E}\cdot{\mathbf B}$ allows to rewrite ${\cal L}_a$
in terms of the electric $\mathbf E$ and magnetic $\mathbf B$
fields. As we will  show  in the following the  coupling ${\cal L}_a$ encapsulates the MEE which characterizes the electromagnetic response of the materials we consider in this work.  Thus we restrict ourselves to a non-dynamical axion field $a(t, {\mathbf x}) \rightarrow \vartheta({\mathbf x})$, to be called the magnetoelectric polarizability (MEP),  which we consider as an additional  electromagnetic  property of the medium, in the same footing as its permittivity and permeability \cite{Essin,King-Smith,Ortiz}. Following a standard convention we now consider the interaction term  ${\cal L}_\vartheta=-(\alpha/4 \pi^2)\, \vartheta({\mathbf x}){\mathbf E}\cdot{\mathbf B}$,  where  $\alpha$ is  the fine structure constant characteristic of the electromagnetic interaction between fermions and photons in the material, which produces this effective term.}
We call $\vartheta$-Electrodynamics ($\vartheta$-ED) this restriction of axion electrodynamics and our purpose is to study the radiation produced by a dipole oriented vertically with respect to the interface between two semi-infinite  planar magnetoelectric media,  having different  constant MEPs.  Excluding important differences in their microscopic structure, we will refer to the medium  as a magnetoelectric, or a $\vartheta$-medium, as long as its macroscopic electromagnetic response  can be described in the framework of the $\vartheta$-ED.

The  paper is organized as follows. In section \ref{THETAED}, we present a review of $\vartheta$-ED which also   contains a summary of the calculation of the  time-dependent Green's function (GF) for the 4-potential $A^{\mu}$  in our setup. 
As the source of the electromagnetic fields, 
in   section  \ref{DIPRAD1} we introduce an oscillating  vertically oriented point-like electric dipole located at  a distance $z_{0}>0$, on the $z$-axis perpendicular to the interface between the two media.  
The convolution of this source with the GF is carried out in  the subsection \ref{4POTENTIAL1} and yields the corresponding electromagnetic potentials  $A^{\mu}$ in terms of closed integrals which are  calculated in the  Appendix \ref{A}. The next step is to obtain  
 the far-field approximation  of those integrals. This is performed using a modified version of the steepest descent method, which is appropriate to the situation where the integrand  is not a smooth function in the vicinity of  the stationary phase due to the appearance of poles in  the steepest descent path at this point. 
 
This approximation,{which heavily relies upon Ref. \cite{Banhos} is explained in detail  and  carefully carried out in the Appendix \ref{B}. These results are  summarized in the subsection \ref{4POTENTIAL1}}. 

 As a consequence of the {presence of the pole} we find that the 4-potential {acquires a contribution from axially symmetric cylindrical  waves (denoted also as surface waves) besides the standard spherically symmetric ones.} A detailed analysis on the former kind of waves allows us to introduce  what we call the  discarding angle $\theta_0$,  which permits us to divide the space in two regions: $\mathcal{V}_1$ where the cylindrical wave contribution can be neglected and $\mathcal{V}_2$ where this contribution  has to be considered {within a certain range of parameters in what is called the intermediate zone in the literature \cite{Banhos}. To characterize the relevance of these cylindrical waves  we introduce a rapidly decreasing function measuring their amplitude and realize that for observation distances  further away from the  intermediate zone in the region ${\cal V}_2$ they turn out to be very much suppressed with respect to the spherical ones. This situation is quantitatively explained in    detail also in the subsection \ref{4POTENTIAL1}}. In subsection \ref{RADFIELD1} the far-field expressions for $\mathbf{E}$ and $\mathbf{B}$ are calculated for each  region. 
In section \ref{SECTIONIV}  we consider the angular distribution, the total radiated power and the energy transport of the dipolar radiation. 
In the subsection \ref{PARAM} we establish the numerical parameters to be used in the subsequent applications. 
Section  \ref{ANGDIST1} comprises a detailed examination of the angular distribution spectrum $dP/d\Omega$ in the  region $\mathcal{V}_1$.  In section \ref{POWER1}, we calculate the power radiated into the  region $\mathcal{V}_1$. Section \ref{ET1} is devoted to the energy transport of the  radiation in the region ${\cal V}_2$.  Here we also give some numerical estimations  considering the  media  already discussed in  Secs. \ref{ANGDIST1} and \ref{POWER1}, plus some additional hypothetical choices.  Finally, Sec. \ref{SUMM} provides a concluding summary and the conclusions from our results. In the Appendix A we derive the exact expressions for the potential $A_\mu$ required to calculate the electromagnetic fields. The  far-field approximation  is carried out in the Appendix B using a modified steepest descent method.  The final  Appendix C includes a brief review of the Faddeeva plasma dispersion function which arises in the discussion of the cylindrical waves. Throughout this paper we use Gaussian units with $\hbar=c=1$, the metric signature is  ($+,-,-,-$) and  $\varepsilon^{0123}=1$. We follow the conventions of Ref. \cite{Jackson}.

\section{$\vartheta $-Electrodynamics}
\label{THETAED}
Let us consider two semi-infinite $\vartheta$-media separated by a planar interface  located at $z=0$, filling the regions $\mathcal{U}_{1}, z<0$ and $\mathcal{U}_{2}, z>0$ of the space. We take both media to be non-magnetic, i.e. $\mu_1=\mu_2=1$ and with the same permittivity  $\epsilon_{1}=\epsilon _{2}=\epsilon$. This condition is motivated by the results of  Ref. \cite{Urrutia-Martin},  which show that  the effects of the MEE are substantially enhanced with respect to the optical contributions when both $\vartheta$-media have the same dielectric constant and permeability. 
Additionally we assume  the parameter $\vartheta$ to be  piecewise constant so that it takes the  values $\vartheta=\vartheta _{1}$ in the region $\mathcal{U}_{1} $ and $\vartheta=\vartheta _{2}$ in the region $\mathcal{U}_{2}$. This is expressed as 
\begin{equation}
\vartheta(\mathbf{x})=\Theta(z)\vartheta _{2}+\Theta(-z)\vartheta _{1}, 
\label{MEP}
\end{equation}
where $\Theta(z)$ is the standard Heaviside function with $\Theta(z)=1,\;\;z\geq 0\;$ and $\Theta(z)=0,\;\;z<0$.  {The dynamics is  governed by  the standard Maxwell equations in a material  medium
 \cite{Jackson,Schwinger}}
 {\beq
\nabla\cdot \mbf{D}=4 \pi \varrho, \qquad \nabla \times \mbf{H}-
\frac{\partial \mbf{D}}{\partial t}=4 \pi \mbf{J}, \qquad \nabla\cdot \mbf{B}=0, \quad  \nabla\cdot \mbf{E}=-\frac{\partial \mbf{B}}{\partial t}, 
\label{GENMAX}
 \eeq}
{which require to specify additional  constitutive relations characterizing the medium under consideration. In the case of the magnetoelectric media the constitutive relations are}  
\begin{equation}
\mathbf{D}=\epsilon \mathbf{E}-\frac{\alpha }{\pi }\vartheta(z)\mathbf{B},\quad 
\mathbf{H}=\mathbf{B}+\frac{\alpha }{\pi }\vartheta(z)\mathbf{E}.
\label{CREL}
\end{equation}
Here  $\alpha$ is the fine-structure constant, $\varrho \;$and $\mathbf{j}$ are the external sources given by the charge and current densities respectively. {Substituting  the constitutive relations (\ref{CREL}) into the inhomogeneous  equations (\ref{GENMAX}) and using the } MEP given in Eq.(\ref{MEP}) yields  our final    equations  
\begin{eqnarray}
\epsilon \nabla \cdot \mathbf{E} &=&4\pi\varrho +\tilde{\theta}\delta (z)%
\mathbf{B}\cdot \mathbf{\hat{u}}\;,  \label{Gauss E} \\
\nabla \times \mathbf{B}-\epsilon \frac{\partial \mathbf{E}}{\partial t}
&=&4\pi \mathbf{j}+\tilde{\theta}\delta (z)\mathbf{E}\times \mathbf{\hat{u}}%
\;.  \label{Ampere}
\end{eqnarray}%
where $\mathbf{\hat{u}}$ is the outward unit normal to the region $\mathcal{U}_{1}$ and 
\begin{equation}\label{TILDE THETA}
\tilde{\theta}=\alpha(\vartheta_{2}-\vartheta _{1})/\pi\;.
\end{equation}
In the case of a TI located in the region $\mathcal{U}_2$   (${\vartheta}_2=\pi$) in front of a regular insulator ($\vartheta=0$) in region $\mathcal{U}_1$,  we have 
$
{\tilde \theta}=\alpha(2\tilde{m}+1), 
$
with $\tilde{m}$ being an  integer depending on the details of the TRS breaking at the interface between the two materials. 

The homogeneous Maxwell equations  still  enable us to define the electromagnetic fields $\mathbf{E}$ and $\mathbf{B}$ in terms the electromagnetic potentials $\Phi$ and $\mathbf{A}$ as
\begin{equation}
\mathbf{E}=-\frac{\partial \mathbf{A}}{\partial t}-\nabla \Phi ,\;\;\;\;\;%
\mathbf{B}=\nabla\times\mathbf{A}, \quad    A^\mu=(\Phi, {\mathbf A}). \label{DEF_POT}
\end{equation}
{We observe that  Eqs. (\ref{Gauss E}) and (\ref{Ampere}), together with the constitutive relations (\ref{CREL}), can also be derived from the  action }
{\beq
S[\Phi, {\mathbf A}]=\int dt \, d^3 x \Big[ \frac{1}{8 \pi}\Big(\epsilon {\mathbf E}^2- {\mathbf B}^2 \Big)-\frac{\alpha}{4 \pi ^2} \vartheta({\mathbf x}) {\mathbf E}\cdot {\mathbf B} -\varrho \Phi+ {\mathbf J}\cdot {\mathbf A} \Big],
\label{ACTION}
\eeq}
{which clearly  incorporates the modified axion term
${\cal L}_\vartheta$ discussed in the Introduction. As usual, the electric and magnetic fields in (\ref{ACTION}) are written in terms of the potentials according to Eq. (\ref{DEF_POT}).}

The Eqs. (\ref{Gauss E}) and (\ref{Ampere}) explicitly show that there are no modifications to the dynamics in the bulk ($z\neq 0$) with respect to standard electrodynamics. {Nevertheless, as it is well known, the solution of  a system of differential equations depends crucially upon de boundary conditions. In this way, the new  physics  induced by $\mathcal{L}_{\vartheta}$ arises from the interface between the media ($z=0$) and will be a consequence of the boundary conditions there. Physically, this is a consequence that TIs behave as normal insulators in the bulk, but possess conducting properties at interfaces, as indicated by  the MEE. Even though we are dealing with a continuous dielectric, ($\epsilon_1=\epsilon_2$), the different MEP of both media  generate effective transmission and reflection coefficients for electromagnetic waves across the interface.  Mathematically, this feature is understood  because $\mathbf{E}\cdot\mathbf{B}$ in $\mathcal{L}_\vartheta$ is a total derivative, so the only allowed modifications to the standard Maxwell equations arise when the integration by parts produces $\partial_\alpha\vartheta\neq0 $, which precisely define the interface in our problem.}

Assuming that the time derivatives of the fields are finite in the vicinity
of the surface $z=0$, the modified Maxwell equations (\ref{Gauss E})
and (\ref{Ampere}) imply the following boundary conditions (BCs)
\begin{equation}
\epsilon \left[ \mathbf{E}_{z}\right] _{z=0^{-}}^{z=0^{+}}=\tilde{\theta}%
\mathbf{B}_{z}|_{z=0}, \quad \left[ \mathbf{B}%
_{\parallel }\right] _{z=0^{-}}^{z=0^{+}}=-\tilde{\theta}\mathbf{E}%
_{\parallel }|_{z=0}\;, \qquad  
\left[ \mathbf{B}_{z}\right] _{z=0^{-}}^{z=0^{+}}=0,  \qquad \left[ \mathbf{E}%
_{\parallel }\right] _{z=0^{-}}^{z=0^{+}}=0\;,  \label{Conditions 2}
\end{equation}%
for vanishing external sources on the surface $z=0$.
The notation is $\left[ \mathbf{V}\right] _{z=0^{-}}^{z=0^{+}}=\mathbf{V}%
(z=0^{+})-\mathbf{V}(z=0^{-})$, $\mathbf{V}\big|_{z=0}=\mathbf{V}(z=0)$,
where $z=0^{\pm }$ indicates the limits $z=0\pm \eta $, with$\;\eta
\rightarrow 0$, respectively. 
 The continuous terms in the right-hand side of the first and second equations in (\ref{Conditions 2}) represent self-induced surface
charge and surface current densities, respectively, which clearly demonstrate  the MEE localized just at the interface between the two media.

A convenient way to deal with the fields produced by arbitrary sources in electrodynamics, in particular in  $\vartheta$-ED,  is by using the corresponding GF $G_{\;\,\nu }^{\mu }(x,x^{\prime })$, which we  briefly revise  below  \cite{Urrutia-Martin-Cambiasso 2, Urrutia-Martin-Cambiasso 3, Urrutia-Martin-Cambiasso 4,Urrutia-Martin-Cambiasso 1,OJF-LFU-ORT-1}. {Before going into the details we comment upon the  advantages provided by the use of GF methods over different alternatives in electrodynamics: the knowledge of the GF of a given physical system allows a direct calculation of the corresponding electromagnetic fields for an  arbitrary sources either  analytically or numerically just by direct substitution. This clearly avoids the guesswork required when using  the image method, which by the way works only in highly symmetrical cases.  Also, it saves a lot of work when one needs to consider different sources in a given system by avoiding to solve the equations for each source. This very useful technique 
extends to many branches of physics like scattering theory, condensed matter physics  and quantum field theory, for example.}

In what follows we restrict ourselves to contributions of free sources $J^\mu=(\varrho, {\mathbf{j}})$ located outside the interface, and to systems without BCs imposed on additional surfaces, except for those at infinity.
A compact formulation of the problem is given in terms of the potentials (\ref{DEF_POT})  expressed in their four dimensional form $(\Phi,\mathbf{A})$ together with 
the GF  
\begin{equation}\label{A and GF}
A^{\mu }(x)=\int d^{4}x^{\prime }G_{\;\,\nu }^{\mu }(x,x^{\prime })J^{\nu
}(x^{\prime })\;,  
\end{equation}
which satisfies the equation 
 \begin{equation}
\left[\Diamond_{\;\;\nu }^{\mu }-\tilde{\theta}\delta
(z)\varepsilon _{\quad \;\nu }^{3\mu \alpha }\partial _{\alpha }\right]
G_{\;\;\sigma }^{\nu }(x,x^{\prime })=4\pi \eta _{\;\;\sigma }^{\mu }\delta
(x-x^{\prime })\;,  \label{Eq GF temporal}
\end{equation}%
in the modified Lorenz gauge
$
\epsilon {\partial \Phi }/{\partial t}+\nabla\cdot\mathbf{A}=0$, 
together with the appropriate BCs. The operator  
$\Diamond_{\;\;\nu }^{\mu }$ is
\begin{equation}
\Diamond_{\;\;\nu }^{\mu }=\left(\epsilon \Box ^{2},\;\Box
^{2}\delta _{\;j}^{i}\right),\;\;\Box ^{2}=\epsilon \partial _{t}^{2}-\nabla ^{2}\;.
\end{equation}
The detailed calculation of the  GF is reported in Sec. III of Ref. \cite{OJF-LFU-ORT-1}. Here we only recall  the results that are written  in terms of $\bar{G}^{\mu}_{\;\;\nu}$, which differs from $G^{\mu}_{\;\;\nu}$ only  in the term $G^{0}_{\;\;0}=\bar{G}^{0}_{\;\;0}/\epsilon$. Since the GF is  time-translational-invariant it is convenient to introduce the corresponding Fourier transform such that 
\begin{equation}
\bar{G}_{\nu }^{\mu }(\mathbf{x},\mathbf{x}^{\prime },t-t')=\int_{-\infty}^{\infty} \frac{d \omega }{2 \pi}e^{-i \omega (t-t')} \, 
\bar{G}_{\nu }^{\mu }(\mathbf{x},\mathbf{x}^{\prime }; \omega)\;.
\end{equation} 
The final result is presented as the sum of three terms, $\bar{G}_{\;\;\nu }^{\mu }(\mathbf{x},\mathbf{x}^{\prime };\omega )=\bar{G}_{ED\;\nu}^{\mu }(\mathbf{x},\mathbf{x}^{\prime };\omega )+\bar{G}_{\tilde{\theta}\;\nu }^{\mu
}(\mathbf{x},\mathbf{x}^{\prime };\omega )+\bar{G}_{\tilde{\theta}^{2}\;\nu }^{\mu}(\mathbf{x},\mathbf{x}^{\prime };\omega )$, whose explicitly form is 
\begin{eqnarray}
\bar{G}_{ED\;\nu }^{\mu }(\mathbf{x},\mathbf{x}^{\prime };\omega )&=&\eta
_{\;\;\nu }^{\mu }4\pi \int \frac{d^{2}\mathbf{k_{\perp }}}{(2\pi )^{2}}e^{i%
\mathbf{k_{\perp }}\cdot \mathbf{R}_{\perp }}\frac{ie^{i\sqrt{\tilde{k}%
_{0}^{2}-\mathbf{k}_{\perp }^{2}}|z-z^{\prime }|}}{2\sqrt{\tilde{k}_{0}^{2}-%
\mathbf{k}_{\perp }^{2}}},  \nonumber \\
\bar{G}_{\tilde{\theta}\;\nu }^{\mu }(\mathbf{x},\mathbf{x}^{\prime };\omega
) &=&i\varepsilon _{\;\;\nu }^{\mu \;\;\alpha 3}\frac{4\pi \tilde{\theta}}{4n^{2}+%
\tilde{\theta}^{2}} \int \frac{d^{2}\mathbf{k_{\perp }}}{(2\pi )^{2}}
e^{i\mathbf{k_{\perp }}\cdot \mathbf{R}_{\perp }}k_{\alpha }\frac{e^{i\sqrt{\tilde{k}_{0}^{2}-\mathbf{k}_{\perp }^{2}}(|z|+|z^{\prime }|)}}{\tilde{k}%
_{0}^{2}-\mathbf{k}_{\perp }^{2}}\;,  \label{GF theta}\\
\bar{G}_{\tilde{\theta}^{2}\;\nu }^{\mu }(\mathbf{x},\mathbf{x}^{\prime
};\omega)&=&\frac{i4\pi\tilde{\theta}^{2}}{4n^{2}+\tilde{\theta}^{2}} \int \frac{%
d^{2}\mathbf{k_{\perp }}}{(2\pi )^{2}}\left[ k^{\mu }k_{\nu }-\left( \eta
_{\;\;\nu }^{\mu }+n^{\mu }n_{\nu }\right)k^{2}\right]
\, e^{i\mathbf{k_{\perp}}\cdot \mathbf{R}_{\perp }}
\frac{e^{i\sqrt{\tilde{k}_{0}^{2}-\mathbf{k}_{\perp }^{2}}(|z|+|z^{\prime }|)}}
{2\left( \tilde{k}_{0}^{2}-\mathbf{k}_{\perp }^{2}\right) ^{3/2}}\;.  \notag
\end{eqnarray}
Here $\mathbf{R}_{\perp }=(\mathbf{x}-\mathbf{x^{\prime }})_{\perp}=(x-x^{\prime },y-y^{\prime })$ and $\mathbf{k_{\perp}}=(k_{x},k_{y})$ is the momentum parallel to the interface, $k^{\alpha}=(\omega,\mathbf{k_{\perp }})$ and $\tilde{k}_{0}=n\omega$ where $n=\sqrt{\epsilon}$ is the refraction index.
We observe that in  the static limit ($\omega=0$), the result (\ref{GF theta}) reduces to the one reported in Ref. \cite{Urrutia-Martin-Cambiasso 2}.


\section{Electric Dipole perpendicular to the interface  }

\label{DIPRAD1}

In this section we  determine the electric field of an oscillating  point-like electric dipole ${\mathbf p}=p \, \mathbf{\hat{z}} $ located at a distance $z_{0} > 0 $ on the $z$-axis and perpendicular to the interface. We restrict ourselves to the far-field approximation (${\tilde k}_0 r\gg 1$) starting from the GF given by Eqs. (\ref{GF theta}).

\subsection{The Electromagnetic Potential $A^\mu$}\label{4POTENTIAL1}
 The charge and current density for this dipole are 
\begin{eqnarray}
\varrho(\mathbf{x}^{\prime};\omega)&=&-p\delta(x^{\prime})\delta(y^{\prime})%
\delta^{\prime}(z^{\prime}-z_{0}), \qquad \mathbf{j}(\mathbf{x}^{\prime};\omega)=-i\omega p\delta(x^{\prime})%
\delta(y^{\prime})\delta(z^{\prime}-z_{0})\mathbf{\hat{z}},
 \label{j mu elec dip}
\end{eqnarray}
respectively, where $\delta^{\prime}(u)= {d \delta(u)}/{du}$.
After convoluting the  sources (\ref{j mu elec dip}) with the GF (\ref{GF theta}) we find the following components of $A^\mu$
\begin{eqnarray}
A^{0}(\mathbf{x};\omega)&=&-\frac{p}{n^2}i\tilde{k}_{0}\cos\theta\frac{e^{i\tilde{k}_{0}(r-z_0\cos\theta)}}{r}-\frac{1}{n^2}\frac{\tilde{\theta}^{2}p}{4n^{2}+\tilde{\theta}^{2}}\mathcal{H}(\mathbf{x},z_0;\omega)\;,\label{A0 e-d almost}\\
 {A^{a}(\mathbf{x};\omega)}&=&-\frac{2\tilde{\theta}p}{4n^{2}+\tilde{\theta}^{2}}\frac{i \varepsilon^{ab}x^b}{\rho}\frac{\partial}{\partial\rho}\mathcal{I}(\mathbf{x},z_0;\omega)+\frac{\tilde{\theta}^{2}p}{4n^{2}+\tilde{\theta}^{2}}\frac{i\omega x^a}{\rho}\frac{\partial}{\partial\rho}\mathcal{J}(\mathbf{x},z_0;\omega)\;,\label{A1 e-d almost}\\
A^{3}(\mathbf{x};\omega)&=&-i\omega p\frac{e^{i\tilde{k}_{0}(r-z_0\cos\theta)}}{r}\;,\label{A3 e-d almost}
\end{eqnarray}
where $\rho=\|\mathbf{x}_\perp\|=\sqrt{x^2+y^2}$, $r=\sqrt{x^2+y^2+z^2}$,  $ \, a,b \in \{1,2\}$,  $\, \varepsilon^{ab}=-\varepsilon^{ba}, \,  \varepsilon^{12} = +1$,  and $\{ x^a \}= \{ x^1, x^2 \}$  with $x^1=x, \,  x^2=y$. 
We also have the functions
\begin{eqnarray}
\mathcal{H}(\mathbf{x},z_0;\omega)&=&\int_{0}^{\infty}\frac{k_{\perp}^{3}dk_{\perp}}{\tilde{k}_{0}^{2}-k_{\perp}^{2}}J_{0}\left(k_{\perp}\rho\right)e^{i\sqrt{\tilde{k}_{0}^{2}-k_{\perp}^{2}}(|z|+z_0)}\;,\label{integral H}\\
\mathcal{I}(\mathbf{x},z_0;\omega)&=&\int_{0}^{\infty}\frac{k_{\perp}dk_{\perp}}{\sqrt{\tilde{k}_{0}^{2}-k_{\perp}^{2}}}J_{0}\left(k_{\perp}\rho\right)e^{i\sqrt{\tilde{k}_{0}^{2}-k_{\perp}^{2}}(|z|+z_0)}\;,\label{integral I}\\
\mathcal{J}(\mathbf{x},z_0;\omega)&=&\int_{0}^{\infty}\frac{k_{\perp}dk_{\perp}}{\tilde{k}_{0}^{2}-k_{\perp}^{2}}J_{0}\left(k_{\perp}\rho\right)e^{i\sqrt{\tilde{k}_{0}^{2}-k_{\perp}^{2}}(|z|+z_0)}\;.\label{integral J}
\end{eqnarray}
The derivation of the above results can be found in the Appendix \ref{A}. 

The next step is to calculate 
the integrals (\ref{integral H})-(\ref{integral J}) in the far-field approximation. {To begin with, we recap the main  ingredients of the calculation carried out in full detail for the function ${\cal H}$  in  Appendix \ref{B}.} We   employ  the steepest descent method \cite{Chew, Mandel,Chew art} and  incorporate some   modifications  based on Refs. \cite{Banhos} and  \cite{Wait}. These modifications are required because  the current integrals (\ref{integral H})-(\ref{integral J}) have poles coinciding with their  stationary point $(k_\perp)_s=\tilde{k}_{0}\sin\theta$ at $\theta=\pi/2$, as can be seen in   Eq. (\ref{u0}) of the Appendix \ref{B}. This means that the factor of the exponential is not a smooth function around the stationary point now, which will prevent  a direct application of the method.   The main idea to overcome this difficulty  is to subtract and add the conflicting pole, as shown in 
 Eqs. (\ref{psi de u}) and (\ref{zeta de u}) of the Appendix \ref{B}. Thereby, we obtain two integrals: one  with the divergence removed  in the vicinity of the stationary point and   another containing the singularity, which can be directly evaluated. The first integral leads to the ordinary stationary phase contributions and the second one gives contributions that are identified as axially symmetric cylindrical  waves \cite{Banhos}. The final results for the integrals $\mathcal{H}, \mathcal{I}$  and $\mathcal{J}$, obtained in full detail in the Appendix \ref{B}, are 
\begin{eqnarray}
\mathcal{H}(\mathbf{x},z_0;\omega)&=&\tilde{k}_0\frac{e^{i\tilde{k}_0r}}{ir}\left\{\frac{\sin^2\theta e^{i\tilde{k}_0z_0|\cos\theta|}}{|\cos\theta|}-\frac{1}{\sqrt{2\left(\sin\theta-\sin^2\theta\right)}}\right\}\nonumber\\
&&+\sqrt{\frac{2}{\pi i\tilde{k}_0r\sin\theta}}\frac{\tilde{k}_0^2}{i}e^{i\tilde{k}_0r\sin\theta}\frac{\pi}{2}\mathrm{erfc}\left[-i\sqrt{i\tilde{k}_0r\left(1-\sin\theta\right)}\right]\;,\label{H final}
\end{eqnarray}
\begin{eqnarray}
\mathcal{J}(\mathbf{x},z_0;\omega)&=&\frac{e^{i\tilde{k}_0r}}{i\tilde{k}_0r}\left\{\frac{e^{i\tilde{k}_0z_0|\cos\theta|}}{|\cos\theta|}-\frac{1}{\sqrt{2\left(\sin\theta-\sin^2\theta\right)}}\right\}\nonumber\\
&&+\sqrt{\frac{2}{\pi i\tilde{k}_0r\sin\theta}}\frac{\pi e^{i\tilde{k}_0r\sin\theta}}{2i}\mathrm{erfc}\left[-i\sqrt{i\tilde{k}_0r\left(1-\sin\theta\right)}\right]\;,\label{J final}
\end{eqnarray}
\begin{eqnarray}
\mathcal{I}(\mathbf{x},z_0;\omega)&=&\frac{e^{i\tilde{k}_0r}}{ir}e^{i\tilde{k}_0z_0|\cos\theta|}\;,\label{I final}
\end{eqnarray}
where $\mathrm{erfc}(z)$ denotes the complementary error function.
The contributions in curly brackets arise from the terms including the subtracted pole, they are 
are well behaved at $\theta=0,\pi/2,\pi$ {and describe the standard spherical waves. }

{On the other hand, the terms from the erfc function in square brackets arise from the pole itself and describe wave propagation corresponding to the amplitude $1/{(r\sin \theta)}^{1/2}\exp[i{\tilde k}_0( r \sin \theta)] \times \exp{(-i{\tilde k}
_0 t)} = 1/{z^{1/2}} \, 
\exp[i{\tilde k}_0 z] \times \exp{(-i{\tilde k}
_0 t)}$. This is clearly a solution of the Helmholtz equation in cylindrical coordinates in the far-zone, thus giving the name of cylindrical waves to this contribution.} 

{The crucial point is that the amplitude of the cylindrical waves  is modulated by the $\mathrm{erfc}$ function,}  
with  argument 
\begin{equation}
i\sqrt{i}\sqrt{\tilde{k}_0r\left(1-\sin\theta\right)}=i \Lambda\equiv i \sqrt{i}\, s,
\label{DEFs}
\end{equation}
{which will provide us with a quantitative way of appraising the relevance of the cylindrical waves. To this end, we now discuss the behavior of the function}
{ \beq
\mathrm{erfc}{(i \Lambda)}=\frac{1}{i \sqrt{\pi}}e^{i{\tilde k}_0 r(1-\sin\theta)}Z(e^{i\pi/4}s), 
\eeq}
{that we rewrite in terms of the  the Faddeeva plasma dispersion function $Z(\Lambda)$ discussed in some detail in the Appendix C.
Up to an irrelevant normalization constant, we define the function}
{
\beq
 F(s)=|Z(e^{i\pi/4} s)|^2/(2\pi)\;,
 \eeq
 }
{ which controls the {amplitude} of  the electromagnetic fields of the cylindrical waves and    can be readily calculated from Eq. (\ref{MODSQUARE})}.
{ As can be seen from the following numerical values  $F(0)=0.5$,  $\, F(1)= 0.112$,  $\, F(3)= 0.0174$,  $\dots$  and $ F(s\gg 1)\rightarrow 0 $, $F(s)$  is a rapidly decreasing function of $s$. In this way, when $s\gg 1$ the cylindrical  wave contribution can be neglected, whereas for $s\rightarrow0$ the  function $F(s)$ contributes maximally and the cylindrical waves should be taken into account.} 

If we recall that  $\rho=r\sin\theta$, $s^2$ can be rewritten as $s^2=\tilde{k}_0(r-\rho)$ and  can be interpreted as a measure of how far  the observer with coordinates $(r,\theta,\phi)$ is  from the interface. In other words, $s^2$ determines how far is the spherical radius $r$ from the cylindrical radius $\rho$ at the observation point. This property enables us to define  what we call the  discarding angle   $\theta_0$, which provides  a condition  to estimate when   we can neglect  or not the cylindrical  wave contribution, according to the magnitude of $F(s)$. {We proceed as follows.}

{For a given observation distance $r_0$, such that ${\tilde k}_0 r_0$ is large enough to describe the far-field regime, we choose 
as an  arbitrary cutoff point the value  $s=s_0$.  At the cutoff we define the discarding angle $\theta_0$ such that    }
\beq
s_0=\sqrt{{\tilde k}_0 r_0 (1-\sin \theta_0)}
\label{cutoff} 
\eeq

More precisely, we can distinguish the upper hemisphere (UH) 
  $\theta\in[0,\pi/2]$ from the lower hemisphere (LH) $\theta\in[\pi/2,\pi]$ by writing
\begin{equation}
\label{theta0 UH}
\theta_0^{UH}=\arcsin\left(1-\frac{s_0^2}{\tilde{k}_0r_0}\right), \qquad \theta_0^{LH}=\frac{\pi}{2}+\arccos\left(1-\frac{s_0^2}{\tilde{k}_0r_0}\right),
\end{equation}
respectively. Let us observe that both discarding angles are  very close to the interface ($\theta=\pi/2$) in the far-field regime.

{For a given observation distance $r_0$,  the rapidly decreasing  behavior of $F(s)$ allows us to adopt the following criterion for estimating the relative  weight of the spherical versus the  cylindrical waves:} 
 {\begin{eqnarray} 
 && {\rm For }  \,\,  s > s_0, \,\,  (0 < \theta <\theta^{UH}_0 \,\,{\rm  and} \,\, \theta^{LH}_0 < \theta < \pi) \,\,  {\rm  \,\, cylindrical \,\,  waves \,\, are \,\, 
 neglected.} \nonumber \\ 
   && {\rm For }  \,\,  s < s_0, \,\,  (\theta^{UH}_0 < \theta <\theta^{LH}_0) \,\, {\rm  \,\, cylindrical \,\,  waves \,\, are \,\, taken \,\, into \,\, account}.
   \label{criterion}
\end{eqnarray}}
{In our case we take $s_0=1$, where the function $F(s)$ has decreased about five times with respect to its maximum at $s=0$. Let us emphasize that $s_0$ can be arbitrarily chosen much larger than one, which will provide a more stringent cutoff.}

\begin{figure}[H]
\centering
\includegraphics[width=0.3\textwidth]{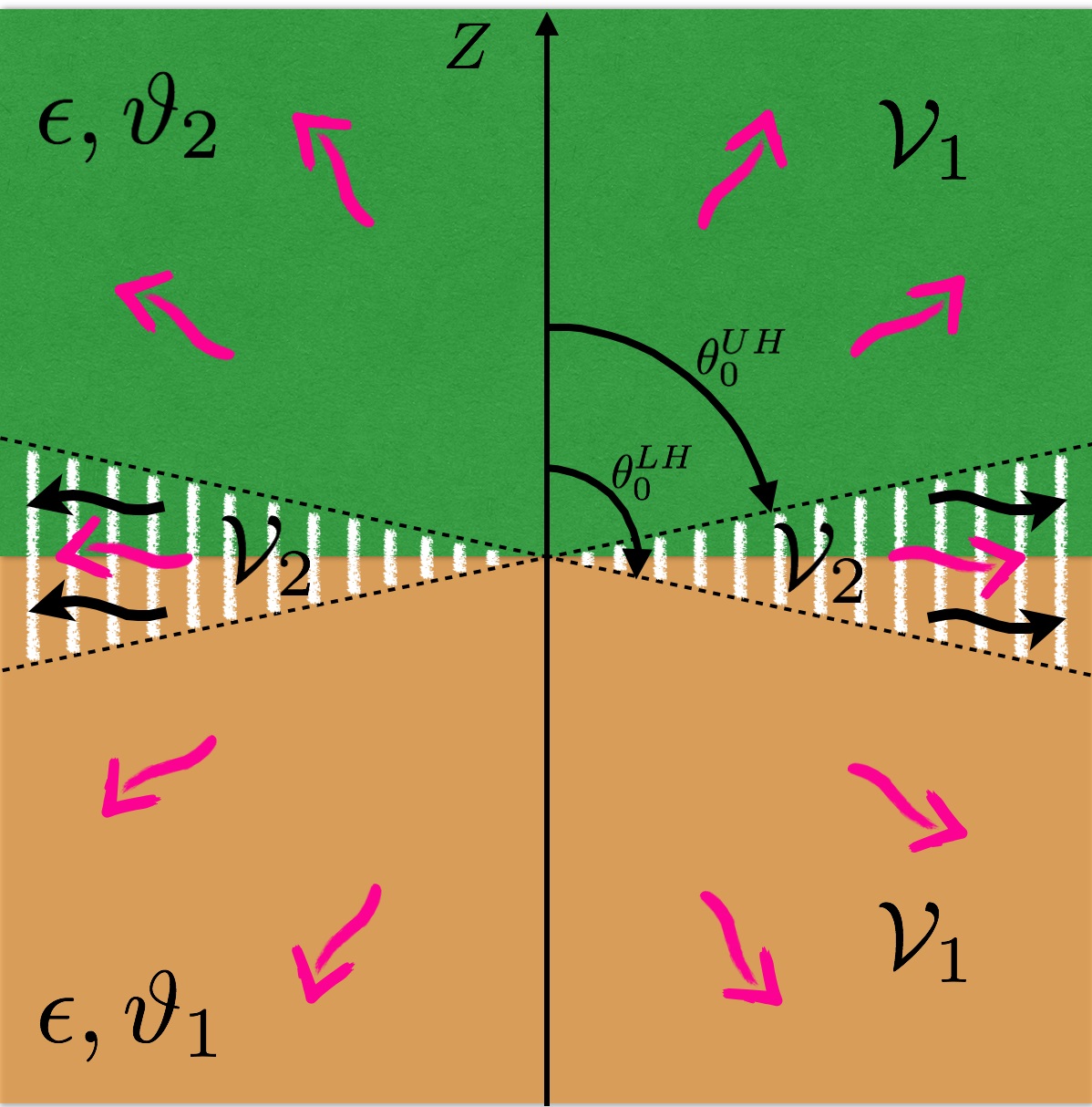} 
\caption{Diagram showing the regions $\mathcal{V}_1$ and $\mathcal{V}_2$ determined through the discarding angles $\theta_0^{UH}$  and $\theta_0^{LH}$. The  region $\mathcal{V}_1$, where only the spherical waves (undulated pink arrows) are taken  into account is defined by $\theta\in[0,\theta_0^{UH})\cup(\theta_0^{LH},\pi]$. {The  region $\mathcal{V}_2$ (white hatched region), defined by $\theta\in[\theta_0^{UH},\theta_0^{LH}]$, contains the intermediate region
($s< s_0=1$) where both  cylindrical waves (undulated black waves) and spherical waves  have to be taken into account. Nevertheless, going further into the radiation zone for each observation point in the intermediate region one can make $s>s_0=1 $, thus making the cylindrical waves unobservable.}}
\label{RAD REGIONS}
\end{figure}

{In other words, for a fixed $r_0$ and a chosen $s_0=1$, 
the discarding angles $\theta_0$ define two  regions, as shown in  Fig. \ref{RAD REGIONS}.  The region   $\mathcal{V}_1$, where $\theta\in[0,\theta_0^{UH})\cup(\theta_0^{LH},\pi]$, is such that $s>1$, while in its complement, the region  $\mathcal{V}_2$ where $\theta\in[\theta_0^{UH}, \theta_0^{LH}]$, we have $s < 1$.}

{To fix ideas let us consider now the UH and examine what happens when we fix the angle  and explore the consequences changing the observation distance to a larger  value  $r_>  > r_0$, i.e., we go  farther into the radiation zone. Suppose that in  the region ${\cal V}_1$ we consider the angle $\theta_1 < \theta_0^{UH}$, where we  have $s_1 =\sqrt{{\tilde k}_0 r_0(1-\sin\theta_1)}> 1 $ according to our choices. Then, keeping $\theta_1$ fixed and going to a larger distance $r_> > r_0$ would only increase the value of $s_>=\sqrt{{\tilde k}_0 r_>(1-\sin\theta_1)}$ such that $s_> > s_1 > s_0=1$.  That is to say, all observation points in ${\cal V}_1$ with $r>r_0$ will have $s>s_0=1$ and the cylidrical waves  will not be relevant there. } 

{On the other hand, the region ${\cal V}_2$  shows a mixed behavior. Again, let us consider an angle $\theta_2 > \theta_0^{UH}$ where we have $s_2 =\sqrt{{\tilde k}_0 r_0(1-\sin\theta_2)} < 1$ by construction. Nevertheless, an increase in the observation distance to $r_> $ can revert the situation yielding  a value $s_> >1$. To this end it is enough to take $r_> > r_0/s_2$. That is to say, the region ${\cal V}_2$ contains the intermediate region where the cylindrical waves are relevant, but going further into the radiation zone these waves can be safely neglected. A similar situation occurs in the LH, which we do not discuss in detail here.}

From the function erfc$\left[-i\sqrt{i{\tilde k}_0 r(1-\sin\theta)}\right]$ appearing in the Eqs. (\ref{H final}) and (\ref{J final}) we identify the analogous of the Sommerfeld numerical distance $ S$ in the standard dipole radiation, which is determined    by rewriting the complementary error function as erfc$(-i \sqrt{S})$ \cite{Banhos}. Thus we have    
\beq
S=i{\tilde k}_0 r(1-\sin\theta)=i s^2,
\eeq 
which varies with the angle $\theta$ for a fixed observation distance $r$. This quantity is closely related to the discarding angle $\theta_0$ according to
\beq
|S|=\frac{1-\sin\theta}{1-\sin \theta_0}.
\eeq 
Going back to the calculation of the electromagnetic field in the far-field approximation, we  now write the corresponding expressions for the  electromagnetic potentials obtained from plugging the results (\ref{H final}-\ref{I final}) into the equations (\ref{A0 e-d almost}-\ref{A3 e-d almost}).   
In the case of the region ${\cal V}_1$ we have
\begin{eqnarray}
A^{0}(\mathbf{x};\omega)&=&-\frac{p}{n^2}i\tilde{k}_{0}\cos\theta\frac{e^{i\tilde{k}_{0}(r-z_0\cos\theta)}}{r}
-\frac{1}{n^2}\frac{\tilde{\theta}^{2}p}{4n^{2}+\tilde{\theta}^{2}}\tilde{k}_{0}\frac{\sin^2\theta}{|\cos\theta|}\frac{e^{i\tilde{k}_{0}(r+z_0|\cos\theta|)}}{ir}\;,\label{A0 e-d V1}\\
A^{a}(\mathbf{x};\omega)&=&\frac{p}{4n^2+{\tilde \theta}^2}\left[-2\tilde{\theta}i\tilde{k}_{0}\left(\frac{\varepsilon^{ab} x^b}{r}\right)+\tilde{\theta}^{2}\frac{i\omega}{|\cos\theta|}\left(\frac{x^a}{r}\right)\right]\frac{e^{i\tilde{k}_{0}(r+z_0|\cos\theta|)}}{r}\;,\label{A1 e-d V1}\\
A^{3}(\mathbf{x};\omega)&=&-i\omega p\frac{e^{i\tilde{k}_{0}(r-z_0\cos\theta)}}{r}, \label{A3 e-d V1}
\end{eqnarray}
where we dropped terms of higher order.
On the other hand, in the { intermediate region $\mathcal{V}_2$, ($s<1$) } the contribution of the cylindrical waves near the interface  is apparent.  The electromagnetic potential now is 
\begin{eqnarray}
A^{0}(\mathbf{x};\omega)&=&\mp\frac{p}{n^2}i\tilde{k}_{0}\xi\frac{e^{i\tilde{k}_{0}(r\mp z_0\xi)}}{r}-\frac{1}{n^2}\frac{\tilde{\theta}^{2}p}{4n^{2}+\tilde{\theta}^{2}}\tilde{k}_{0}\frac{e^{i\tilde{k}_{0}r}}{ir}\left(i\tilde{k}_0z_0-\frac{\xi}{8}\right) \nonumber \\
&&-\frac{1}{n^2}\frac{\tilde{\theta}^{2}p}{4n^{2}+\tilde{\theta}^{2}}\tilde{k}_{0}^2
\sqrt{\frac{2}{\pi i\tilde{k}_0r}}e^{i\tilde{k}_0r}\left(\frac{\pi}{2i}+\sqrt{\frac{\pi i\tilde{k}_0r}{2}}\xi\right)\;,\label{A0 e-d V2}\\
A^{a}(\mathbf{x};\omega)&=&\frac{p}{4n^2 +{\tilde \theta}^2}\left[-2 i\tilde{\theta}\tilde{k}_{0}\frac{\varepsilon^{ab}x^b}{\rho}\frac{e^{i\tilde{k}_{0}(r+z_0\xi)}}{r}+\tilde{\theta}^{2}i\omega \frac{x^a}{\rho}\frac{e^{i\tilde{k}_{0}r}}{r}\left(i\tilde{k}_0z_0-\frac{\xi}{8}\right) \right.\nonumber\\
&&\left. -\tilde{\theta}^{2}\omega\tilde{k}_0 \frac{x^a}{\rho}\sqrt{\frac{2}{\pi i\tilde{k}_0r}}e^{i\tilde{k}_0r}\left(\frac{\pi}{2i}+\sqrt{\frac{\pi i\tilde{k}_0r}{2}}\xi\right)\right]\;,\label{A1 e-d V2}\\
A^{3}(\mathbf{x};\omega)&=&-i\omega p\frac{e^{i\tilde{k}_{0}(r\mp z_0\xi)}}{r},\label{A3 e-d V2}
\end{eqnarray}
where we again dropped terms of higher order.
The minus (plus) sign in the first term of the  right-hand side in Eq. 
(\ref{A0 e-d V2})  corresponds to the UH (LH),  respectively.  We have also  performed a first order power expansion around $\pi/2$ in the complementary error function arising from Eqs. (\ref{H final}) and (\ref{J final}) in terms of the variable 
$\xi < 1 $ given by 
\begin{equation}
 {\xi}^{UH}\equiv \pi/2 - {\theta}^{UH}, \qquad  
   {\xi}^{LH} = {\theta}^{LH}+ \pi/2,
 \label{DEFXI}
\end{equation} 
for the UH and LH, respectively.  Let us observe that for both hemispheres we have ${\xi} >0$  and also that $(1-\sin \theta)=\xi^2/2$. In analogous way it is convenient to introduce  the corresponding variable $\xi_0$ related to the  discarding angle $\theta_0$ just by replacing $\xi \rightarrow \xi_0$ and $\theta \rightarrow \theta_0$ in Eq. (\ref{DEFXI}). Using also  Eq. (\ref{theta0 UH}) we obtain  {{ $\xi_0=\sqrt{2/(n \omega r_0)}$.} In this way  
$|S|=\xi^2/\xi_0^2  < 1 $ for both hemispheres. The expansion in powers of $\xi$ is only valid {in the intermediate zone where we have  $\xi \ll \xi_0$. }

\subsection{The Electric  Field}
\label{RADFIELD1}
Since Faraday law  yields   $\mathbf{B}=\sqrt{\epsilon} \, {\mathbf {\hat n}}\times {\mathbf E}$ in the  far-field approximation we need to calculate only the electric field $\mathbf{E}(\mathbf{x};\omega)$ to get a complete description of the radiation regime, where  ${\mathbf {\hat n}}\cdot {\mathbf B=0}$ is satisfied. The components of the electric field are calculated through 
\begin{equation}\label{E and Amu}
\mathbf{E}(\mathbf{x};\omega)=-i\tilde{k}_0\mathbf{\hat{n}}A^{0}(\mathbf{x};\omega)+i\omega\mathbf{A}(\mathbf{x};\omega)\;.
\end{equation}
\subsubsection{The  region $\mathcal{V}_1$}\label{RADFIELD1_PERP_V1}

Substituting in Eq. (\ref{E and Amu}) the previous expressions for  $A^{\mu}(\mathbf{x};\omega)$ in  the region $\mathcal{V}_1$ we obtain the following electric field:
 \begin{eqnarray}
E^{a}(\mathbf{x};\omega)&=&\left[- f_{\perp}(\theta,z_0,\omega)\frac{z x^a}{r^2}
+\frac{2\tilde{\theta} n}{4n^{2}+\tilde{\theta}^{2}} e^{i\tilde{k}_{0}z_{0}|\cos\theta|} \frac{\varepsilon^{ab} x^b}{r} \right] \omega^2  p\frac{ e^{i\tilde{k}_{0} r}}{r},\nonumber  \\
E^{3}(\mathbf{x};\omega)&=&\left[ \frac{}{} \sin^{2}\theta \, f_{\perp}(\theta,z_0,\omega) \right] \omega^{2} p \frac{e^{i\tilde{k}_{0}r}}{r},
\label{E e-d perp V1}
\end{eqnarray}
with 
\begin{equation}\label{f perp}
f_{{\perp}}(\theta,z_0,\omega)=e^{-i\tilde{k}_{0} z_{0}\cos\theta}+\mathrm{sgn}\left(\cos\theta\right)\frac{\tilde{\theta}^{2}}{4n^{2}+\tilde{\theta}^{2}}e^{i\tilde{k}_{0}z_{0}|\cos\theta|}.
\end{equation}
Here $\mathrm{sgn}$ denotes the sign function with the additional condition $\mathrm{sgn}(0)=0$. It is possible to verify that $\mathbf{\hat{n}}\cdot\mathbf{E}=0$ as required for the electric field in the far-zone regime. 
The  main feature in the components of the electric field   (\ref{E e-d perp V1})  is the presence of  two different phases in the exponential related to the source variables ${\mathbf x}'=z_0\,\mathbf{\hat{z}}$, which are specified by $\cos\theta$ and $|\cos\theta|$ as shown in Eq. (\ref{f perp}). The first exponential contributes with the term $\exp\left[i\tilde{k}_{0}\left(r-z_{0}\cos\theta\right)\right]$ having the characteristic phase of dipole radiation in standard electrodynamics \cite{Jackson, Schwinger}. On the other hand, the contributions arising from  the new terms involving the  MEP, which are  proportional to $\tilde \theta$ and $\tilde \theta^2$,  yield  the exponential $\exp\left[i\tilde{k}_{0}\left(r+z_{0}|\cos\theta|\right)\right]$.
  As we will show in the next subsection, the modifications in the power spectrum of the dipolar radiation in our setting  arise precisely due to the contribution $z_0|\cos\theta|$ in the phase of the electric field.  
The dependence on the sign of $\cos\theta$ enforces two cases, which we denote as Case $(-)$ and Case $(+)$. The former case occurs when $|\cos\theta|=-\cos\theta$ , i.e. when $\theta\in(\theta_0^{LH},\pi]$  is in the LH. In this situation  the three components of the electric field  will have the same phase and  we do not expect 
significant  changes with respect to the usual angular   dependence  of the  dipolar radiation because the phase of the electric field  is that of standard electrodynamics. By contrast, the Case $(+)$ takes place when $|\cos\theta|=\cos\theta$,  which is realized  for $\theta\in[0,\theta_0^{UH})$ in the UH. In this case the electric field presents  two different phases which will  interfere  yielding  new effects different  from  those in the usual dipolar radiation. 

Finally, we analyze the function $f_\perp(\theta,z_0,\omega)$ given by Eq. (\ref{f perp}), which codifies the  different phases of the electric field (\ref{E e-d perp V1}). For the Case $(+)$, $f_\perp$ takes the following form
\begin{equation}\label{f perp +}
f_\perp^+(\theta,z_0,\omega)=f_\perp(\theta,z_0,\omega)|_{UH}=e^{-i\tilde{k}_{0} z_{0}\cos\theta}+\frac{\tilde{\theta}^{2}}{4n^{2}+\tilde{\theta}^{2}}e^{i\tilde{k}_{0}z_{0}\cos\theta}.
\end{equation}
On the other hand, for the Case $(-)$,  the function $f_\perp$ is
\begin{eqnarray}
f_\perp^-(\theta,z_0,\omega)&=&f_\perp(\theta,z_0,\omega)|_{LH}= \frac{4n^{2}}{4n^{2}+\tilde{\theta}^{2}}e^{-i\tilde{k}_{0}z_{0}\cos\theta}.
\label{f perp -}
\end{eqnarray}
In the following we show that the factors $\tilde{\theta}^{2}/(4n^{2}+\tilde{\theta}^{2}), \,\, 4n^{2}/(4n^{2}+\tilde{\theta}^{2})$  and  $2\tilde{\theta}/(4n^{2}+\tilde{\theta}^{2})$  correspond to some reflection and transmission coefficients at the interface. Let us start with  the radiation fields in the UH by considering the general expression for the reflected electric field discussed in the  Appendix B of  Ref.  \cite{Crosse-Fuchs-Buhmann}, where the authors calculate  the GF of a planar interface separating two semi-infinite  TIs. The reflective part of such GF is written as
\begin{equation}\label{GF Buhmann}
G^{ij}(\mathbf{x},\mathbf{x}';\omega)=\int\frac{d^2\mathbf{k}_\perp}{(2\pi)^2}e^{i\mathbf{k_{\perp }}\cdot \mathbf{R}_{\perp }}\mathcal{R}^{ij}(\mathbf{k}_\perp,k_z,z,z')\;,
\end{equation}
which connects the electric field components $E^i$  directly with the current density $j^k$ through the equation
\begin{equation}
E^i(\mathbf{x};\omega)=-4\pi i\omega \int d^3\mathbf{x}' G^{ik}(\mathbf{x},\mathbf{x}';\omega)j^k(\mathbf{x}';\omega)\;.
\end{equation}
In our case only  $j^3=-i\omega p \delta (x') \delta(y') \delta (z'-z_0)$ is different from zero so that the only contributions to ${\cal R}^{ij}$ come from ${\cal R}^{i3}$, which can be read from 
Eqs.  (B8), (B10) and (B12) of  Ref.  \cite{Crosse-Fuchs-Buhmann} for an incident  TM  polarized plane wave, and which we rewrite here
\begin{eqnarray}
\mathcal{R}^{13}(\mathbf{k}_\perp,k_z,z,z_0)&=&\frac{ie^{ik_{z}(z+z_0)}}{2k_z}\left[-\frac{k_xk_z}{k^2}R_{TM,TM}+\frac{k_y}{k}R_{TE,TM}\right]\;,\\
\mathcal{R}^{23}(\mathbf{k}_\perp,k_z,z,z_0)&=&\frac{ie^{ik_{z}(z+z_0)}}{2k_z}\left[-\frac{k_yk_z}{k^2}R_{TM,TM}-\frac{k_x}{k}R_{TE,TM}\right]\;,\\
\mathcal{R}^{33}(\mathbf{k}_\perp,k_z,z,z_0)&=&\frac{ie^{ik_{z}(z+z_0)}}{2k_z}\left[\frac{k^2_\perp}{k^2}R_{TM,TM}\right].
\end{eqnarray}
Incidentally, the above equations show that the TM and TE polarizations  are mixed as a consequence of the  MEE. The explicit expressions for the reflection coefficients are given in  Eqs. (44)-(46) of Ref. \cite{{Crosse-Fuchs-Buhmann}}. The notation in Ref. \cite{{Crosse-Fuchs-Buhmann}}  $\{k, k_p, k_z\}$ is equivalent to ours $\{{\tilde k}_0, |{\mathbf k}_\perp|, \sqrt{{{\tilde k}}_0^2- {\mathbf k}_\perp^2}\}$, respectively. 
In this way, the components of the electric field
are
\begin{equation}
E^{1}(\mathbf{x};\omega )=-4\pi i\omega ^{2}p\;\;\int \frac{d^{2}\mathbf{k}%
_{\perp }}{(2\pi )^{2}}\frac{e^{ik_{z}z_{0}}}{2k_{z}}\left[ -\frac{k_{x}k_{z}%
}{k^{2}}R_{TM,TM}+\frac{k_{y}}{k}R_{TE,TM}\right] e^{i\mathbf{k_{\perp }}%
\cdot \mathbf{R}_{\perp }}e^{ik_{z}z},
\label{E11}
\end{equation}%
\begin{equation}
E^{2}(\mathbf{x};\omega )=-4\pi i\omega ^{2}p\;\;\int \frac{d^{2}\mathbf{k}%
_{\perp }}{(2\pi )^{2}}\frac{e^{ik_{z}z_{0}}}{2k_{z}}\left[ -\frac{k_{y}k_{z}%
}{k^{2}}R_{TM,TM}-\frac{k_{x}}{k}R_{TE,TM}\right] e^{i\mathbf{k_{\perp }}%
\cdot \mathbf{R}_{\perp }}e^{ik_{z}z},
\label{E22}
\end{equation}%
\begin{equation}
E^{3}(\mathbf{x};\omega )=-4\pi i\omega ^{2}p\;\;\int \frac{d^{2}\mathbf{k}%
_{\perp }}{(2\pi )^{2}}\frac{e^{ik_{z}z_{0}}}{2k_{z}}\left[ \frac{k_{\perp
}^{2}}{k^{2}}R_{TM,TM}\right] e^{i\mathbf{k_{\perp }}\cdot \mathbf{R}_{\perp
}}e^{ik_{z}z}.
\label{E33}
\end{equation}
To compute  the far-field approximation of the electric field written above,  which is   necessary  to compare with our expressions (\ref{E e-d perp V1}) and (\ref{f perp}), we make use of the angular spectrum representation method which we briefly review \cite{Novotny-Hecht}.  For fields satisfying the Helmholtz equation $(\nabla^2+ \kappa^2 ){\mathbf E}=0$, with $\kappa^2=\epsilon\omega^2$, which can be  written as 
\begin{equation}
E^i(x,y,z)= \int d^2 {\mathbf k}_\perp {\hat E}^i(k_x,k_y, z)
e^{i{\mathbf k}_\perp\cdot {\mathbf x}_\perp},
\label{E1} 
\end{equation}
one can show that 
\begin{equation}
{\hat E}^i(k_x,k_y, z)={\hat E}^i(k_x,k_y, z=0)e^{\pm i k_z z},  \qquad k_z=\sqrt{\kappa^2- {{\mbf k}_\perp}^2}, \quad \mathrm{Im}(k_z) \geq 0,
\label{E2}
\end{equation}
choosing the $+, -$ signs according to $z>0$ or $z<0$, respectively.
Substituting Eq. (\ref{E2}) into Eq.(\ref{E1}) yields the so-called angular spectrum representation of the electric field. One of the notable consequences of this approach is that the far-field approximation
 of the electric field is given in terms of the function ${\mbf {\hat E}}(k_x,k_y, z=0)$. According to Ref.\cite{Novotny-Hecht} we have  
\begin{equation}
\begin{split}
&\mathbf{E}_{{\tilde k}_0 r\rightarrow \infty }\left( \frac{x}{r},
\frac{y}{r},\frac{z}{r} \right)=-2 \pi i\tilde{%
k}_{0}s_{z}\mathbf{\hat{E}}\Big(k_{x}=\tilde{k}_{0}s_{x},\;k_{y}=\tilde{k}%
_{0}s_{y},\;z=0 \Big)\frac{e^{i\tilde{k}_{0}r}}{r}, \\
& s_{x} =\sin \theta \cos \varphi, \qquad s_{y}=\sin \theta \sin \varphi, \qquad s_{z}=\cos \theta, \qquad  k_{z}=\tilde{k}_{0}s_{z}=%
\tilde{k}_{0}\cos \theta.
\end{split} 
\label{ASR}
\end{equation}
Our next step is to identify the respective functions ${\hat E}^i(k_x,k_y, z=0)$ in each of the components (\ref{E11})-(\ref{E33}), so that we can apply the relation (\ref{ASR}). Making the required substitutions we find that
\beq
{\hat E}^i(k_x,k_y, z=0)=-i\frac{\omega^2 p}{2\pi k_z} e^{i k_z z_0} \Big[ \frac{}{} \quad  \Big]^i,
\eeq
where each square bracket $[\,\, \,]^i$ denotes the corresponding one in Eqs. (\ref{E11})-(\ref{E33}). Substituting in (\ref{ASR}) yields 
\begin{eqnarray}
E_{{\tilde k}_0r \rightarrow \infty }^{1}  &=&  \left[ \frac{%
xz}{r^{2}}R_{TM,TM}-\frac{y}{r}R_{TE,TM}\right] p\omega ^{2}\frac{e^{i\tilde{%
k}_{0}r}}{r}e^{i\tilde{k}_{0}\cos \theta z_{0}}, \\
E_{{\tilde k}_0r \rightarrow \infty }^{2} &=&  \left[ \frac{%
yz}{r^{2}}R_{TM,TM}+ \frac{x}{r}R_{TE,TM}\right] p\omega ^{2}\frac{e^{i\tilde{%
k}_{0}r}}{r}e^{i\tilde{k}_{0}\cos \theta z_{0}}, \\
E_{{\tilde k}_0r \rightarrow \infty }^{3}  &=&  \left[- \frac{|%
\mathbf{x}_{\perp }|^{2}}{r^{2}}R_{TM,TM}\right] p\omega ^{2}\frac{e^{i%
\tilde{k}_{0}r}}{r}e^{i\tilde{k}_{0}\cos \theta z_{0}}, \qquad \frac{|\mathbf{x}%
_{\perp }|^{2}}{r^{2}}=\sin ^{2}\theta. \qquad 
\end{eqnarray}
Comparing  the above results with our expressions (\ref{E e-d perp V1}) and (\ref{f perp}) for $f_\perp^+$ yields the identifications
\beq
R_{TM,TM}=\frac{{\tilde \theta}^2}{4n^2 + {\tilde \theta}^2} \qquad R_{TE, TM} =  - \frac{2 {\tilde \theta} n}{4n^2 + {\tilde \theta}^2}.
\label{RCOEFF}
\eeq
Carrying the analogous calculation for the LH with $f_\parallel^-$,  we read the transmission coefficients 
\beq
T_{TM,TM}=\frac{4 n^2}{4n^2 + {\tilde \theta}^2}, \qquad T_{TE,TM}=R_{TE,TM}.
\label{TCOEFF}
\eeq
We immediately verify that  $R_{TM,TM}+T_{TM,TM}=1$ as expected. Let us observe that the expressions for the transmission and reflection coefficients obtained  from our calculation can be verified from the general expressions (43)-(46) in Ref. \cite{Crosse-Fuchs-Buhmann}, after the following restrictions are made: $\epsilon_1=\epsilon_2= \epsilon=n^2, \,  \mu_1=\mu_2=1, \,  k_{z1}=k_{z2}=k_z$ and  $ \Delta= {\tilde \theta}$.


\subsubsection{The  {intermdiate zone in the} region $\mathcal{V}_2$}\label{RADFIELD1_PERP_V2}

{Let us recall that for any observation point in ${\cal V}_2$ with $s<s_0=1$ we can go farther into the radiation zone and find $s_> > 1$, thus eliminating the cylindrical waves. In this subsection we deal only with the intermediate region having $s<s_0=1$}. This region corresponds to what is normally called the intermediate region in the literature \cite{Banhos} and is characterized by the condition that the Sommerfeld numerical distance {$S=i s^2$} satisfies  $|S |<1$,
in spite of ${\tilde k}_0 r $ being  large. Clearly this is possible because we are in the limit $\theta \rightarrow \pi/2, \, ( \xi \rightarrow 0$),  i.e. very close to the interface.  

{A precise characterization of the intermediate zone in the UH  is provided as  follows:  According to our criterion (\ref{criterion}), which is written there for an arbitrary $s_0$, we choose $s_0=1$ as the specific value for this parameter. Then, for $r_0$ in the radiation zone (${\tilde k}_0r_0\gg1$) the discarding angle $\theta_0$ is determined such that $s_0=\sqrt{{\tilde k}_0r_0 (1- \sin \theta^{UH}_0)}$. For angles $\theta^{UH}_0 < \beta < \pi/2  $ we have $0 < s_\beta= \sqrt{{\tilde k}_0r_0 (1- \sin \beta)}< s_0$. Within this angular range {we still can move further into the radiation zone to $r_\beta$, within the interval}  $r_0 < r_\beta < r_0 \, (s_0/s_\beta) $, where cylindrical waves are still present and which define the intermediate zone.} 

So, in this region the electric field is 
\begin{eqnarray}
E^{a}(\mathbf{x};\omega)&=&\left[\mp \xi \frac{x^a}{\rho} \,  e^{\mp i\tilde{k}_{0} z_0\xi}
+\frac{2\tilde{\theta} \, n }{4n^{2}+\tilde{\theta}^{2}}\frac{\varepsilon^{ab} x^b}{\rho} e^{i\tilde{k}_{0}z_0 \xi}
\right] p \,  \omega^2 \, \frac{e^{i\tilde{k}_{0}r}}{r}\;,\nonumber\\
E^{3}(\mathbf{x};\omega)&=& \left[\frac{e^{\mp i\tilde{k}_{0} z_0 \xi}}{r}
\pm \frac{\tilde{\theta}^{2}}{4n^{2}+\tilde{\theta}^{2}} \tilde{k}_0 \xi \left(\frac{i z_0}{r}
+ \frac{\pi}{2}
\sqrt{\frac{2}{\pi i\tilde{k}_0r}}\right) \right] \, p \, \omega^2 e^{i\tilde{k}_0r},
\label{E e-d perp V2}
\end{eqnarray}
where we have dropped terms $\mathcal{O}(\xi^2)$. The variable $\xi$ was previously defined in Eq. (\ref{DEFXI}) for each hemisphere.
 Also we  verified that $\mathbf{\hat{n}}\cdot\mathbf{E}=0$ using  the approximations $\sin \theta\approx 1$ and $\cos \theta \approx \xi$, which are adequate for the region ${\cal V}_2$.
From  Eq. (\ref{E e-d perp V2}) we observe the presence of cylindrical waves,  codified in the term proportional to $e^{i{\tilde k}_0 r}/\sqrt{r}$ \cite{Sommerfeld 1}, {where we notice that $r \approx \rho$ close to the interface}. They are also present in the standard case of dipolar radiation when two different electromagnetic media are separated by a planar interface. Nevertheless, in our case they  only contribute when  at least one of the media is magnetoeletric, i.e. when $\tilde \theta \neq 0$ defines the interface. This is because we have chosen two non-magnetic media with the the same permittivity $\epsilon$, which means that setting  $\tilde \theta=0$ yields  an infinite media with no interface at all. {The subject of cylindrical waves in dipole radiation  has been exhaustively discussed in the literature been a highly controversial} topic.  An authoritative discussion of  this case, including an historical perspective, {can be found  in Section 4.10} of Ref. \cite{Banhos}.

{We finalize this section by presenting plots for the real part of the electric fields (\ref{E e-d perp V1}) and (\ref{E e-d perp V2}) in their corresponding regions $\mathcal{V}_1$ and $\mathcal{V}_2$. These plots will provide a quantitative behavior of the electromagnetic field and reinforce the space splitting exposed in Fig. \ref{RAD REGIONS}. Figs. \ref{Electric Field Patterns 1}(a) and \ref{Electric Field Patterns 1}(b) show the $x$ and $y$ components, respectively, and Figs. \ref{Electric Field Patterns 1}(c) and \ref{Electric Field Patterns 1}(d) are devoted to the $z$ component. All the figures represent the real part of the electric field (equivalent to the time dependent field at $t=0$) in the $x-z$ plane. Here the interface is constituted by a normal insulator with $\epsilon=4$ and $\vartheta_2=0$ in the UH and a medium with $\epsilon=4$ and $\vartheta_1=5$ in the LH to make evident the new effects. The dipole has a strength $p= 2.71 \times 10^{3} \, \,  {\rm eV}^{-1}, \,\,  $ a frequency $ \omega= 1.5 \, $ eV and is located at $z_0= 25$ eV$^{-1}$ (an explanation of this parameters choice will be given in the Sec. \ref{PARAM} immediately below). The field patterns for the $x$ and $z$ components in Figs. \ref{Electric Field Patterns 1}(a), \ref{Electric Field Patterns 1}(c) and \ref{Electric Field Patterns 1}(d) show different behaviors at both sides of the interface. Indeed, in the upper semi-space the effect of the interference between the two phases of the electric field associated to the Case $(+)$ is quite appreciable. Regarding the lower semi-infinite space, the features of the Case $(-)$ are visible, because one observes clearly the absence of an interference pattern and the same behavior of an electric dipole field. Remarkably the $y$ component in Fig. \ref{Electric Field Patterns 1}(b) is different from zero in comparison with the usual electric dipole radiation, results proportional to $\tilde{\theta}$ and does not exhibit an interference pattern due the vanishing of the first term of Eq. (\ref{E e-d perp V1}) at the $x-z$ plane.} 

{Recalling from Eq. (\ref{E e-d perp V2}) that only the $z$ component of the electric field contributes to the cylindrical  waves, we need to employ the discarding angles given by Eqs. (\ref{theta0 UH}) to split the space into the regions $\mathcal{V}_1$ and $\mathcal{V}_2$ of Fig. \ref{RAD REGIONS}. Fig. \ref{Electric Field Patterns 1}(c) illustrates this splitting for the $z$ component in the UH for angles in the range $\theta\in[0,\theta_0^{UH})$ with $\theta_0^{UH}\simeq 1.53918 \simeq 88.19^{\circ}$ and Fig. \ref{Electric Field Patterns 1}(d) shows the same but in the LH for angles in the range $(\theta_0^{LH},\pi]$ with $\theta_0^{LH}\simeq 1.60241 \simeq 91.81^{\circ}$. Conversely, Figs. \ref{Electric Field Patterns 2}(a) and \ref{Electric Field Patterns 2}(b) show the behavior of the $z$ component in the UH for angles in the range $\theta\in[\theta_0^{UH},\pi/2)$ and in the LH for angles in the range $\theta\in[\pi/2,\theta_0^{LH}]$, respectively. Here the appearance of the axially symmetric cylindrical waves is clear, although our plots show that they are confined to a finite distance range and decay rapidly for large distances parallel to the interface. The discarding angles are {{ not to  scale }} for the purpose of  {making evident}    the appearance of cylindrical waves. { Our approximation yields zero   for the $x$ and $y$ components of the electric field in  the region $\mathcal{V}_2$. }}
\begin{figure}[H]
\centering
\subfloat[]{
\label{g:1}
\includegraphics[width=0.49\textwidth]{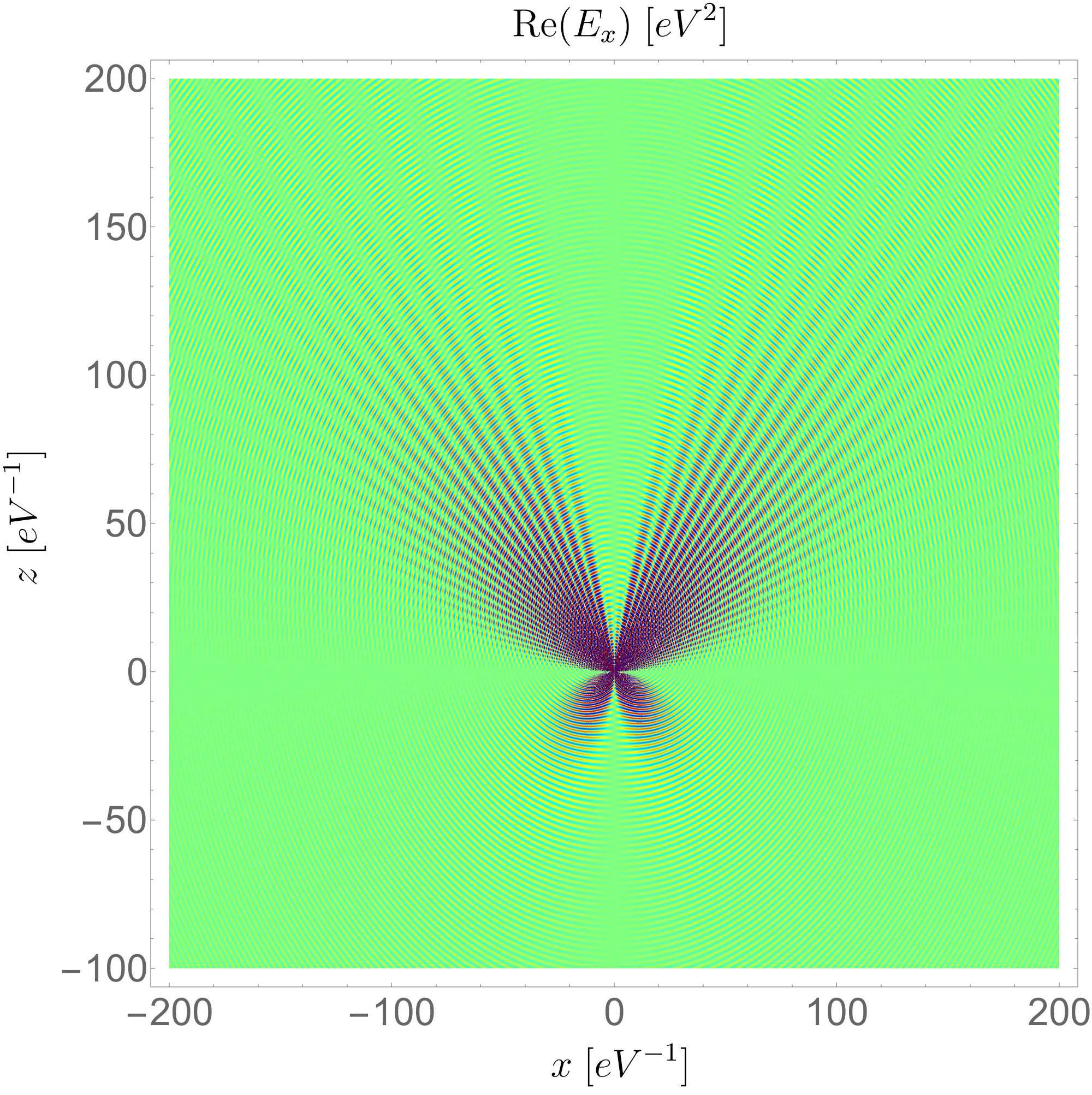}}
\subfloat[]{
\label{g:2}
\includegraphics[width=0.49\textwidth]{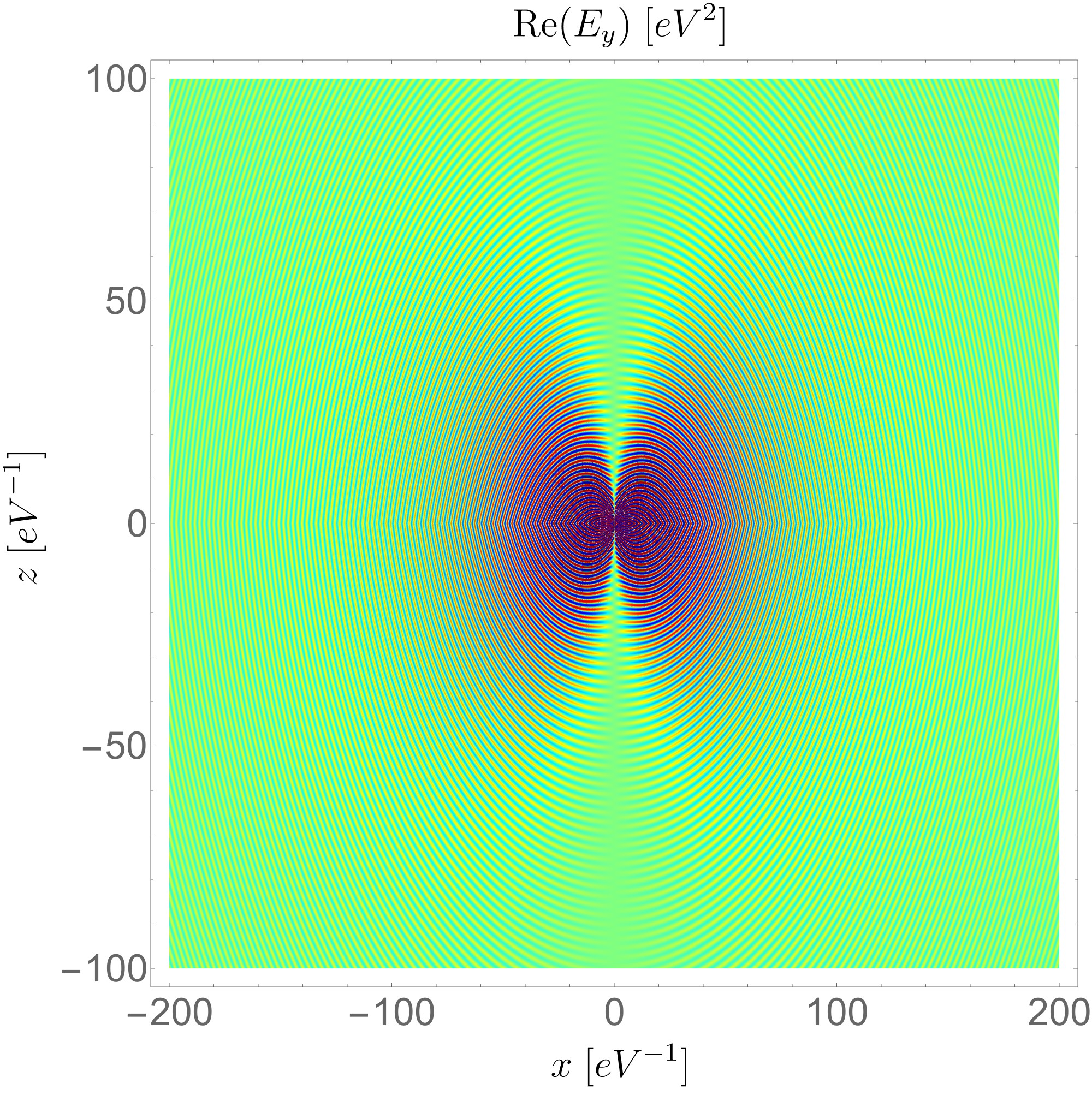}}

\subfloat[]{
\label{g:3}
\includegraphics[width=0.49\textwidth]{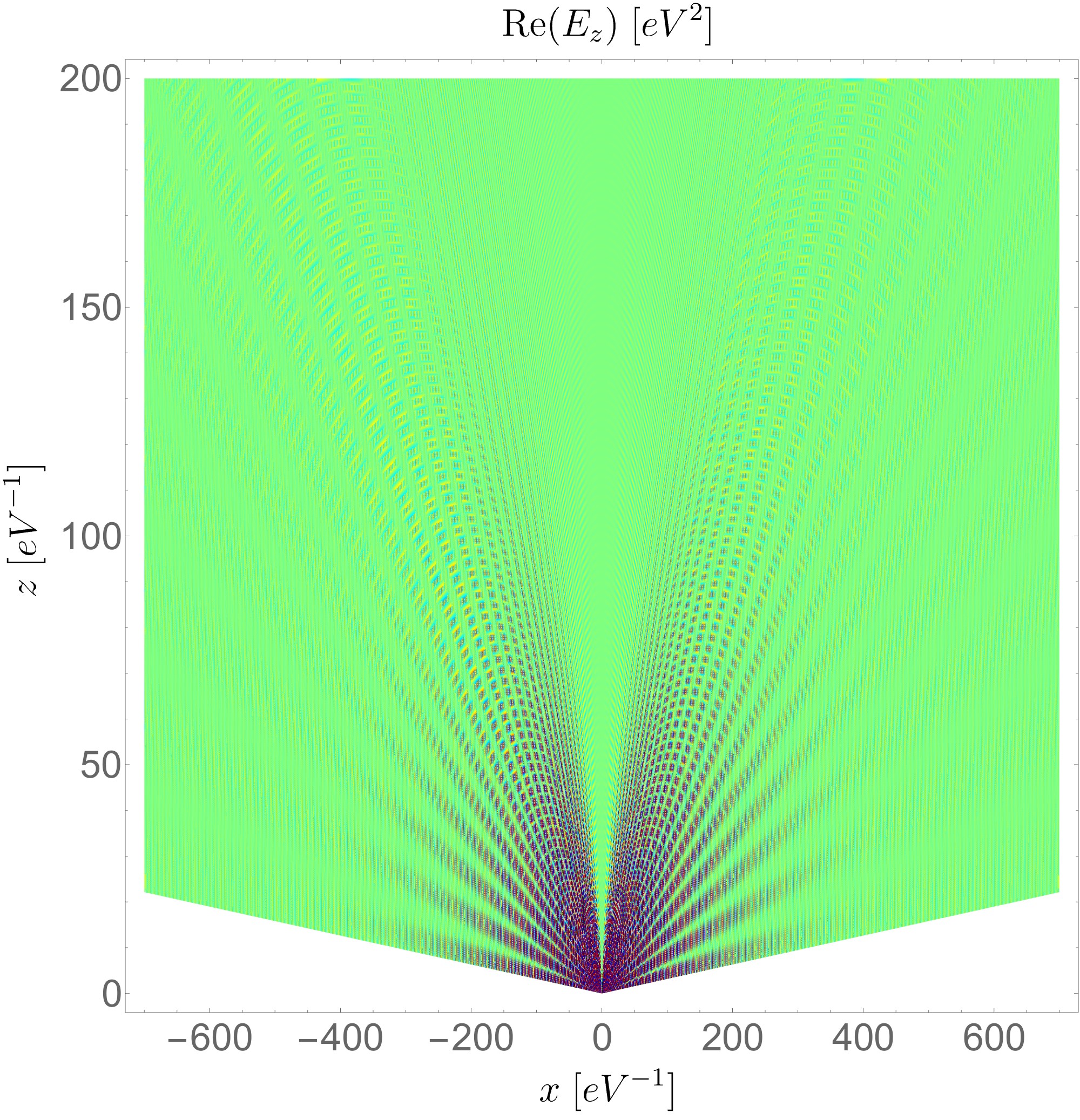}}
\subfloat[]{
\label{g:4}
\includegraphics[width=0.49\textwidth]{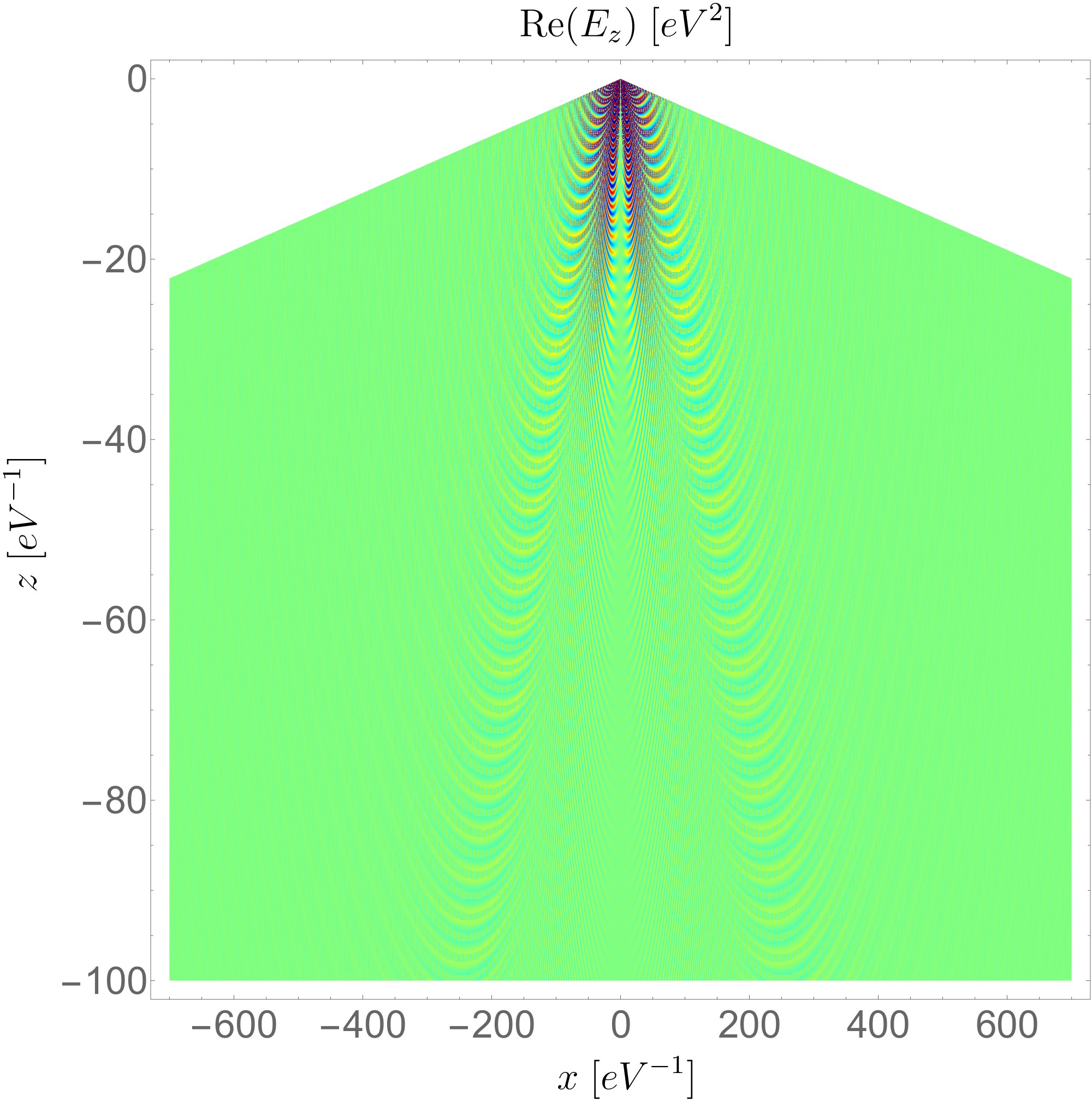}}
\caption{ {The electric field pattern (real part) in the $x-z$ plane for a vertical oriented point dipole with single frequency $ \omega= 1.5 \, $ eV, strength $p= 2.71 \times 10^{3} \, \,  {\rm eV}^{-1}$ and located at $z_0= 25$ eV$^{-1}$ close to a magnetoelectric 
interface, for the region $\mathcal{V}_1$ where only the spherical waves are significant. Here the UH is a normal insulator with $\epsilon=4$ and $\vartheta_2=0$ and the LH is a medium with $\epsilon=4$ and $\vartheta_1=5$. The plots  (a) and (b) are the $x$ and $y$ components respectively. The plots (c) and (d) are the $z$ components in the UH with discarding angle $\theta_0^{UH}\simeq 1.53918 \simeq 88.19^{\circ}$ and in the LH with discarding angle $\theta_0^{LH}\simeq 1.60241 \simeq 91.81^{\circ}$ respectively. } }
\label{Electric Field Patterns 1}
\end{figure}

\begin{figure}[H]
\centering
\subfloat[]{
\label{g:11}
\includegraphics[width=0.49\textwidth]{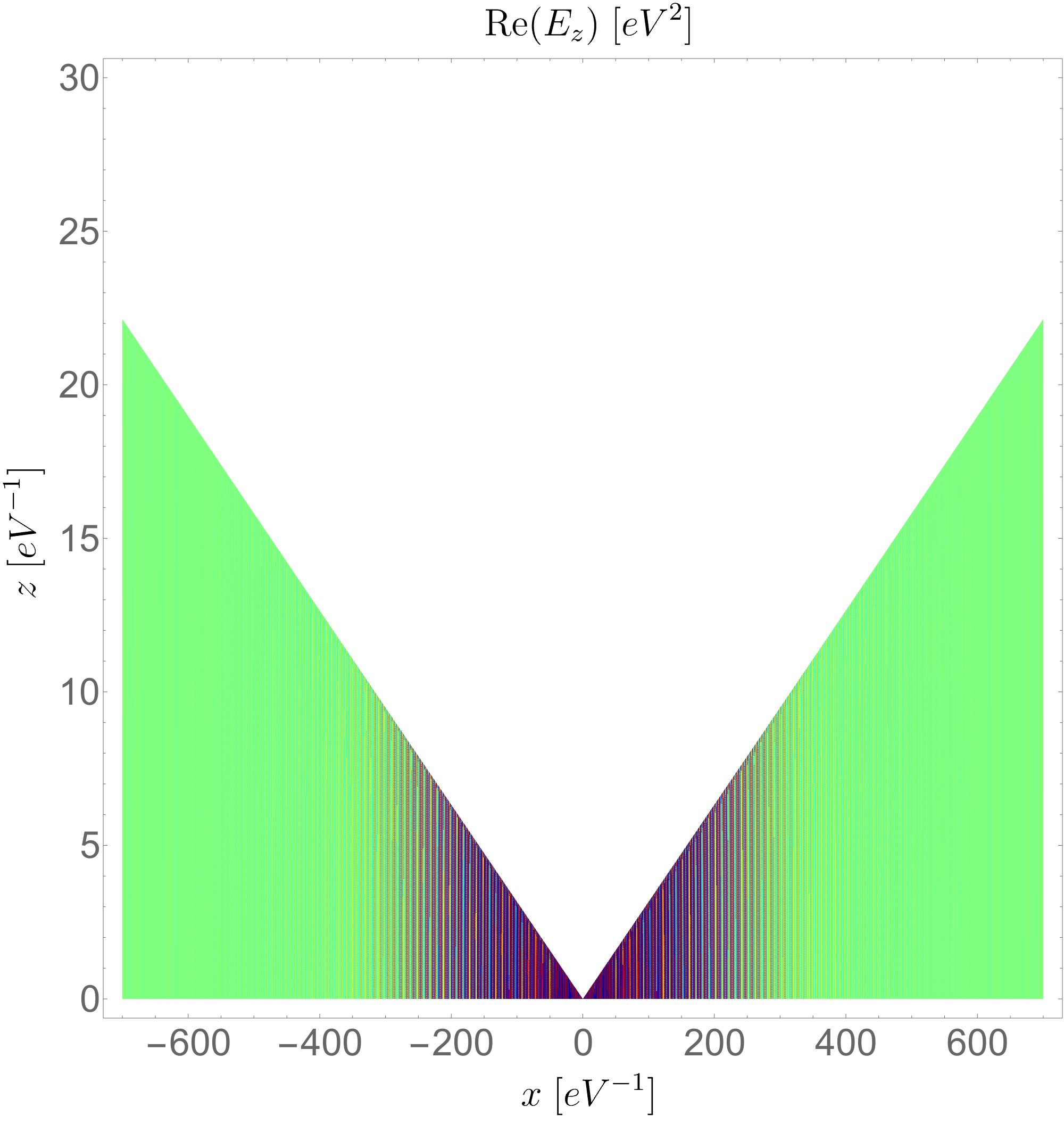}}
\subfloat[]{
\label{g:21}
\includegraphics[width=0.49\textwidth]{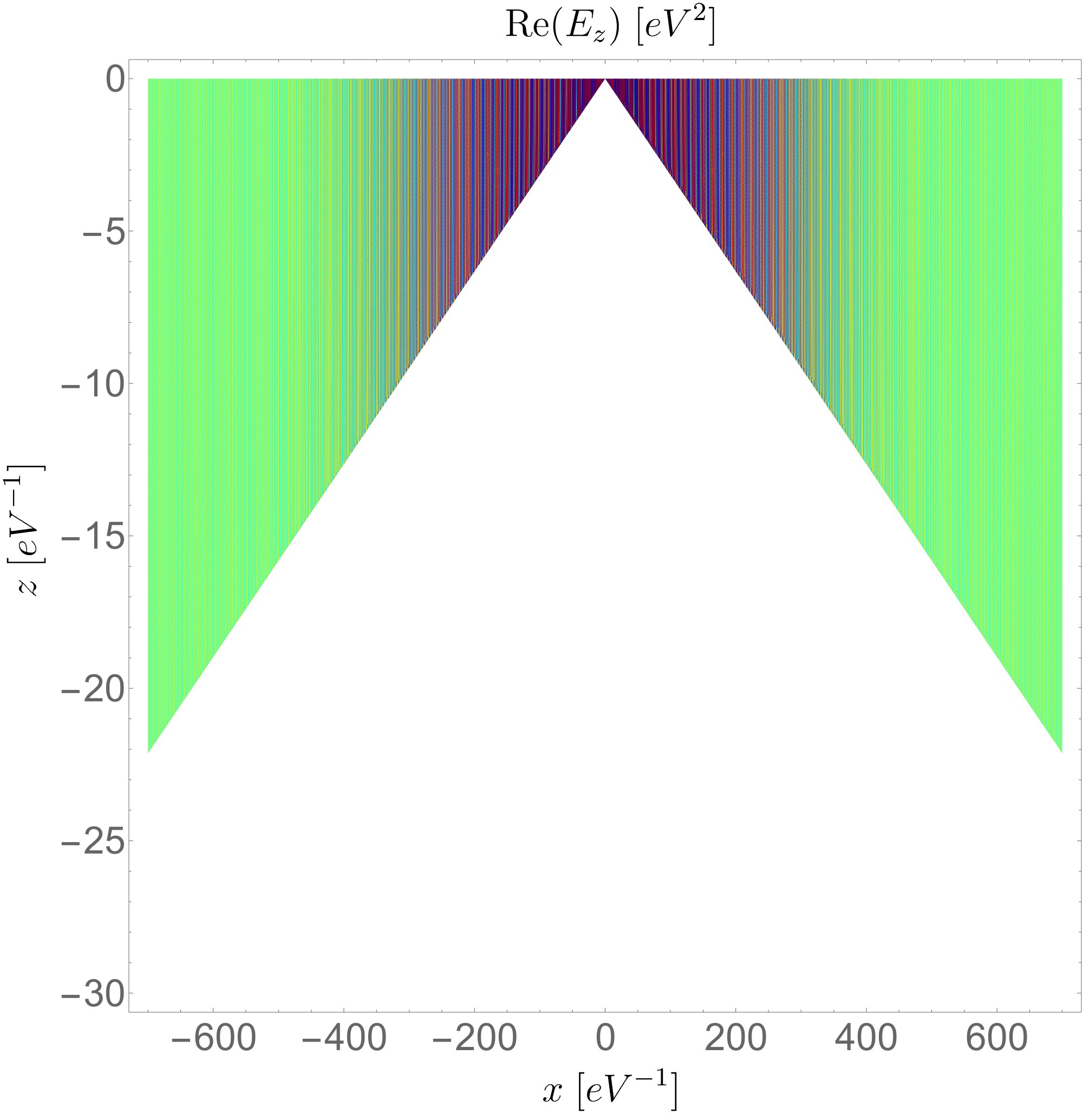}}
\caption{ { The electric field pattern (real part) in the $x-z$ plane for a vertical oriented point dipole with single frequency $ \omega= 1.5 \, $ eV, strength $p= 2.71 \times 10^{3} \, \,  {\rm eV}^{-1}$ and located at $z_0= 25$ eV$^{-1}$ close to a magnetoelectric 
interface,  for the region $\mathcal{V}_2$ where the spherical waves and the cylindrical  waves are significant. Here the UH is a normal insulator with $\epsilon=4$ and $\vartheta_2=0$ and the LH is a medium with $\epsilon=4$ and $\vartheta_1=5$. The plots (a) and (b) are the $z$ components in the 
UH and in the LH respectively. The discarding angles are $\theta_0^{UH}=88.19^{\circ}$ and  $\theta_0^{LH}=91.81^{\circ}$, which are  not to  scale  for the purpose of making evident the presence of cylindrical waves. } }
\label{Electric Field Patterns 2}
\end{figure}

\section{Angular distribution, total radiated power and energy transport}

\label{SECTIONIV}

\subsection{The parameters}

\label{PARAM}

With the purpose of illustrating our results with some numerical estimations  we need to fix the parameters defining our setup.
Our choice  is motivated by the fact that current magnetoelectric media are of great interest in atomic physics and optics,
therefore we think as a dipolar source an atom with a given dipolar moment $p$, whose emission spectrum goes from the near infrared to the near ultraviolet.  Furthermore,  the magnetoelectric coupling is usually very small (of the order of the fine structure constant for TIs), so that  appreciable effects will appear  near the interface. In this way we have chosen the distance between the dipole and the observer  to be  lesser than 1 mm. For all cases in the following numerical estimations, with the exception of Fig. \ref{P+PERP}, we  take  
the frequency $\omega=1.5$eV (362.7 THz or $\lambda= 826.6$ nm) in the near infrared, the observer distance {$r=667$ eV$^{-1}$} (0.131  mm), and  the dipole location at $z_0=25$ eV$^{-1}$ (4.94 $\mu$m). The far-field condition is well satisfied with  $n \omega r \approx 1000 \,n $. The remaining free parameters are ${\tilde \theta}$ and $n$, which characterize the medium. {This setup provides a microscopic antenna in front of a magnetoelectric medium. The additional  boundary conditions at the interface  drastically modify the dominant dipolar radiation. These changes can be directly observed by measuring the angular distribution of the radiation, which looks feasible having in mind similar techniques developed in Refs. \cite{ANGD1,ANGD2}.
Another possibility is to observe the modified radiative lifetime of the atom, which must change due to the dominance of the modified dipolar radiation  \cite{LT}. These effects have been already demonstrated in experiments \cite{LT1,LT2}.}

\subsection{Angular distribution for the radiation  in the region ${\cal V}_1$}

\label{ANGDIST1}

In this subsection we  obtain the angular distribution of radiated power associated to the electric field given by Eqs. (\ref{E e-d perp V1}) for the  region $\mathcal{V}_1$. 
%
Recalling that the electromagnetic  fields   satisfy 
 $\mathbf{\hat{n}}\cdot\mathbf{E}=0$ and $\mathbf{B}=\sqrt{\epsilon} \, {\mathbf {\hat n}}\times {\mathbf E}$ we obtain  the  standard Poynting vector  in a material media with $\mu=1$, 
\begin{equation}
\mathbf{S}=\frac{1}{4\pi}\mathbf{E}\times\mathbf{H}=\frac{\sqrt{\epsilon}}{4\pi}\|\mathbf{E}\|^{2}\mathbf{\hat{n}},
\label{PRAD}
\end{equation}
where $\mathbf{\hat{n}}$ coincides with the direction of the phase velocity of the outgoing wave.
According to Refs.  \cite{Jackson,Schwinger}, the time-averaged power radiated per unit solid angle solid by a localized source is
\begin{eqnarray}
\frac{dP}{d\Omega}&=&\frac{r^{2}}{2}\mathrm{Re}%
\left[\frac{\mathbf{E}(\mathbf{x};\omega)\times\mathbf{H}^{*}(\mathbf{x};\omega)%
}{4\pi}\right]= \frac{nr^{2}}{8\pi}\mathbf{E}(\mathbf{x};\omega)\cdot\mathbf{E}^{*}(%
\mathbf{x};\omega).
\end{eqnarray}
 The result for our dipole $\mathbf{p}$ is
 \begin{eqnarray}
\frac{dP}{d\Omega}&=&\frac{n\omega^{4}p^{2}}{8\pi}\sin^{2}\theta\left\{1+%
\Upsilon \,  \mathrm{sgn}^{2}\left(\cos\theta\right) 
+ 2 \Upsilon  \,  \mathrm{sgn}\left(\cos\theta\right)\cos\left[\tilde{k}_{0}z_{0}\left(|\cos\theta|+\cos\theta\right)\right]\right\},\;\; 
\label{dP dOmega dipole perp}
\end{eqnarray}
where $\Upsilon={\tilde \theta}^2/(4n^2+ {\tilde \theta}^2)$.
Some comments regarding this angular distribution are now in order.  The expression (\ref{dP dOmega dipole perp}) is an even function of the MEP $\tilde{\theta}$ as well as of  the angle $\theta$. Furthermore, the last term in Eq. (\ref{dP dOmega dipole perp})   arises from the interference between the two different phases exhibited by the  electric field in  Eq. (\ref{E e-d perp V1}) and  could or could not contribute depending on the sign of $\cos\theta$. 

At this stage it is important to emphasize  that our result in  Eq. (\ref{dP dOmega dipole perp}) shows that  $\Upsilon$
sets the scale in the magnitude of the  power radiated  in the region ${\cal V}_1$. This parameter has the 
 relevant property  of being bounded within the interval
 $\, 0 < \Upsilon < 1\, $, independently of the values which ${\tilde \theta}$ and $n$ might take. This will severely constrain the response of the $\vartheta$-medium with respect to the output produced  
 by an electric dipole in an infinite media with refraction index $n$ . We  refer to the latter   reference setup  as the standard electrodynamics  (SED) case, which is obtained  setting  $\tilde{\theta}=0$ in Eq. (\ref{dP dOmega dipole perp}) yielding the well known dipolar angular distribution \cite{Jackson}.

Now, we analyze the  angular distribution of the radiated power (\ref{dP dOmega dipole perp}) for the  Case $(-)$ discussed in Sec. \ref{RADFIELD1_PERP_V1}, when the electric field  has a single phase. Making this choice in Eq. (\ref{dP dOmega dipole perp}), we obtain   
\begin{eqnarray}\label{dP dOmega dipole perp -}
\frac{dP_{(-)}}{d\Omega}=\frac{n\omega^{4}p^{2}}{8\pi}\sin^{2}\theta\Big(1- \Upsilon\Big).
\end{eqnarray}
 Notice that the factor  $(1-\Upsilon)= 4n^2/(4n^2 + {\tilde \theta}^2 )$ is always positive which confirms a basic property of the radiated power. We observe that the angular dependence of the radiated power  remains unchanged with respect to the SED case, {confirming what we found at the level of the electric field in Figs. \ref{Electric Field Patterns 1}(a) and \ref{Electric Field Patterns 1}(d).} Nevertheless, the magnitude of the radiation turns out to be smaller for a fixed angle, which provides a fundamental difference with respect to this reference setup. Surprisingly,  in the highly hypothetical situation where  $\Upsilon \rightarrow 1$, the radiation in the LH  would be completely canceled  i.e. the setup would behave as a perfect mirror.   
 
On the other hand, for the Case $(+)$,
when  the electric field includes two different phases, we obtain the angular distribution
\begin{equation}
\label{dP dOmega dipole perp +}
\frac{dP_{(+)}}{d\Omega}=\frac{n\omega^{4}p^{2}}{8\pi}\sin^{2}\theta
\left[1+\Upsilon + 2\Upsilon \cos\left(2\tilde{k}_{0} z_{0}\cos\theta\right)\right],
\end{equation}
which present additional contributions to the angular distribution with respect to those in SED. They arise from the last term  in Eq. (\ref{dP dOmega dipole perp +}). Furthermore, as opposed to the previous case,  the angular distribution now depends explicitly on the dipole position $z_{0}$. Let us observe that the minimum value $-1$ of 
$\cos(2 {\tilde k}_0 z_0 \cos \theta )$ produces the factor 
$(1-\Upsilon)$ in the square bracket, which  was discussed above.

\begin{figure}
\centering
\subfloat[]
{
\label{f:5}
\includegraphics[width=0.3\textwidth]{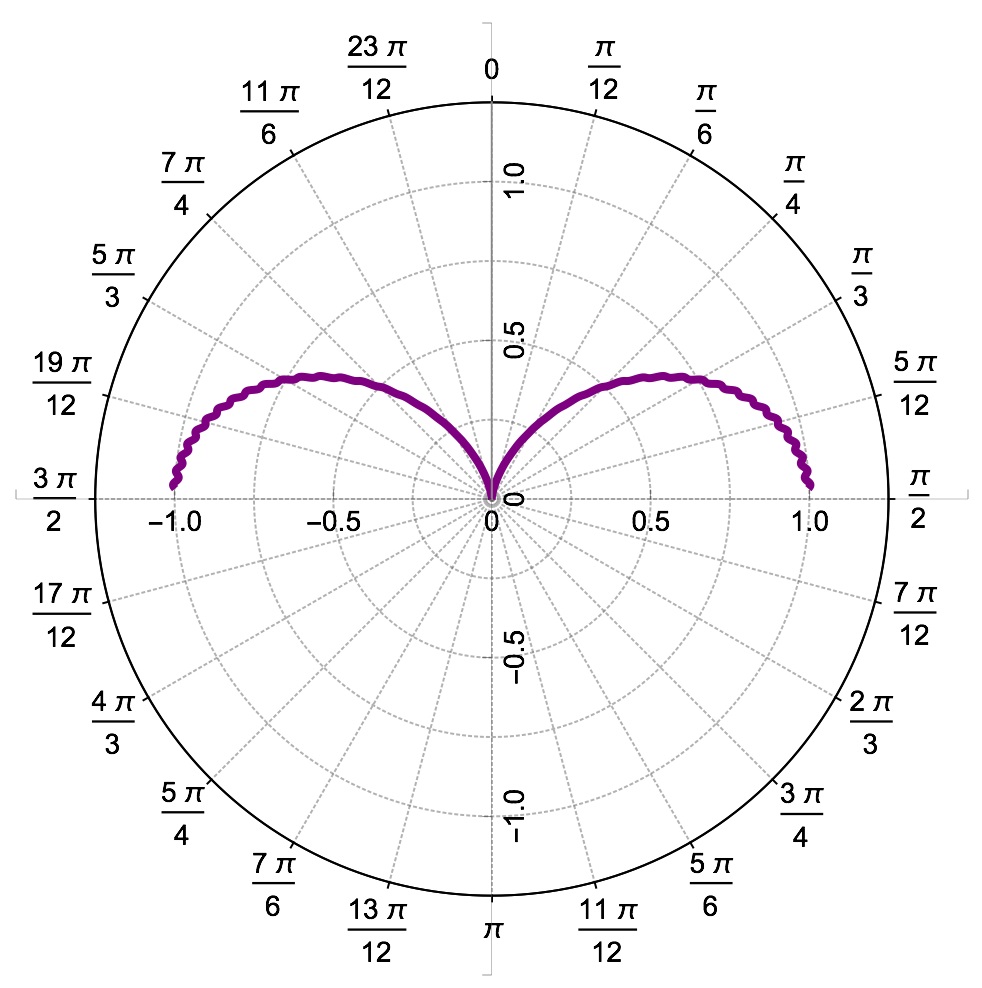}
}
\subfloat[]
{
\label{f:9}
\includegraphics[width=0.3\textwidth]{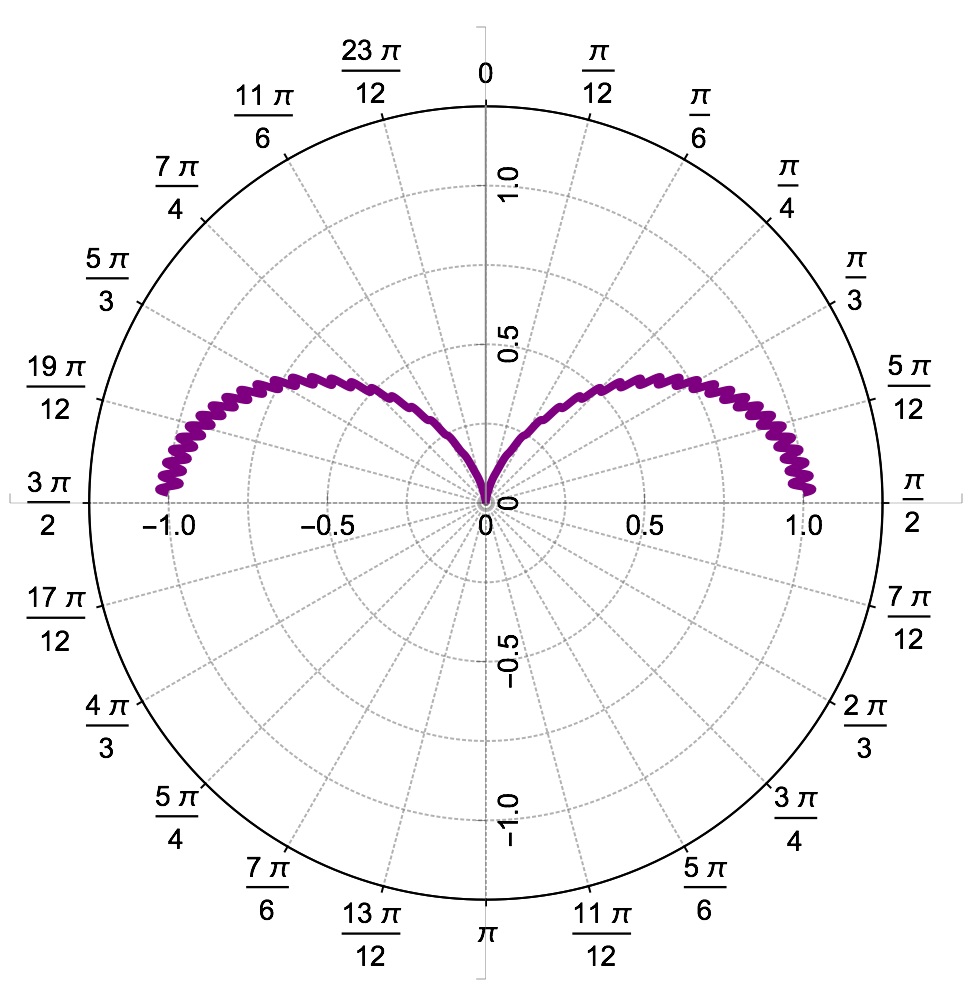}
}
\subfloat[]
{
\label{f:13}
\includegraphics[width=0.3\textwidth]{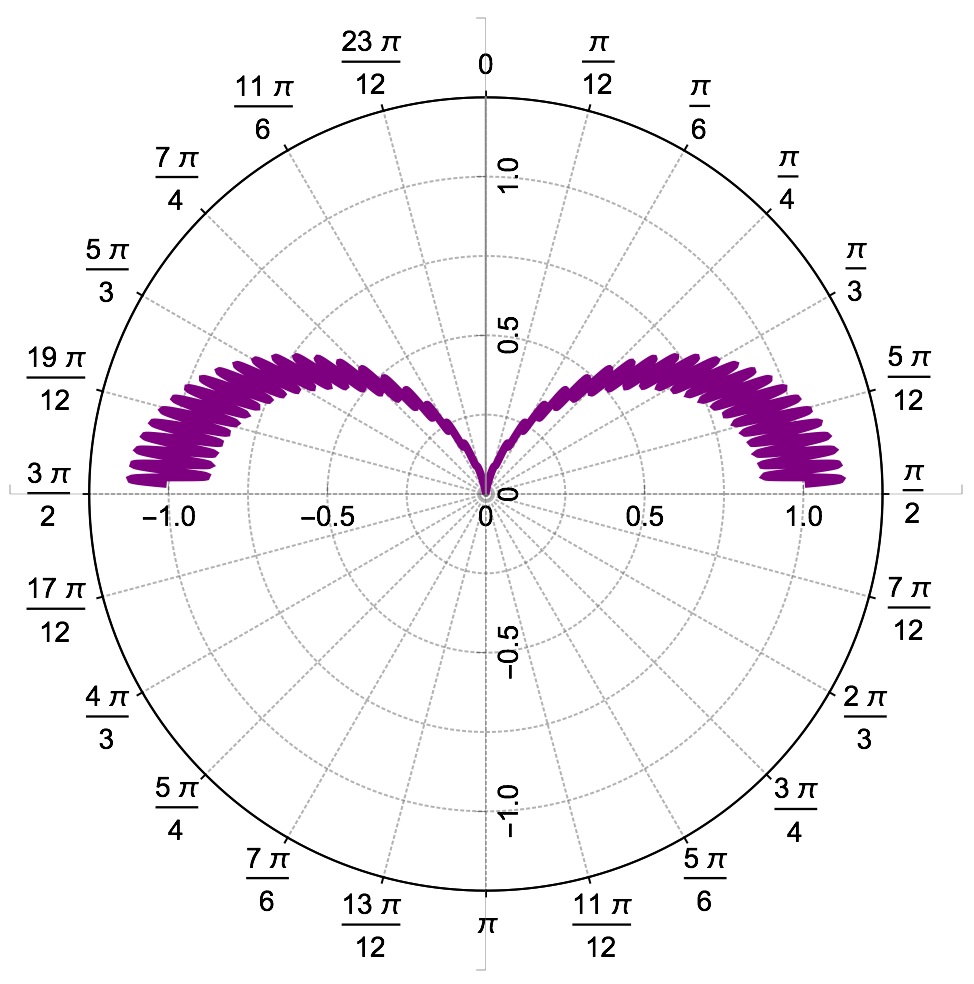}
}
\caption{Angular distribution of the radiated power $dP_{(+)}/d\Omega$.
(a) Polar plot  with $\tilde{\theta}=0.22$, $n=1.87$,
(b) Polar plot  with $\tilde{\theta}=0.5$, $n=2$. 
(c) Polar plot  with $\tilde{\theta}=1$, $n=2$.
The scale is normalized by multiplying $dP_{(+)}/d\Omega$  by $4\pi/\omega^{4}p^{2}$. The remaining parameters $\omega=1.5 \,  {\rm eV}, \,  r=667 \, {\rm eV}^{-1}, \,  z_0=25 \, {\rm eV}^{-1} \, $ are common to this and all subsequent  figures.}
\label{PATTERN CASE +1}
\end{figure}
The behavior of the angular distribution (\ref{dP dOmega dipole perp +}) is shown in Fig. \ref{PATTERN CASE +1}. In each case, the electric dipole is located at   $z_0>0$ and the interface corresponds  to the line ($3\pi/2-\pi/2$) defining $z=0$. 
The Fig. \ref{f:5} is plotted for the $\vartheta$-medium TbPO$_4$   with $n=1.87$ \cite{TbPO4} and  $\tilde{\theta}=0.22$ \cite{Rivera}. After comparing  with   the SED case  we  appreciate  only weak signals of interference.
The Fig.  \ref{f:9} corresponds to an hypothetical material with   $\tilde \theta=0.5$ and  $n=2$. Finally, in Fig. \ref{f:13} we see a  clear enhancement in the interference pattern for our electric dipole radiating in front of  an another  hypothetical material with $\tilde{\theta}=1$ and $n=2$. An increasing value of the parameter  $\tilde \theta$ in Fig. \ref{PATTERN CASE +1} makes  evident the interference effects,  which are  expected to be more pronounced  in the
vicinity of $\theta= \pi/2$ where the last term in Eq. (\ref{dP dOmega dipole perp +}) oscillates maximally. {This interference} effect agrees {with our results plotted in Figs. \ref{Electric Field Patterns 1}(a) and \ref{Electric Field Patterns 1}(c).}

\begin{figure}[H]
\centering
\subfloat[]
{
\label{PHASEA}
\includegraphics[scale=0.3]{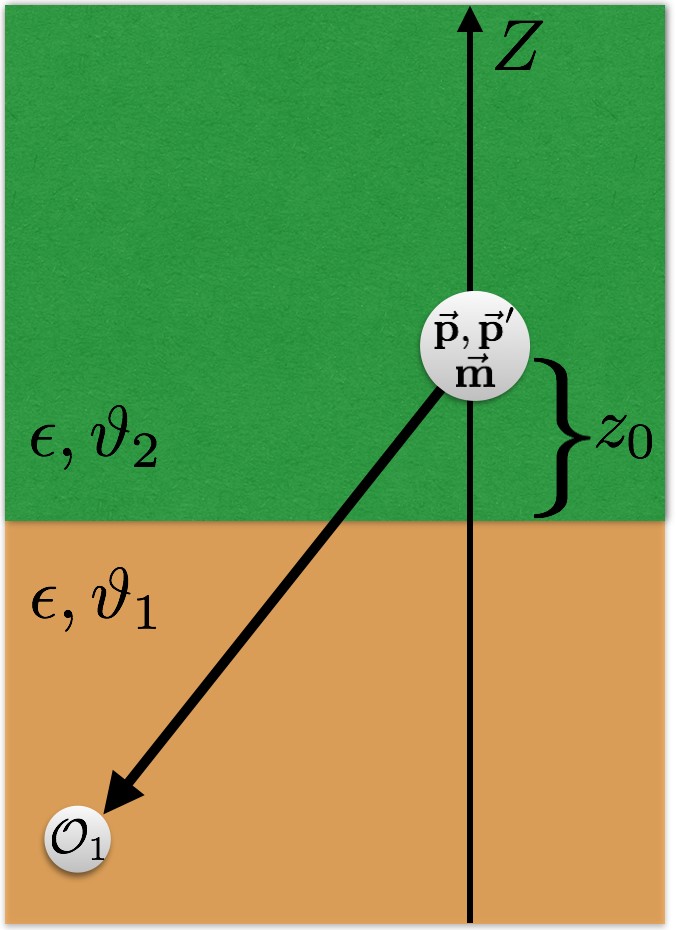}
}
\hspace{3cm}
\subfloat[]
{ 
\label{PHASEB}
\includegraphics[scale=0.3]{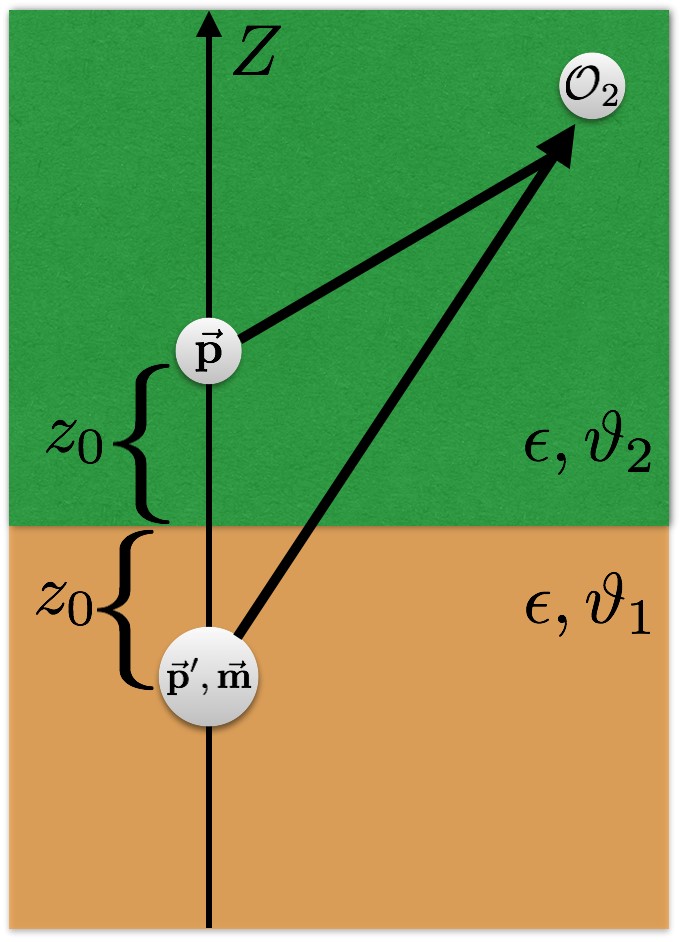}
}
\caption{{
(a) Illustration of the Case ($-$), where the radiation (black thick arrows) has a single phase, which is the one of standard ED, when the observer ${\cal O}_1$ is in a different semi-space with respect to the dipole ${\mathbf p}$  at $z_0$. 
(b) In the Case ($+$)(observer ${\cal O}_2$ in the same semi-space of the dipole) the radiation seen by ${\cal O}_2$ includes two different phases: the one arising from ${\mathbf p}$ at $z_0$
and the other that comes from the images ${\mathbf p}'$  and ${\mathbf m}$  at $-z_0$. Recall that we have two non-magnetic media with $\epsilon_1=\epsilon_2$, which yield no refraction at the interface. 
}}
\end{figure}

{Even though the method of images does not generalize to the time dependent case, a qualitative interpretation for the radiation patterns described above  can be given by extending to 
the quasi-static approximation
the characterization of a point charge located in front of a magnetoelectric medium in terms of electric and magnetic images presented  in detail in Ref. \cite{Qi Science}}. In this way, the full description of the  MEE  of a dipole located in front of a planar medium includes electric and magnetic image dipoles. 
Let us first deal with  the Case $(-)$, which corresponds to the situation when the electric dipole and the observer are in different semi-spaces, i.e. the dipole is located at $\mathbf{r}_0=(0,0,z_0)$ with $z_0>0$ and the observer's angle $\theta$ is in the LH. The MEE is mimicked by introducing an image dipole $\mathbf{p}^{\prime}$ and an image magnetic dipole $\mathbf{m}$ both located at $\mathbf{r}_0$, i.e. in the same semi-space and in the same position of the electric dipole $\mathbf{p}$. So, the observer will measure the same phase of the radiation from $\mathbf{p}$, $\mathbf{p}^{\prime}$ and $\mathbf{m}$, which is given by choosing  $|\cos\theta|=-\cos\theta$ in the phase of the electric field (\ref{E e-d perp V1}). This sign choice affects the angular distribution (\ref{dP dOmega dipole perp}) by canceling  the interference term, as Eq. (\ref{dP dOmega dipole perp -}) shows. Therefore, the angular dependence  of the radiation that the observer detects will not present a substantial difference from that  of SED, \textcolor{red}{as Figs. \ref{Electric Field Patterns 1}(a) and \ref{Electric Field Patterns 1}(d) can also confirm.} Let us recall that we have chosen the two non-magnetic  media having  the same permittivity, which eliminates the optical  refraction and reflection phenomena when passing from one magnetoelectric medium to the other. On the other hand, the  Case $(+)$ can be understood in a similar way. Here both objects, electric dipole and observer, are in the same semi-space  and the observer's angle $\theta$ is in the UH.  Again we emulate the MEE by inserting an image dipole $\mathbf{p}^{\prime}$  and the same  image magnetic dipole $\mathbf{m}$ \cite{Qi Science}, both localized at $-\mathbf{r}_0$. In this way, the observer will detect radiation with two different phases: one from the source electric dipole and another
from the image objects $\mathbf{p}^{\prime}$ and $\mathbf{m}$, which corresponds to the choice  $|\cos\theta|= + \cos\theta$ in the phase of the electric field (\ref{E e-d perp V1}).
The plus sign selection impacts significantly the angular distribution (\ref{dP dOmega dipole perp}) because the interference term is now non-zero and contributes to  observable quantities as shown in  Eq. (\ref{dP dOmega dipole perp +}) together with {Figs. \ref{Electric Field Patterns 1}(a), \ref{Electric Field Patterns 1}(c) and \ref{PATTERN CASE +1}.}  This interference arises between  the radiation coming from bottom to top, generated by the image objects, and  the direct  signal from the dipole source and generates a   different  angular dependence in the UH  when  compared with  the electric dipolar radiation of SED. {Both cases are  schematically illustrated in  Figs. \ref{PHASEA} and \ref{PHASEB}, respectively}.

\subsection{Power radiated  in the  region $\mathcal{V}_1$}\label{POWER1}

In order to compare  the magnitude of the  radiation in the $\vartheta$-medium with respect to the SED case it is convenient to introduce what we call the enhancement factor ${\cal R}_{\pm}$ defined as ${\cal R}_{\pm}=2P_{\pm}/P_0$,  where $P_{0}=n\omega^{4}p^{2}/3$ is the total power radiated by an electric dipole in the SED case \cite{Jackson}. 

Next  we calculate  the power radiated for the angular distributions (\ref{dP dOmega dipole perp -}) and (\ref{dP dOmega dipole perp +}) in  the region $\mathcal{V}_1$. 
Let us begin with the angular distribution of the Case $(-)$ given by Eq. (\ref{dP dOmega dipole perp -}). Integrating over the solid angle $\Omega$ for $\theta\in(\theta_0^{LH},\pi]$ and $\phi\in[0,2\pi]$, we find the radiated  power 
\begin{equation}\label{P total dipole perp -}
P_{(-)}=\frac{P_{0}}{2}\Big(1 -
 \Upsilon \Big)\left[1+\frac{9}{8}\cos\theta_0^{LH}-\frac{1}{8}\cos3\theta_0^{LH}\right].
\end{equation}
A good estimation of the enhancement factor is obtained writing  $\theta_0^{}= \pi/2 +\xi_0$ and recalling that $\xi_0 < 1$. We obtain
\beq
{\cal R}_{(-)}= \Big(1-\Upsilon \Big)\left(1+\frac{3 \xi_0}{2}\right).
\eeq
which can be larger (smaller)  than one according to $\Upsilon < 3 \, \xi_0/2 $  ($\Upsilon > 3 \,  \xi_0/2 $ ), respectively. In the hypothetical limit ${\tilde \theta} \gg 2n $  we have $\Upsilon=1$ and there is no radiated power in the LH, which tells us that the setup behaves like a perfect mirror as discussed in the previous section. 

Now, we repeat the  calculation  for the angular distribution of the Case $(+)$ given by Eq. (\ref{dP dOmega dipole perp +}), which is more interesting. After integrating over the solid angle $\Omega$ for $\theta\in[0,\theta_0^{UH})$, the power radiated $P_{(+)}$ is
\begin{eqnarray}
P_{(+)}&=&\frac{P_{0}}{2}\Big(1+ \Upsilon \Big)\left[1-\frac{9}{8}\cos\theta_0^{UH}+\frac{1}{8}\cos3\theta_0^{UH}\right]\nonumber\\
&&+\frac{P_{0}}{2} 3\Upsilon \left\{\frac{\sin(2\varkappa)}{4 \varkappa^3}-\frac{\cos(2\varkappa)}{2\varkappa^2}+\frac{\cos\theta_0^{UH}\cos(2\varkappa \cos\theta_0^{UH})}{2\varkappa^2}\right.\nonumber\\
&&-\left.\frac{\left[1+\varkappa^{2}-\varkappa^{2}\cos(2\theta_0^{UH})\right]\sin(2\varkappa \cos\theta_0^{UH})}{4\varkappa^3}\right\}, \qquad \varkappa\equiv \tilde{k}_{0} z_{0}. 
\label{P total dipole perp +}
\end{eqnarray}
The main difference with respect to the power radiated in the LH  is that now $P_{(+)}$ depends on the position of the dipole through the variable $\varkappa$. The power radiated $P_{(+)}$ is  positive definite and due to the term in braces  we expect to find new  effects in comparison with the previous Case $(-)$.
Moreover, from Eq. (\ref{P total dipole perp +}) and retaining $n$ and $\omega$ fixed we find the following interesting limits for $z_0$
\begin{eqnarray}
P_{(+)}\left(z_{0}\rightarrow\infty\right)&=&\frac{P_{0}}{2}\Big(1+\Upsilon \Big)\left(1-\frac{9}{8}\cos\theta_0^{UH}+\frac{1}{8}\cos3 \theta_0^{UH}\right),\label{P total dipole perp + z0}\\
P_{(+)}\left(z_{0}\rightarrow 0\right) &=&\frac{P_{0}}{2}\Big( 1+3\, \Upsilon\Big) \;\left( 1-\frac{9}{8}\cos\theta_0^{UH} +\frac{1}{8}\cos
3 \theta_0^{UH} )\right)   \nonumber \\
&&+\frac{P_{0}}{2}\; \frac{2}{5} \Upsilon \;\varkappa^{2} \Big( 5 \cos^3 \theta_0^{UH}-3
\cos^5 \theta_0^{UH}-2\Big), \quad \varkappa \ll 1.
\label{P total dipole perp + z0 null}
\end{eqnarray}%
The  Eq. (\ref{P total dipole perp + z0 null}) tell us that there are no divergences in Eq. (\ref{P total dipole perp +}) when the electric dipole is very close to the interface. Since for all practical purpose $\theta_0^{UH}$ is very close to $\pi/2$, we obtain a very good approximation of  $P_{(+)}$ in the intricate Eq. (\ref{P total dipole perp +}) by setting $\theta_0^{UH}=\pi/2$, which yields  the simplified expression
\begin{eqnarray}
 P_{(+)}=\frac{P_{0}}{2}\left( 1+ 
 \Upsilon
 \;\left[ 1+\frac{3\sin
\left( 2 \varkappa\right) }{4\varkappa^{3}}-\frac{3\left( \cos 2\varkappa \right)}{2 \varkappa^{2}} \right] \right).
\label{P+APP}
\end{eqnarray} 
 As we can calculate from Eq. (\ref{P+APP}), the enhancement factor $ {\cal R}_{(+)} =2 P_{(+)}/P_0 $  has  the following properties. The maximum occurs when the dipole is at the interface ($\varkappa=0$) and yields
$
{\cal R}^{\rm max}_{(+)}=(1+ 3\Upsilon)
$. Also we found an absolute minimum located at $\varkappa \approx 2.88$ where  $
{\cal R}^{\rm min}_{(+)}=(1+ 0.83 \,\Upsilon)
$. The limit for very large $\varkappa$ is $
{\cal R}^{\infty}_{(+)}=(1+\Upsilon)
$.
In the Fig. \ref{P+PERP} we plot the ratio $P_{(+)}/P_0$ in the  approximation of Eq. (\ref{P+APP}), as a function of $\varkappa$ for different choices of the parameter $\Upsilon$, which provides a qualitative confirmation  of the behavior of ${\cal R}_{(+)}$ discussed above.  

\begin{figure}[H]
\centering
\includegraphics[width=0.5\textwidth]{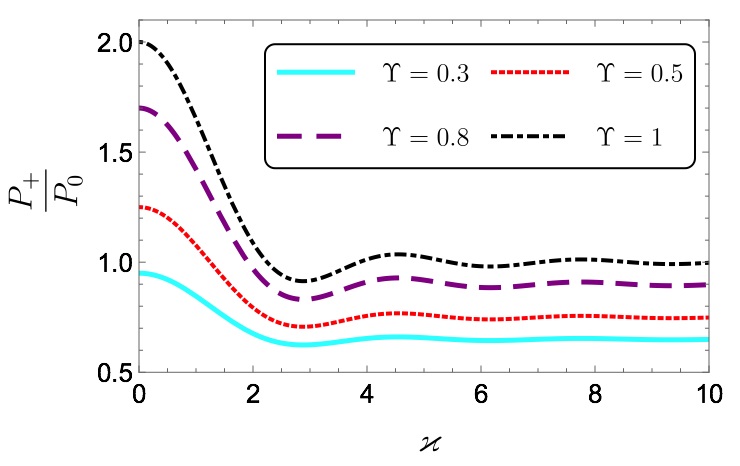}
\caption{Plot of $P_{(+)}/P_{0}$ as a function of $\varkappa$ for different choices of $\Upsilon$. The enhancement factor is ${\cal R}_{(+)}=2P_{(+)}/P_0.$}
\label{P+PERP}
\end{figure}

\subsection{Energy transport in the {intermediate region of} $\mathcal{V}_2$}\label{ET1}
In this subsection we discuss  the energy flux in the region $\mathcal{V}_2$ {when $s> s_0=1$  and   the spherical and cylindrical waves coexist as shown in  Eq. (\ref{E e-d perp V2}). This region was fully characterized previously in the paragraph before  Eq. (\ref{E e-d perp V2})}

As pointed out in Ref. \cite{Wait}, the separation of the electric field (\ref{E e-d perp V2}) in these two components is not significant since they do not constitute independent solutions of Maxwell equations.  Recalling Eq. (\ref{PRAD}) for the 
 time-averaged Poynting vector $\langle\mathbf{S}\rangle$
we obtain
\begin{eqnarray}
\langle\mathbf{S}\rangle_{\mathcal{V}_2}^{UH}&=&\mathbf{\hat{n}}\frac{np^2\omega^4}{8\pi}\left\{\frac{1}{r^2}\left[1+\frac{4\tilde{\theta}^{2}n^2}{(4n^{2}+\tilde{\theta}^{2})^2} \right]
 +\frac{\omega^{1/2}}{r^{3/2}}\frac{\tilde{\theta}^{2}}{4n^{2}+\tilde{\theta}^{2}} \sqrt{n \,\pi}  \, \xi 
 \right\}  +\mathcal{O}(\xi^2) \boldsymbol{\; , }  \label{Poynting V21} \\ 
 \langle\mathbf{S}\rangle_{\mathcal{V}_2}^{LH}&=&\mathbf{\hat{n}}\frac{np^2\omega^4}{8\pi}\left\{\frac{1}{r^2}\left[1+\frac{4\tilde{\theta}^{2}n^2}{(4n^{2}+\tilde{\theta}^{2})^2} \right]
 - \frac{\omega^{1/2}}{r^{3/2}}\frac{\tilde{\theta}^{2}}{4n^{2}+\tilde{\theta}^{2}} \sqrt{n \,\pi}  \, \xi
 \right\}  +\mathcal{O}(\xi^2) \boldsymbol{\; , }
\label{Poynting V2}
\end{eqnarray}
to first order in $\xi$. In the above linear expansion  we require 
\beq
n\omega z_0 \xi \;\ll\; 1.
\label{LCOND}
\eeq
 We observe  that these fluxes are independent of the position of the dipole. 
In Eqs. (\ref{Poynting V21}) and (\ref{Poynting V2}) we encounter   two different terms: one modulated by $r^{-2}$ which contains the energy flux coming from the spherical wave contribution, and  another one proportional to $r^{-3/2}$ that encodes the interference between the spherical and cylindrical waves. The contribution of the cylindrical wave itself  is of order $\xi^2 $  which we have  consistently neglected  in our approximation. 

Two remarks are now in order: (i) For a fixed set of parameters, Eqs.  (\ref{Poynting V21}) and (\ref{Poynting V2}) yields $\langle\mathbf{S}\rangle_{\mathcal{V}_2}^{UH}- \langle\mathbf{S}\rangle_{\mathcal{V}_2}^{LH} > 0$.  (ii) The full expression for the energy flux must be positive definite, but we are dealing only with a linear approximation in Eq. (\ref{Poynting V2}). This  forces us to establish  an additional bound  for the validity of our results. The dangerous contribution is in $\langle\mathbf{S}\rangle_{\mathcal{V}_2}^{LH}$, where the relative minus sign might produce a negative value. Recalling that 
{$\xi_0=\sqrt{2/(n \omega r_0)}$}, defined after Eq. (\ref{DEFXI}),   we rewrite the resulting condition from Eq. (\ref{Poynting V2}) as 
\beq
\frac{\xi}{\xi_0} <  \frac{1}{\sqrt{2 \pi}}\frac{(4n^2+ {\tilde \theta}^2)^2+4n^2 {\tilde \theta}^2}{{\tilde \theta}^2(4n^2+ {\tilde \theta}^2)} {\left(\frac{r_0}{r}\right)^{1/2}} <  \frac{1}{\sqrt{2 \pi}}\frac{1+\Upsilon-\Upsilon^2}{\Upsilon}\equiv Q(\Upsilon),  \quad   {\tilde \theta}\neq 0,
\label{INEQ}
\eeq
{since $r>r_0$ in the intermediate zone.  Recalling that $0 < \Upsilon < 1$, the function $Q(\Upsilon)$ is a decreasing function having its minimum value $Q(\Upsilon=1)= 0.40$.  }
This means that for any value $\xi/\xi_0 < 0.40$, the energy fluxes are always positive in the whole range of $\Upsilon $ {since the inequality (\ref{INEQ}) is always satisfied}. 

On the  contrary, when  $0.40 < \xi/\xi_0 \equiv  \zeta  < 1 $ we have to determine the maximum allowed value $\Upsilon_{{\rm max}}$ by solving $Q(\Upsilon_{{\rm max}})=\zeta$, so that  the energy fluxes are positive only in the  range {$0 < \Upsilon <   \Upsilon_{{\rm max}}$}. Let us notice that the lowest values observed for $\tilde \theta$ are of the order of the fine structure constant $\alpha=1/137$, which  effectively  replaces  the theoretical lower  limit $\Upsilon=0$ by the more realistic one $\Upsilon_{{\rm min}}=1.3 \times 10^{-5} /n^2$. 

\begingroup
\setlength{\tabcolsep}{3.8pt} 
\renewcommand{\arraystretch}{1} 
\begin{table}
\begin{center}
\begin{tabular}{|c|c|c|c|c|c|c|}
\hline
$\vartheta$-medium & \,\, n \,\, & \,\, \footnotesize$\tilde{\theta}$ \,\,  & \,\, $\Upsilon$ \,\,  &  \,\, $\langle S\rangle^{UH}  \,\, [{\rm eV^4]}$ \,\, & \,\,$\langle S\rangle^{LH} \,\, [{\rm eV^4]}$ \,\,& \,\,$\langle S\rangle^{SED} \,\, [{\rm eV^4]} \,\,$  \\
\hline \hline
TlBiSe$_2 $ & \,\, 2 \,\, & \,\, $11\alpha$ \,\, & \,\,4.0$\times10^{-4}$ \,\, & \,\, 6.64 \,\, & \,\, 6.64 \,\, & \,\, 6.63 \,\,  \\
\hline
TbPO$_4$ & 1.87 & 0.22 & 3.5$\times10^{-3}$  & 6.21 & 6.21 & 6.18   \\
\hline
Hyp. I & 2 & 0.5 & 1.5$\times10^{-2}$  & 6.74 & 6.73 & 6.63   \\
\hline
 Hyp. II  & 2  & 1  & 5.9$\times10^{-2}$   & 7.03 &  6.97 & 6.63 \\
\hline
Hyp. III  &  2  &  5  & 6.1$\times 10^{-1}$ &  8.53 &  7.89 & 6.63  \\
\hline
\end{tabular}
\caption{The energy flux very close to the interface ($\xi=10^{-3}$) between a normal insulator ($n=2  \, (1.87), {\tilde \vartheta}=0, \, \mu=1 $) and different magnetoelectrics with the same $n$ and $\mu$. The radiating  dipole has $p= 2.71 \times 10^{3} \, \,  {\rm eV}^{-1}, \,\,  $ $ \omega= 1.5 \, $ eV and  $z_0= 25$ eV$^{-1}$.  The observer distance  is 
 $r=667 \, {\rm eV}^{-1}$.
}
\par
\label{TABLE1}
\end{center}
\end{table}
\endgroup

Next we perform some numerical estimations of Eq. (\ref{Poynting V2}) shown in Table \ref{TABLE1}. 
There we refer to  the setup described in  the beginning of section \ref{THETAED}, for the case where medium 1 is a regular insulator with $n=2 \, (1.87) $, $\mu=1$ and $\vartheta=0$, while medium 2  corresponds to different magnetoelectric media with the same refraction index and permeability and whose  value of ${\tilde \theta}$ is indicated in the third column. {Since we are  interested only in the magnetoelectric response of the real materials listed in Table \ref{TABLE1} it is enough to say that   TbPO$_4$ is an antiferromagnet exhibiting a linear MEE, whose relevant properties have been extensively studied in Ref.   \cite{TbPO4,Rivera}. On the other hand TlBiSe$_2 $ has been experimentally identified as a TI admitting MEPs given by  $\tilde{\theta}=(2n+1)\pi$ \cite{TlBiSe2A,TlBiSe2B,TlBiSe2GRUSHIN}. Its electronic properties are presented in \cite{TlBiSe2C}}.
The remaining entries correspond to  hypothetical materials aiming to illustrate the effects of increasing
the  strength of the MEE. {For this reason we take them with the same refraction index $n=2$.}  We  compare these fluxes  with  the magnitude of   $\langle\mathbf{S}\rangle^{SED}$  written in the last column. 
We recall the dipole  characteristics  $p= 2.71 \times 10^{3} \,  {\rm eV}^{-1} \, \,  (10^{-21} \,  {\rm C\cdot cm}) $, $ \omega= 1.5 \, $ eV, $z_0= 25$ eV$^{-1}$ and the observer distance 
 $r=667 \, {\rm eV}^{-1}$.  In this case $\xi_0= 3.2 \times 10^{-2}$. 
We present the magnitude of the  Poynting vector $\langle\mathbf{S}\rangle_{\mathcal{V}_2}$ for both hemispheres evaluated at $\xi=10^{-3}$, which  we choose as a representative value satisfying the condition (\ref{LCOND}) with  $n\omega z_0 \xi= 7.5 \times 10^{-2}$, as well as $\xi/\xi_0= 3.1 \times 10^{-2}$, for $n \approx 2$. This latter number indicates that  the condition (\ref{Poynting V2}) is fulfilled for all values of $\Upsilon$ in this case.

\section{Summary and Conclusions}
\label{SUMM}
We  discuss the radiation  produced by a point-like electric dipole oriented perpendicular to and at a distance $z_0$ from the  interface    which   separates  two planar semi-infinite non-magnetic  magnetoelectric media with the same permittivity, whose electromagnetic response obeys the modified Maxwell equations (\ref{Gauss E}) and (\ref{Ampere}) of  $\vartheta$-electrodynamics. The choice $\epsilon_1= \epsilon_2$ is made to highlight and isolate  the 
purely magnetoelectric effects on the radiation, which depend on the parameter ${\tilde \theta}= \alpha(\vartheta_2-\vartheta_1)/\pi$. As a consequence of a careful calculation of the far-field approximation in the electric field  we discover the additional generation of axially symmetric cylindrical (surface) waves close to the interface,  as shown in Eqs. (\ref{E e-d perp V2}).
The analysis of the cylindrical waves leads us to introduce  two discarding angles $\theta^{UH}_0$  and $\theta^{LH}_0$, defined in Eq. (\ref{theta0 UH}) and  shown in Fig. \ref{RAD REGIONS}, which allow to distinguish two separate regimes: i)  the region $\mathcal{V}_1$, ($0 < \theta < \theta_0^{UH}, \,\,  \theta_0^{LH} < \theta < \pi$), where only the spherical waves are relevant and ii) the region $\mathcal{V}_2$,  ($\theta_0^{UH} < \theta < \theta_0^{LH}$), where both the cylindrical and the spherical waves must be taken into account. {The behavior of the electric field in region $\mathcal{V}_1$ and $\mathcal{V}_2$ is illustrated in Figs. \ref{Electric Field Patterns 1} and \ref{Electric Field Patterns 2}, respectively. }.

Due to the presence of the $\vartheta$-media we find modifications   in the angular distribution of the radiation given by  Eq. (\ref{dP dOmega dipole perp}) and illustrated in Fig. \ref{PATTERN CASE +1}. Noticeable interference effects are manifest in the upper hemisphere when the observer is in the same region of the dipole. On the contrary, no interference occurs when the observer and the dipole are in the same region, in which case the angular distribution looks similar to that of a dipole in a homogeneous media, except for important  changes in its magnitude. Such different interference effects say that the system distinguishes whether the electric dipole and the observer are in the same semi-space or not, corresponding to the Cases $(+)$ and $(-)$ respectively, discussed  at the end of  Sec. \ref{ANGDIST1}. 

Starting from the  far-field approximation of the electric field we have correctly identified the Fresnel coefficients at the interface by making use of the angular spectrum representation  \cite{Novotny-Hecht} together with the results of  Ref. \cite{Crosse-Fuchs-Buhmann} dealing with wave propagation in layered  topological insulators.

The modifications of the angular distribution in the region ${\cal V}_1$ produce new expressions for the  total radiated power $P_{(\pm)}$, which were calculated  in Eqs. (\ref{P total dipole perp -}) and (\ref{P total dipole perp +}).
The result $P_{-}$ for the lower hemisphere is independent of the dipole's  location  $z_0$ and shows a behavior similar  to the {standard electrodynamics configuration,} but  modulated by two amplitudes. The amplitude  depending on the discarding angle ${\theta_0}^{LH}$ is very close to one, because for all practical purposes ${\theta_0}^{LH}= \pi/2$. The second amplitude depends on $\Upsilon$ and induces an unexpected behavior yielding  an enhancement factor ${\cal R}_{(-)}$ that  can be  less  than one in some cases. Further, in the limiting case when  $\Upsilon \rightarrow 1$ the radiation in the lower hemisphere  would be completely canceled, such that the setup behaves as a perfect mirror. The result for $P_{(+)}$ is more intricate since the dependence upon $z_0$ now survives in  the angular distribution of Eq. (\ref{dP dOmega dipole perp}). Again, the discarding angle is very close to $\pi/2$ and we  take this approximation to obtain some general features of the enhancement factor. We find the maximum value ${\cal R}_{(+)}=1+3 \Upsilon$,  when the dipole is located at the interface ($z_0=0$). In the limit  $\varkappa =n \omega z_0$ very large we have  ${\cal R}_{(+)} \rightarrow 1+ \Upsilon$. Also we find an absolute minimum at $\varkappa \approx 2.88$ 
where ${\cal R}_{(+)}=1+0.83 \,  \Upsilon$. Thus, in this case we have an  enhancement factor larger than one, which nevertheless is limited to the  maximum  value of   ${\cal R}^{\rm max}_{(+)}=4$, independently of our choice of the parameters  $\tilde \theta$ and $n$ of the medium. In Fig. \ref{P+PERP} we plot  the ratio $P_{(+)}/P_0$ as a function of $\varkappa$, for different choices of $\Upsilon$, {where  $P_0$ stands for the total power radiated by the dipole in standard electrodynamics}.

{Regarding the region ${\cal V}_2$, we have carefully characterized along the text the conditions under which } the cylindrical  waves arise.  The cylindrical  waves are present in the whole interval $ \, 0 < \xi <\xi_0$,  {$|s|<1$  so that  $|S| = \xi^2/\xi_0^2 < 1$ for both hemispheres in this region.}
Our linear approximation in $\xi$, {carried out in  Eqs. (\ref{Poynting V21}) and (\ref{Poynting V2})},   is only valid when $\xi \ll \xi_0$, and the effect of the $\vartheta$-medium is again  codified in the parameter $\Upsilon$. 
As expected, the effects of the magnetoelectric become more evident for large ${\tilde \theta}$. The fluxes in  both  hemispheres are larger than in the {standard electrodynamics configuration} and the excess of radiation  in the upper hemisphere with respect to the lower hemisphere is evident in Table \ref{TABLE1}. 

In order to stress their similarities and  differences we  give some comparison between the dipolar radiation studied in this work which includes  a magnetoelectric medium, and that produced in the presence of two standard insulators with a planar interface and different permittivities $\epsilon(\mathbf{x})=\Theta(z)\epsilon _{2}+\Theta(-z)\epsilon _{1}$ with $\epsilon_1\neq\epsilon_2$ and $\vartheta_1=\vartheta_2=0$.  In the latter case the radiation  of a vertically oriented dipole picks up only the TM polarization as shown in  Refs. \cite{Novotny-Hecht,Chew}. In our case we have identified these contributions to the electric field through the corresponding transmission $T_{TM,TM}$ and reflection $R_{TM,TM}$ coefficients in Eqs. (\ref{RCOEFF}) and (\ref{TCOEFF}).
However, due to the magnetoelectric effect, the electric field gets  an additional input  arising from the mixing of TM and TE modes described by the reflection coefficient $R_{TE, TM}$ in Eq. (\ref{RCOEFF}). While in purely  dielectric configuration  these coefficients have an angular dependence, in our case they turn out to be constants depending  only on the parameters of the  media. This is  a consequence of our  choice $\epsilon_1=\epsilon_2$, which  forbids the existence of reflection and refraction at the interface.
On the other hand, both configurations share the generation of axially symmetric cylindrical waves at the interface of the two media,  as shown in  Sec. \ref{RADFIELD1_PERP_V2} {and particularly in Fig. \ref{Electric Field Patterns 2}.} Again, in our case the physical origin of such cylindrical  waves  relies on  the change in the magnetoelectric polarizability across the two media and not because of a difference in the permittivity constant as it happens in the 
purely dielectric configuration.

{Let us emphasize once again that our methods can be applied to  study the radiation in all materials whose macroscopic electromagnetic response is described by $\vartheta$-electrodynamics.  This  includes any magnetoelectric medium, which 
can be found among a wide range of ferromagnetic, 
ferroelectric, multiferroic materials and topological insulators, for example. Our results contribute to the list of  uncovered consequences of the magnetolelectric effect, which still remains far from experimental confirmation}. 
Generally speaking, the parameter $ {\tilde \alpha} \equiv \alpha\vartheta/\pi $ (in Gaussian units)  which sets the scale for the magnetoelectric effect  via the constitutive relations (\ref{CREL}) is very small. {Then is is clear that to enhance such effects, materials with much  higher magnetoelectric polarizabilities are required.}  Besides those values previously mentioned in the text, some typical values are: $2.8 \times 10^{-2}$ for MgO/Fe \cite{63} and  $7.2$  for Gd$_2$O$_3$/Co \cite{27}. Nonetheless, the search for a giant magnetoelectric polarizability continues recently  in composite materials reaching values as high as $9.0 \times 10^2$  for BaTiO$_3$/Co$_{60}$Fe$_{40}$, for example \cite{186}. { Among the numerous  technological applications envisaged as a consequence of the magnetoelectric effects we mention just a few: electric field control of magnetism, low-energy-consumption non-volatile magnetoelectronic  memory devices, high sensitivity  magnetometers, microwave frequency transducers and  spintronics for future photonic devices \cite{APP1,APP2,APP3}. Nevertheless, all these possibilities  crucially depend on finding materials with higher and higher  magnetoelectric polarizabilities. }

\acknowledgements
OJF has been supported by the postdoctoral fellowship CONACYT-770691 and the doctoral fellowship CONACYT-271523. OJF and LFU acknowledge support from the CONACYT project \#237503 as well as the project DGAPA-UNAM-IN103319. LFU acknowledges also support from the project CONACyT (M\'exico) \# CF-428214. LFU and OJF thank Professor Rub\'en Barrera, Prof. Hugo Morales-T\'ecotl, Dr. Alberto Mart\'in-Ruiz and Dr. Omar Rodr\'iguez-Tzompantzi for useful discussions and suggestions. {OJF thanks Dr. J. A. Crosse for his useful help to plot the electric fields. }

\appendix
\section{The electromagnetic potential  and the integrals ${\mathcal H}$, ${  \mathcal I}$ and ${ \mathcal J}$ }\label{A}
In this appendix we  derive the electromagnetic potential $A_\mu$ due to the vertically oriented  dipole, together with  the  integrals in Eqs.  (\ref{integral H}),  (\ref{integral I}) and  (\ref{integral J}) of Sec. \ref{4POTENTIAL1}. The procedure is based on the Appendix of Ref. \cite{OJF-LFU-ORT-1}. To this aim we begin from the expression (\ref{A and GF}) in the frequency-space and from the current defined in Eq. (\ref{j mu elec dip}). Let us start with the component $A^0(\mathbf{x};\omega)$, which takes the  form 
\begin{eqnarray}
\hspace{-1cm}A^0(\mathbf{x};\omega) &=&-\frac{ip}{\epsilon}\int dz'\int_0^\infty\frac{k_{\perp }dk_{\perp }}{\sqrt{\tilde{k}_{0}^{2}-k_{\perp}^{2}}}J_{0}(k_{\perp}\rho)e^{i\sqrt{\tilde{k}_{0}^{2}-k_{\perp }^{2}}|z-z'|}\delta^{\prime}(z^{\prime}-z_{0})\nonumber\\
\hspace{-1cm} &&-\frac{i}{\epsilon}\frac{\tilde{\theta}^{2}p}{4n^{2}+\tilde{\theta}^{2}}\int dz'\int_0^\infty\frac{k_{\perp }^3dk_{\perp }}{\left(\tilde{k}_{0}^{2}-k_{\perp}^{2}\right)^{3/2}}J_{0}(k_{\perp}\rho)e^{i\sqrt{\tilde{k}_{0}^{2}-k_{\perp }^{2}}(|z|+|z'|)}\delta^{\prime}(z^{\prime}-z_{0}),
\label{A0 e-d Appendix}
\end{eqnarray}
after making the convolution of the GF (\ref{GF theta}) with the current (\ref{j mu elec dip}).  
In obtaining the above relation  we have expressed the area element $d^{2}\mathbf{k_{\perp}}=k_{\perp}dk_{\perp}d\varphi$ in polar coordinates and we have chosen the $k_{\perp x}$ axis in the direction of the vector $\mathbf{x}_{\perp}=(x,y,0)$. This defines the coordinate system ${\cal S}$ to be repeatedly used in the following. Next we write  $\mathbf{k_{\perp}}\cdot\mathbf{x}_{\perp}=k_{\perp}\rho\cos\varphi$ with $\rho=\|\mathbf{x}_\perp\|=\sqrt{x^2+y^2}$ and recall that the angular integral of $\exp (ik\rho\cos \varphi)$ provides a representation of the Bessel function $J_{0}(k_{\perp}\rho)$ \cite{Abramowitz}.
The first integral in Eq. (\ref{A0 e-d Appendix}) can be carried out by recalling  the Sommerfeld identity \cite{Sommerfeld 1,Sommerfeld 2,Sommerfeld}
\begin{equation}\label{SOMM_ID0}
i\int_{0}^{\infty}\frac{ k_{\perp} dk_{\perp}}{\sqrt{\tilde{k}_{0}^{2}-k_{\perp}^{2}}} J_{0}( k_{\perp}R_{\perp})
e^{i\sqrt{\tilde{k}_{0}^{2}-k_{\perp}^{2}} |z-z'|}=\frac{e^{i \tilde{k}_{0}R}}{R}\;,
\end{equation}
where $R=\sqrt{\rho^{2}+(z-z')^{2}}$. Then, we impose the coordinate conditions appropriate to the far-field approximation
\begin{eqnarray}
\|\mathbf{x}\|\gg\|\mathbf{x}^{\prime}\|, \quad \,\,\,   R_{\perp}=\|\left(\mathbf{x}-\mathbf{x}%
^{\prime}\right)_{\perp}\|\simeq\|\mathbf{x}_{\perp}\|=\rho, \quad  |z-z^{\prime}|\simeq|z|, \label{FFC}
\end{eqnarray}
which yields the well-known result of standard ED,
${e^{i \tilde{k}_{0}R}}/{R}\rightarrow {e^{i\tilde{k}_{0}(r-\hat{\mathbf{n}}\cdot\mathbf{x}^{\prime})}}/{r}$,
with $\hat{\mathbf{n}}$ being a unit vector in the direction of $\mathbf{x}$ and where $\|\mathbf{x}\|=r$ \cite{Jackson, Schwinger}. Substituting this approximation into the first integral of Eq. (\ref{A0 e-d Appendix}) and integrating $\delta'(z'-z_0)$ we obtain 
\begin{equation}
A^0(\mathbf{x};\omega)=-\frac{p}{\epsilon}i\tilde{k}_{0}\cos\theta\frac{e^{i\tilde{k}_{0}(r-z_0\cos\theta)}}{r}
-\frac{1}{\epsilon}\frac{\tilde{\theta}^{2}p}{4n^{2}+\tilde{\theta}^{2}}\int \frac{k_{\perp }^{3}dk_{\perp }}{\tilde{k}_{0}^{2}-k_{\perp }^{2}}J_{0}(k_{\perp}\rho)e^{i\sqrt{\tilde{k}_{0}^{2}-k_{\perp }^{2}}(|z|+z_0)}.
\label{A0 e-d Appendix1}
\end{equation}
Therefore, after making $n^2=\epsilon$ and identifying the integral $\cal H$ defined  in Eq. (\ref{integral H}) we find the expression (\ref{A0 e-d almost}).

For convenience we proceed now to calculate simultaneously the components $A^1(\mathbf{x};\omega)$ and $A^2(\mathbf{x};\omega)$, which can be written together as 
\begin{equation}
\label{A i barra e-d}
A^{a}(\mathbf{x};\omega)=-\frac{2\tilde{\theta}p}{4n^{2}+\tilde{\theta}^{2}}\varepsilon _{\;\;0}^{a\;\;b3}I_{b}(\mathbf{x},z_0;\omega)-\frac{\tilde{\theta}^{2}p}{4n^{2}+\tilde{\theta}^{2}}Q_{a}(\mathbf{x},z_0;\omega),
\end{equation}
where we define the  integrals
\begin{eqnarray}
I_{a}(\mathbf{x},z_0;\omega)&=&-\int_{0}^{\infty}\frac{k_{\perp}^{2}dk_{\perp}}{\sqrt{\tilde{k}_{0}^{2}-k_{\perp}^{2}}}e^{i\sqrt{\tilde{k}_{0}^{2}-k_{\perp}^{2}}(|z|+z_0)}J_{0}(k_{\perp}\rho) \, v_{a},\\
Q_{a}(\mathbf{x},z_0;\omega)&=&\int_{0}^{\infty}\frac{k_{\perp}^{2}dk_{\perp}}{\tilde{k}_{0}^{2}-k_{\perp}^{2}}e^{i\sqrt{\tilde{k}_{0}^{2}-k_{\perp}^{2}}(|z|+z_0)} J_{0}(k_{\perp}\rho)\, v_{a}\;,
\end{eqnarray}
with $v_{a}=(\cos\varphi,\sin\varphi, 0)$.
Here $a,b=1,2$ and $k_a=(\mathbf{k}_\perp, 0)$.
Choosing the  coordinate system ${\cal S}$ we find  $I_2=0$ and $Q_2=0$, which tells us that  both vectors ${\mathbf I}$ and  ${\mathbf Q}$   point in the direction of  ${\mathbf x}_\perp$. Thus we can write 
\begin{eqnarray}
{\mathbf I}(\mathbf{x},z_0;\omega)&=&\mathbf{x}_{\perp} \, \frac{i}{\rho}\frac{\partial}{\partial\rho}\int_{0}^{\infty}\frac{k_{\perp} dk_{\perp}}{\sqrt{\tilde{k}_{0}^{2}-k_{\perp}^{2}}}J_{0}\left(k_{\perp}\rho\right)e^{i\sqrt{\tilde{k}_{0}^{2}-k_{\perp}^{2}}(|z|+z_0)},\label{I barra j}\\
{\mathbf Q}(\mathbf{x},z_0;\omega)&=&- \mathbf{x}_\perp \, \frac{i }{\rho}\frac{\partial}{\partial\rho}\int_{0}^{\infty}\frac{k_{\perp} dk_{\perp}}{\tilde{k}_{0}^{2}-k_{\perp}^{2}}J_{0}\left(k_{\perp}\rho\right)e^{i\sqrt{\tilde{k}_{0}^{2}-k_{\perp}^{2}}(|z|+z_0)}\;,\label{Q barra j}
\end{eqnarray}
where we identify the integrals $\mathcal{I}$  and $\mathcal{J}$  previously introduced in Eqs. (\ref{integral I}) and (\ref{integral J}). Plugging the latter forms of $\mathcal{I}$  and $\mathcal{J}$  into Eq. (\ref{A i barra e-d}) we find the expression (\ref{A1 e-d almost}).
Finally, we find
\begin{equation}
A^3(\mathbf{x};\omega)=\omega p\int_{0}^{\infty}\frac{ k_{\perp} dk_{\perp}}{\sqrt{\tilde{k}_{0}^{2}-k_{\perp}^{2}}} J_{0}( k_{\perp}R_{\perp})e^{i\sqrt{\tilde{k}_{0}^{2}-k_{\perp}^{2}} |z-z_0|}\;.
\end{equation}
After employing the Sommerfeld identity (\ref{SOMM_ID0}) together with the  far-field approximation (\ref{FFC}) we obtain Eq. (\ref{A3 e-d almost}).

\section{Evaluation of   the integrals (\ref{integral H})-(\ref{integral J}) by the modified steepest descent method}\label{B}
In  this appendix  we closely follow Ref. \cite{Banhos} and  apply a modified steepest descent method to find the far-field approximation of  the integrals ${\mathcal H}$, ${  \mathcal I}$ and ${ \mathcal J}$ whose results are given in Eqs. (\ref{H final})-(\ref{I final}). {This Appendix is divided in two sections. In the first one, we present in full detail the method by solving the integral $\mathcal{H}$ defined in  Eq.  (\ref{integral H}). In the second  section we only indicate  a summary  of the method  leading to  the integrals $\mathcal{J}$  and  $\mathcal{I}$, respectively. } 

\begin{figure}
\centering
\subfloat[]
{
\label{SOMMERFELD PATH}
\includegraphics[width=8CM]{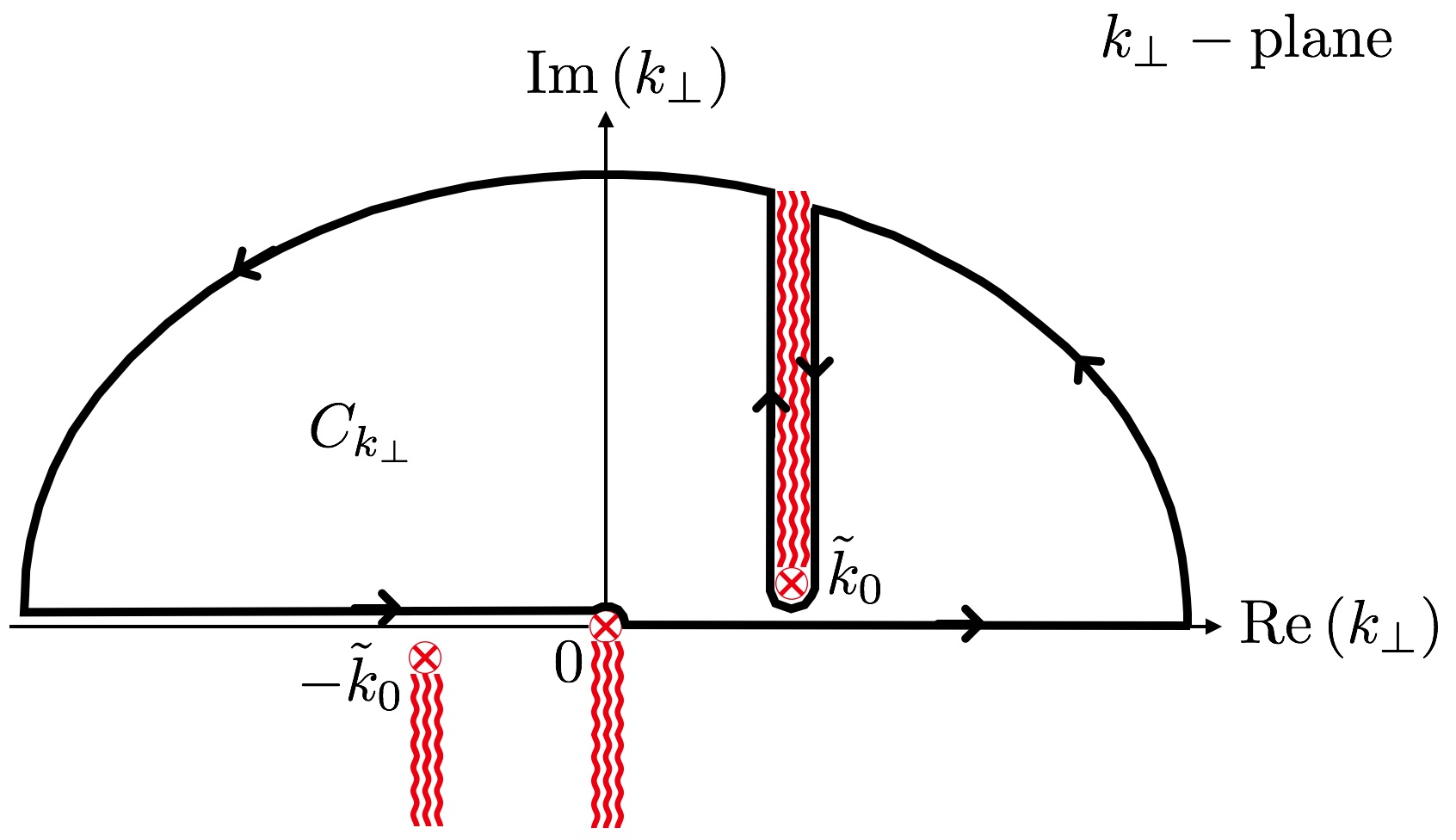} 
}
\subfloat[]
{
\label{ALPHA PATH}
\includegraphics[width=8CM]{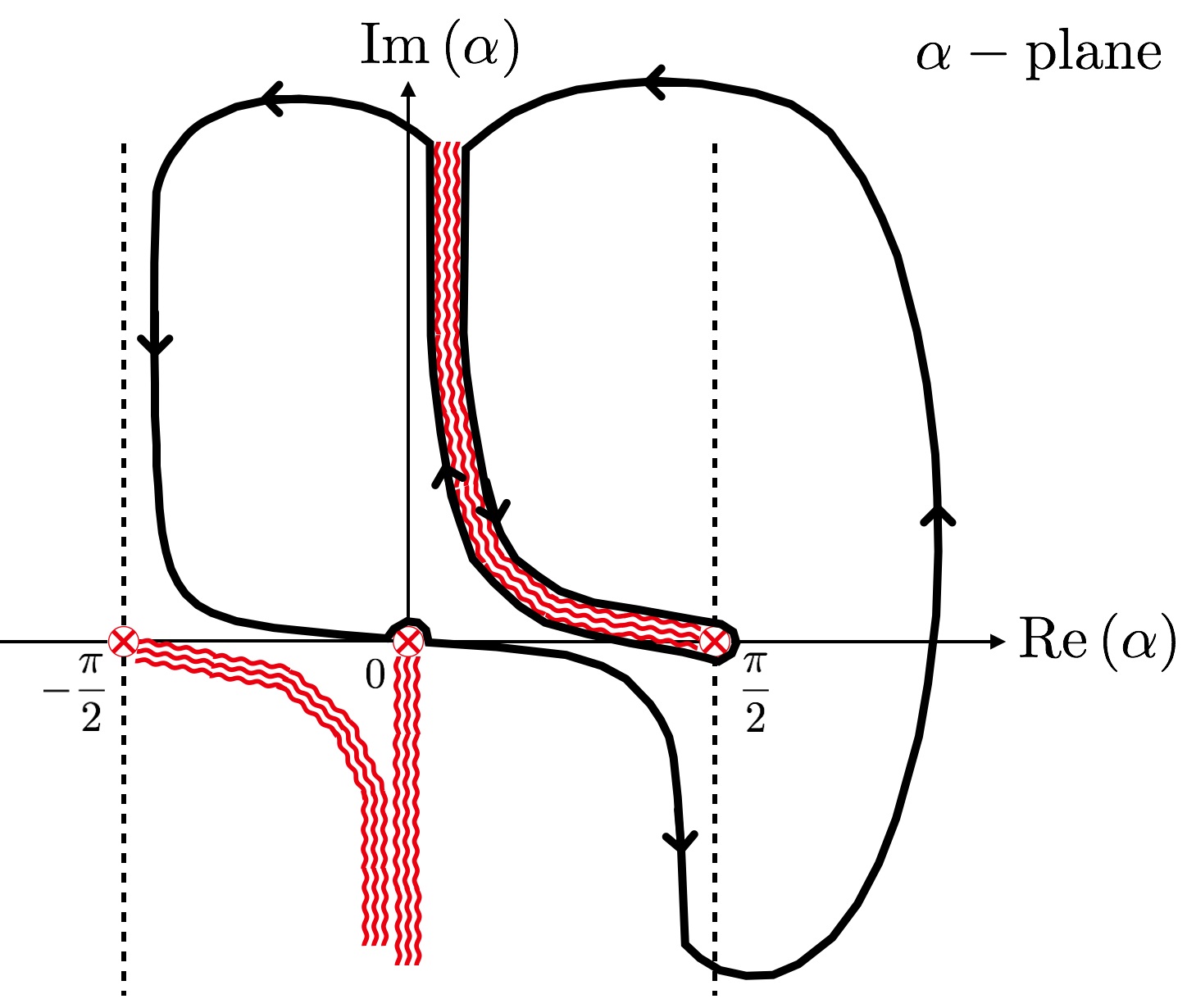}
}
\caption{(a) The Sommerfeld path of integration $C_{k_{\perp}}$ in the ${k}_\perp$-plane showing  the branch cuts originating from  the branch points $\pm\tilde{k}_{0}$ and $0$. (b) A permissible deformation $C_\alpha$  of the path of integration obtained by the transformation $k_\perp=\tilde{k}_{0} \sin \alpha$. The branch  points in the $\alpha$-plane are  $\alpha_\pm(\pm\tilde{k}_{0})=\pm\pi/2$ and $\alpha(0)=0$.}
\label{PATHS}
\end{figure}
\subsection{The integral $\mathcal{H}$ }
\label{Integral H Appendix}
It will prove  convenient to rewrite $\mathcal{H}$ in terms of Hankel functions. We start from
\begin{equation}\label{J and H}
J_{0}(x)=\frac{1}{2}\left[H_{0}^{(1)}(x)+H_{0}^{(2)}(x)\right],
\end{equation}
where $H_{0}^{(1)}(x)$ and $H_{0}^{(2)}(x)$ are the Hankel functions, together with the reflection formula $H_{0}^{(1)}(e^{i\pi}x)=-H_{0}^{(2)}(x)$ \cite{Arfken},
which allows us to extend the integration interval in Eq.  (\ref{integral H}) to $-\infty$. The result is 
\begin{equation}\label{H Sommerfeld}
\mathcal{H}(\mathbf{x},z_0;\omega)=\frac{1}{2}\oint_{C_{k_{\perp}}}\frac{k_{\perp}^{3}dk_{\perp}}{\tilde{k}_{0}^{2}-k_{\perp}^{2}}H_{0}^{(1)}\left(k_{\perp}\rho\right)e^{i\sqrt{\tilde{k}_{0}^{2}-k_{\perp}^{2}}(|z|+z_0)}\;,
\end{equation}
where $C_{k_{\perp}}$ is the so-called Sommerfeld path of integration defined in Fig. \ref{SOMMERFELD PATH}, which  avoids  the  branch cuts dictated by the Hankel function $H_0^{(1)}$ and by the square root $\sqrt{\tilde{k}_{0}^{2}-k_{\perp}^{2}}$. At this point is worth mentioning that henceforth we will retain some dissipation in the medium ($1\gg\mathrm{Im}[\tilde{k}_{0}]>0$) for convergence purposes and to avoid troublesome questions of convergence that arise when $\mathrm{Im}[\tilde{k}_{0}]=0$.
This guarantees that $\mathrm{Re}\left[i\sqrt{\tilde{k}_{0}^{2}-k_{\perp}^{2}}\right]<0$, i.e., the exponential argument will be negative implying the rapidly exponential decay that assures the convergence of the integral ${\cal H}$  \cite{Banhos}.

First, we apply the conformal transformation $k_\perp=\tilde{k}_0\sin\alpha$ obtaining \cite{Banhos,Wait} 
\begin{equation}
\mathcal{H}(\mathbf{x},z_0;\omega)=\frac{\tilde{k}_{0}^{2}}{2}\oint_{C_{\alpha}}d\alpha\frac{\sin^3\alpha}{\cos\alpha}H_{0}^{(1)}\left(\tilde{k}_0\rho\sin\alpha\right)e^{i\tilde{k}_0\cos\alpha(|z|+z_0)},
\end{equation}
where $C_\alpha$ is given in Fig. \ref{ALPHA PATH}. Following Ref.  \cite{Wait}, now  it is convenient to use the asymptotic expansion of the Hankel function \cite{Abramowitz}
\begin{equation}\label{asymptotic Hankel}
H_{0}^{(1)}\left(\tilde{k}_0\rho\sin\alpha\right)\sim\sqrt{\frac{2}{\pi \tilde{k}_0\rho\sin\alpha}}%
e^{i\tilde{k}_0\rho\sin\alpha-i\frac{\pi}{4}}\;,
\end{equation}
which is allowed because we are focusing on the far-field approximation required  for the  analysis of radiation. In this way, we have
\begin{equation}
\mathcal{H}(\mathbf{x},z_0;\omega)=\tilde{k}_{0}^{2}\sqrt{\frac{1}{2\pi\tilde{k}_0R_\perp}}e^{-i\pi/4}\oint_{C_{\alpha}}d\alpha\frac{\sin^{5/2}\alpha}{\cos\alpha}e^{i\tilde{k}_0\rho\sin\alpha+i\tilde{k}_0\cos\alpha(|z|+z_0)}\;.
\end{equation}
For the moment we restrict ourselves only to the UH ($\cos\theta>0$). The  calculation for the LH is  sketched after  Eq. (\ref{FFINALH}). So, we write $|z|=r\cos\theta$ and $\rho=r\sin\theta$, i.e., $r=\sqrt{\rho^2+z^2}$. Thereby, we find
\begin{equation}\label{H theta phase}
\mathcal{H}(\mathbf{x},z_0;\omega)=\tilde{k}_{0}^{2}\sqrt{\frac{1}{2\pi\tilde{k}_0r\sin\theta}}e^{-i\pi/4}\oint_{C_{\alpha}}d\alpha\frac{\sin^{5/2}\alpha}{\cos\alpha}e^{i\tilde{k}_0r\cos\left(\alpha-\theta\right)+i\tilde{k}_0z_0\cos\alpha}\;.
\end{equation}
Next we determine the saddle-point of (\ref{H theta phase}) by choosing the stationary phase as only  $\varphi\left(\alpha\right)=i\tilde{k}_0r\cos\left(\alpha-\theta\right)$, according to Ref.  \cite{Banhos}.  The saddle-point $\alpha_s$ is determined through $\varphi^{\prime} \left(\alpha_s\right)=0$, which gives $\alpha_s=\theta$. This yields the full stationary phase to be $i{\tilde k}_0r  \cos \theta$.  At this stage, the steepest descent path is specified on the $\alpha$-plane by demanding the condition 
\begin{equation}\label{SD condition 1}
\mathrm{Im}\left[\varphi\left(\alpha\right)\right]=\mathrm{Im}\left[\varphi\left(\alpha_s\right)\right]\Rightarrow\mathrm{Im}\left[i\tilde{k}_0r\cos\left(\alpha-\theta\right)\right]=\mathrm{Im}\left[i\tilde{k}_0r\right]
\end{equation}
over $C_\alpha$, as sketched in Fig. \ref{ALPHA SD PATH}.
\begin{figure}[H]
\centering
\subfloat[]
{
\label{ALPHA SD PATH}
\includegraphics[width=8CM]{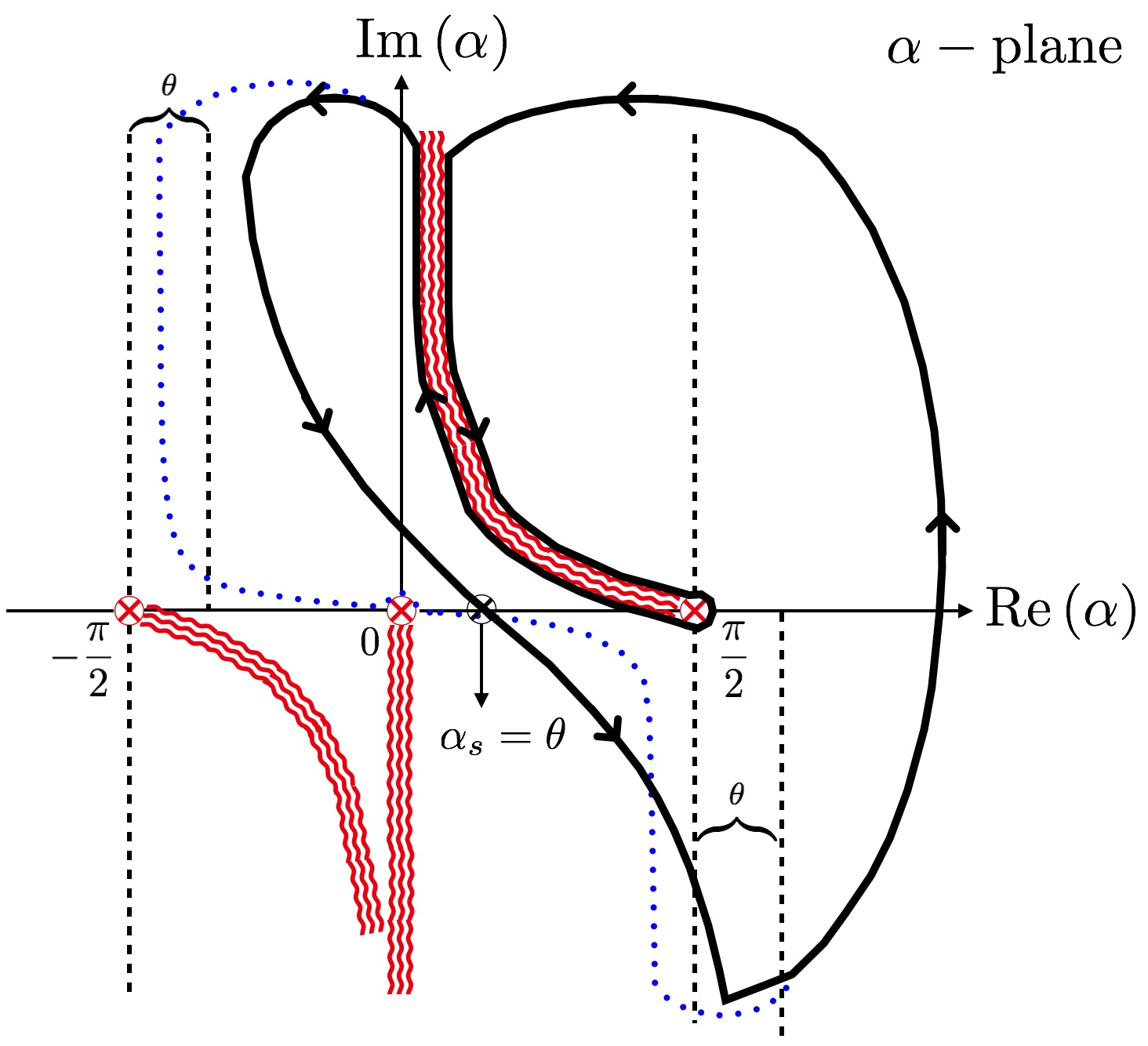} 
}
\subfloat[]
{
\label{W PATH}
\includegraphics[width=8CM]{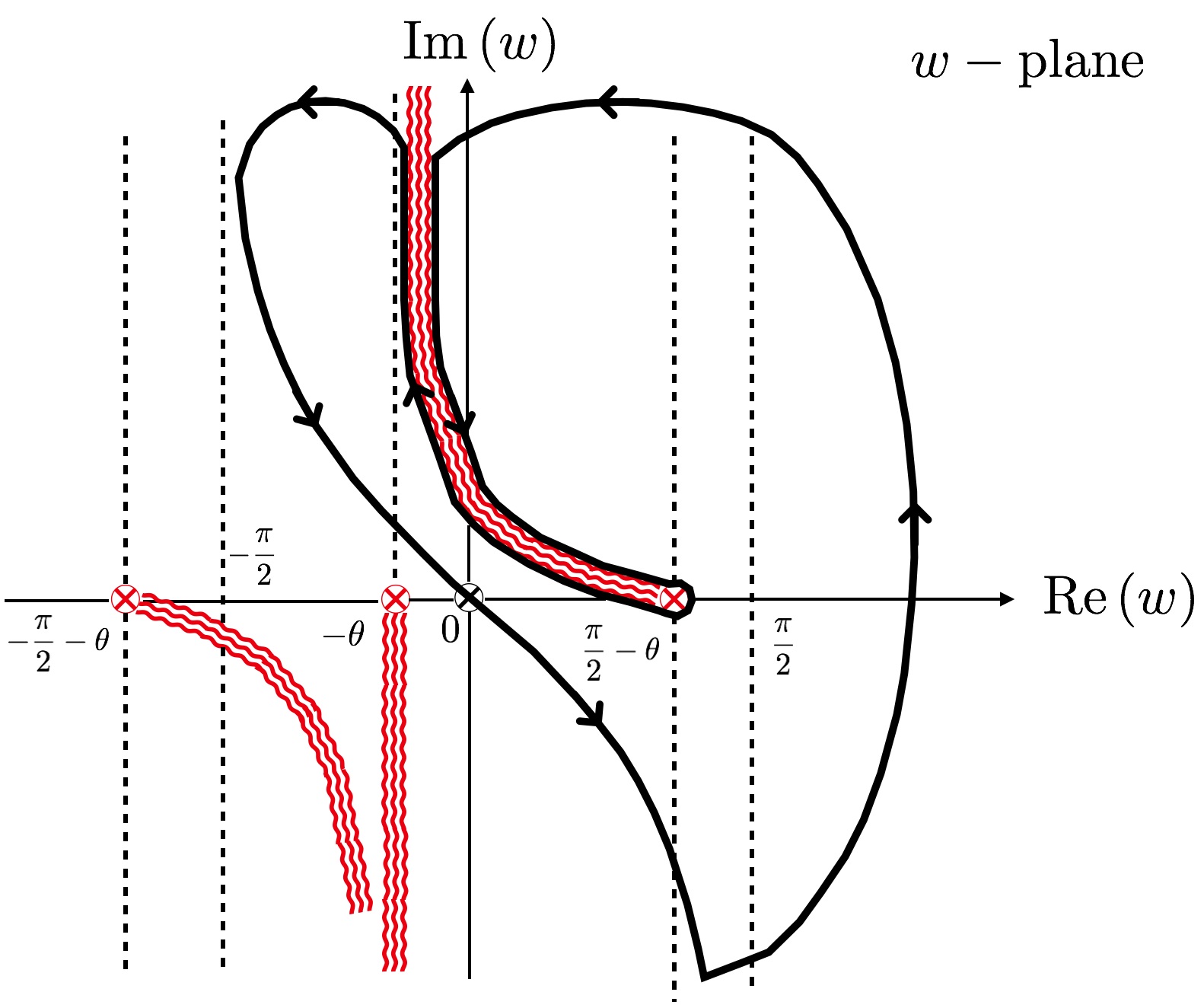} 
}
\caption{(a) The path of integration $C_\alpha$ with the steepest descent condition (\ref{SD condition 1}) in the $\alpha$-plane. The path of steepest descent has the asymptotes  $\mathrm{Re}(\alpha)=-\pi/2+\theta$,  $\pi/2+\theta$, and crosses the real axis of the $\alpha$-plane at the saddle-point $\alpha_s=\theta$. The previous path $C_\alpha$ of Fig. \ref{ALPHA PATH} is sketched here in the blue dotted line. (b)  The path of integration $C_w$ in the $w$-plane obtained by the shift $w=\alpha-\theta$. The resulting path of steepest descent now has the asymptotes $\mathrm{Re}(w)=\pm\pi/2$ and crosses the real axis of the $w$-plane at $w=0$. }
\label{PATHS1}
\end{figure}
Now we shift the origin to coincide with the saddle point  by setting  $w=\alpha-\theta$ in $\mathcal{H}$,  which yields 
\begin{equation}\label{H theta phase w}
\mathcal{H}(\mathbf{x},z_0;\omega)=\tilde{k}_{0}^{2}\sqrt{\frac{1}{2\pi\tilde{k}_0r\sin\theta}}e^{-i\pi/4}\oint_{C_{w}}dw\frac{\sin^{5/2}\left(\theta+w\right)}{\cos\left(\theta+w\right)}e^{i\tilde{k}_0r\cos w+i\tilde{k}_0z_0\cos\left(\theta+w\right)}\;.
\end{equation}
The reparametrized  path  $C_w$, shown in Fig. \ref{W PATH}, now satisfies the following steepest descent condition
\begin{equation}\label{SD condition 2}
\mathrm{Im}\left[i\tilde{k}_0r\cos w\right]=\mathrm{Im}\left[i\tilde{k}_0r\right].
\end{equation}
The next step is to introduce the  conformal transformation \cite{Banhos} 
\begin{equation}\label{u transformation}
\frac{u^{2}}{2}=\varphi\left(0\right)-\varphi\left(w\right)=i\tilde{k}_0r\left(1-\cos w\right)\;,
\end{equation}
whose purpose is to map the path of steepest descent into the real axis. This requires the change of variables $\cos w=1-{u^2}/{2i\tilde{k}_0r}$  in Eq.(\ref{H theta phase w}), after which we obtain
\begin{equation}
\mathcal{H}(\mathbf{x},z_0;\omega)=\frac{\tilde{k}_0}{i}\sqrt{\frac{1}{2\pi\sin\theta}}\frac{e^{i\tilde{k}_0r}}{r}\oint_{C_{u}}duF_1(u)e^{-u^2/2}, \quad F_1(u)=\frac{\sin^{5/2}\left[\theta+w(u)\right]e^{i\tilde{k}_0z_0\cos\left[\theta+w(u)\right]}}{\cos\left[\theta+w(u)\right]\sqrt{1-\frac{u^2}{4i\tilde{k}_0r}}}.
\label{HAF}
\end{equation}
The path  $C_u$ is sketched in Fig. \ref{U PATH}.
\begin{figure}
  \centering
\includegraphics[width=8CM]{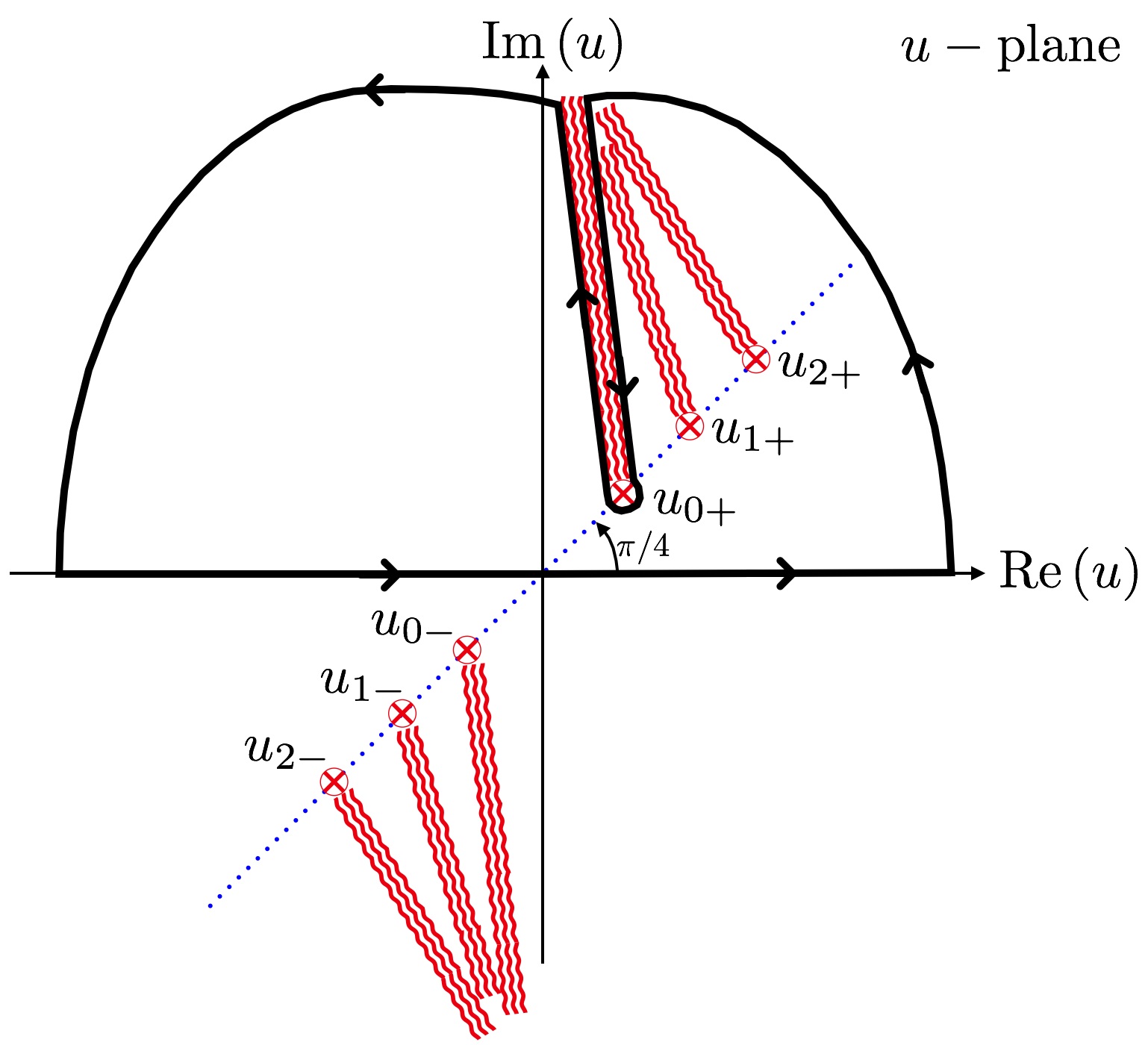} 
\caption{The path of integration $C_u$ obtained by the conformal transformation given in Eq. (\ref{u transformation}) when $\pi/2>\theta\gg0$. The path of steepest descent is mapped into the real axis of the $u$-plane. Here the branch points are $u_{0 \pm}=u(\pi/2-\theta)=\pm\sqrt{2i\tilde{k}_0r\left(1-\sin\theta\right)}$, $u_{\pm 1}=u(-\pi/2-\theta)=\pm\sqrt{2i\tilde{k}_0r\left(1+\sin\theta\right)}$ and  $u_{2 \pm}=u(-\theta)=\pm\sqrt{2i\tilde{k}_0r\left(1-\cos\theta\right)}$. The branch cuts converge at $\infty\times e^{i\pi/2}$ and $\infty \times e^{-i\pi/2}$. When $\theta\approx0$ the branch cuts lie over the blue dotted line of slope $\pi/4$ and converge at the points $\infty \times  e^{i\pi/4}$ and $\infty \times e^{i5\pi/4}$.}\label{U PATH}
\end{figure}
Then we look at  the behavior of $F_1(u)$ and find that it has  poles in the $u$-plane located at $u_0$ given by  
\begin{equation}\label{general poles}
\sqrt{1-\frac{u_0^2}{4i\tilde{k}_0r}}=0, \qquad \cos\left[\theta+w(u_0)\right]=0\;.
\end{equation}
We will consider only those poles  in the second equation above, because the poles from the first equations will only matter when  seeking  for correction  terms of higher order than $r^{-1}$, which  we will not pursue here. Recalling the last change of variables $w \rightarrow u$ we find that the poles are
\begin{equation}\label{u0}
u_{0 \pm} =\pm\sqrt{2i\tilde{k}_0r\left(1-\sin\theta\right)}\equiv \pm \sqrt{2} \Lambda, \quad \Lambda= \sqrt{i\tilde{k}_0r\left(1-\sin\theta\right)}.
\end{equation}
 Since our integration path is in the upper-half plane we require only $u_{0+}$.

From Eq. (\ref{HAF}) we realize that  $F_1(u)$ is not a smooth function around  the stationary phase $i{\tilde k}_0z_0 \cos[\theta+\omega(u)]$ precisely due to the pole contribution, which prevents a direct application of the method.  The main idea to overcome this difficulty  is to subtract and add the conflicting pole as we will do next \cite{Banhos, Michalski-Mosig}.  This extraction procedure was mathematically justified by van der Waerden \cite{van der Waerden}. Due to the symmetry of the conformal transformation $u$ in Eq. (\ref{u transformation}) we will focus on the upper $u$-semi-plane. For this extraction, we need the residue of $F_1(u)$ at $u_{0+}$ which is 
\begin{equation}\label{residue}
\mathrm{Res}\left(F_1;u_{0+}\right)=\frac{\sqrt{\tilde{k}_0r}}{i\sqrt{i}}\;.
\end{equation}
Then, we introduce the function
\begin{equation}\label{psi de u}
\psi(u)=\underbrace{F_1(u)-\frac{\mathrm{Res}\left(F_1;u_{0+}\right)}{u-u_{0+}}}_{
\psi_1(u)}+\underbrace{\frac{\mathrm{Res}\left(F_1;u_{0+}\right)}{u-u_{0+}}}_{
\psi_2(u)}\;.
\end{equation}
In this way, $\psi_1(u)$ is analytic at $u_{0+}$, so that we can apply the standard steepest descent method. Meanwhile, $\psi_2(u)$ will contain the simple pole contribution that will be analyzed later. At this stage we rewrite $\mathcal{H}$ in Eq. (\ref{HAF}) as
\begin{equation}\label{H_SD and P}
\mathcal{H}(\mathbf{x},z_0;\omega)=\frac{\tilde{k}_0}{i}\sqrt{\frac{1}{2\pi\sin\theta}}\frac{e^{i\tilde{k}_0r}}{r}\left[\mathcal{H}_{SD}(\mathbf{x},z_0;\omega)+\mathcal{H}_{P}(\mathbf{x},z_0;\omega)\right]\;,
\end{equation}
where we define
\begin{equation}
\mathcal{H}_{SD}(\mathbf{x},z_0;\omega)=\oint_{C_{u}}du\,\psi_1(u)e^{-u^2/2}, \qquad \mathcal{H}_{P}(\mathbf{x},z_0;\omega)=\oint_{C_{u}}du\,\psi_2(u)e^{-u^2/2}.
\end{equation}
The first contribution provides  the standard steepest descent integral,  and  the second term results from  the integration of the simple pole. For simplicity, we omit the explicit dependence of $\mathcal{H}_{SD}$ and $\mathcal{H}_{P}$ in what follows. For the moment, let us focus on $\mathcal{H}_{SD}$, whose detailed form is
\begin{equation}\label{H SD explicit}
\mathcal{H}_{SD}=\oint_{C_{u}}du\left\{\frac{\sin^{5/2}\left[\theta+w(u)\right]e^{i\tilde{k}_0z_0\cos\left[\theta+w(u)\right]}}{\cos\left[\theta+w(u)\right]}-\frac{\sqrt{\tilde{k}_0r}}{i\sqrt{i}(u-u_{0 +})}\right\}e^{-u^2/2}\;.
\end{equation}
As we mentioned previously, the $u$-transformation has already mapped the path of steepest descent into the real axis on the $u$-plane. Thus, we are able to approximate $\mathcal{H}_{SD}$ with standard calculus techniques. Since  we are interested only in the dominant term of $\mathcal{H}_{SD}$, it is enough to consider the zeroth order term in the expansion of    $\psi_1(u)$ in its Taylor series around $u=0$ ($w=0$), because most of the contribution arises from its  vicinity due to the presence of the Gaussian function $e^{-u^2/2}$. Performing this, we obtain
\begin{equation}
{\mathcal H}_{SD}=\left\{\frac{\sin^{5/2}\theta e^{i\tilde{k}_0z_0\cos\theta}}{\cos\theta}-\frac{1}{\sqrt{2\left(1-\sin\theta\right)}}\right\}\sqrt{2\pi}\;,
\end{equation}
where we have already introduced the expression  of  $u_{0+}$ from Eq. (\ref{u0}). 
Substituting back  this result in Eq. (\ref{H_SD and P}), we find 
 \begin{eqnarray}
\mathcal{H}&=&\tilde{k}_0\frac{e^{i\tilde{k}_0r}}{ir}\left\{\frac{\sin^2\theta e^{i\tilde{k}_0z_0\cos\theta}}{\cos\theta}-\frac{1}{\sqrt{2\left(\sin\theta-\sin^2\theta\right)}}\right\}
+\frac{\tilde{k}_0}{i}\sqrt{\frac{1}{2\pi\sin\theta}}\frac{e^{i\tilde{k}_0r}}{r}\mathcal{H}_{P}.\label{H almost}
\end{eqnarray}
We observe that the first term inside the curly brackets  is just the same term without the contribution of the simple pole that we reported in \cite{OJF-LFU-ORT-1}. The second contribution appears to introduce divergences at  $\theta=0,\pi$,  due to  the term $1/\sqrt{\sin\theta}$. However, these singularities are artificial  and arise from  the insertion of the Hankel function $H^{(1)}_0$ in Eq. (\ref{J and H}) \cite{Banhos}. One can trace back these artificial divergences to the branch cuts represented by the ray that starts at the origin of Fig. \ref{SOMMERFELD PATH} and Fig. \ref{ALPHA PATH}, by the ray that begins at $-\theta$ in Fig. \ref{W PATH} in the $w$-plane and by the ray that starts in  $u_2$ in Fig. \ref{U PATH} in the $u$-plane, after the successive transformations are performed. Nevertheless, one can prove   that these divergences are apparent  as long as we work within the far-field approximation $\tilde{k}_0r\rightarrow\infty$.
It only remains to determine $\mathcal{H}_{P}$. As mentioned above, the $u$-transformation maps the path of steepest descent to the real axis on the $u$-plane. So, we only need to compute $\mathcal{H}_{P}$ along that axis. Explicitly, we have that
\begin{equation}
\mathcal{H}_{P}=\frac{\pi \sqrt{\tilde{k}_0 r}}{\sqrt{i}}\frac{1}{i \pi}\int_{-\infty}^{\infty}du\frac{e^{-u^2/2}}{u-u_{0 +}} \equiv \frac{\pi \sqrt{\tilde{k}_0 r}}{\sqrt{i}} \,  {\mathcal{\tilde  W}}(u_{0 +})= \frac{\sqrt{\pi {\tilde k}_0 r}}{i \sqrt{i}} Z(\Lambda)
\label{HPFIN},
\end{equation}
recalling  that  $u_{0 +}=\sqrt{2} \Lambda$ is given by Eq. (\ref{u0}).
 We  discuss some   basic properties of the Faddeeva  function  $Z(\Lambda)$ and the  function ${\mathcal{\tilde  W}}$ in the Appendix C.

Substituting   Eqs. (\ref{HPFIN}) and (\ref{erfc 1}) in Eq. (\ref{H almost}) yields our final expression for ${\mathcal H}$
\begin{eqnarray}
&&\mathcal{H}=\tilde{k}_0\frac{e^{i\tilde{k}_0r}}{ir}\left\{\frac{\sin^2\theta e^{i\tilde{k}_0z_0\cos\theta}}{\cos\theta}-\frac{1}{\sqrt{2\left(\sin\theta-\sin^2\theta\right)}}\right\} \nonumber\\
&&+\sqrt{\frac{2}{\pi i\tilde{k}_0r\sin\theta}}\frac{\tilde{k}_0^2}{i}e^{i\tilde{k}_0r\sin\theta}\frac{\pi}{2}\mathrm{erfc}\left[-i\sqrt{i\tilde{k}_0r\left(1-\sin\theta\right)}\right], \nonumber \\
\label{FFINALH}
\end{eqnarray}
where the second term shows the presence of cylindrical waves that arise directly from the simple pole contribution  as Refs.  \cite{Banhos,Michalski-Mosig}  establish. The expression for ${\bar {\cal W}}$ is given in Eq. (\ref{erfc 1}).
Analogously, we obtain the form of ${\cal H}$ in the LH, which indicates that the whole expression valid for both hemispheres is obtained by replacing $\cos\theta$ with $|\cos\theta|$ in Eq. (\ref{FFINALH}), yielding  Eq. (\ref{H final}). For the sake of completeness  we show that Eq. (\ref{H final}) converges around $\pi/2$. To deal with $\theta=\pi/2$ we find it convenient to set  $\theta=\pi/2-\xi$ in the UH and expand (\ref{H final})  in a power series of $\xi$ around $\xi=0$.  We obtain 
\begin{eqnarray}
\mathcal{H}(\xi\rightarrow0)&\simeq&\tilde{k}_0\frac{e^{i\tilde{k}_0r}}{ir}\left\{i\tilde{k}_0z_0-\frac{\xi}{8}\right\} 
+\sqrt{\frac{2}{\pi i\tilde{k}_0r}}\tilde{k}_0^2e^{i\tilde{k}_0r}\left(\frac{\pi}{2i}+\sqrt{\frac{\pi i\tilde{k}_0r}{2}}\xi\right)+\mathcal{O}(\xi^2),\label{H cerca pi 2}
\end{eqnarray}
where we observe that the divergence disappeared. Lack of space prevent us to write down the proofs that ${\cal H}(\theta=0)=0={\cal H}(\theta=\pi).$

\subsection{The integrals $\mathcal{J}$ and ${\cal I}$ }
\label{Integral J Appendix}

 After introducing  the Hankel function $H_{0}^{(1)}(k_{\perp}\rho)$ in   Eq.  (\ref{integral J}) we  observe that $\cal J$ has almost the same form as (\ref{H Sommerfeld}) except for the $k_\perp^2$ extra factor in the integrand. 
Performing   the same  chain  of transformations from the $k_\perp$-plane to the $u$-plane as done in the Appendix \ref{Integral H Appendix} we obtain

\begin{equation}
\mathcal{J}(\mathbf{x},z_0;\omega)=\frac{e^{i\tilde{k}_0r}}{i\tilde{k}_0r}\sqrt{\frac{1}{2\pi\sin\theta}}\oint_{C_{u}}duF_2(u)e^{-u^2/2}, \quad F_2(u)=\frac{\sin^{1/2}\left[\theta+w(u)\right]e^{i\tilde{k}_0z_0\cos\left[\theta+w(u)\right]}}{\cos\left[\theta+w(u)\right]\sqrt{1-\frac{u^2}{4i\tilde{k}_0r}}},
\label{F2}
\end{equation}
where again we restrict  ourselves to the UH. Notice that $F_2(u)$ differs form  $F_1(u)$ only in the exponent of the function $\sin[\theta +\omega(u)]$, which is a consequence of the distinct powers of $k_\perp$ in the  definitions of $ \cal H$ and $\cal J$. Since the singularities are the same,  the separation of the pole yields 
\begin{equation}\label{zeta de u}
\zeta(u)=\underbrace{F_2(u)-\frac{\mathrm{Res}\left(F_2;u_{0 +}\right)}{u-u_{0 +}}}_{
\zeta_1(u)}+\underbrace{\frac{\mathrm{Res}\left(F_2;u_{0 +}\right)}{u-u_{0 +}}}_{
\zeta_2(u)}\;,
\end{equation}
with $u_
{0 +}$ given by Eq. (\ref{u0}). Since  $\zeta_1(u)$ is already analytic, we  approximate its integral through the standard steepest descent method. Meanwhile, $\zeta_2(u)$ will contain the contribution of the simple pole. In this way, we rewrite $\mathcal{J}$ as 
\begin{equation}\label{J J SD J P}
\mathcal{J}(\mathbf{x},z_0;\omega)=\frac{e^{i\tilde{k}_0r}}{i\tilde{k}_0r}\sqrt{\frac{1}{2\pi\sin\theta}}
\left[\mathcal{J}_{SD}(\mathbf{x},z_0;\omega)+\mathcal{J}_P(\mathbf{x},z_0;\omega)\right]\;,
\end{equation}
where 
\begin{equation}\label{J SD explicit}
\mathcal{J}_{SD}(\mathbf{x},z_0;\omega)=\oint_{C_{u}}du\,\zeta_1(u)e^{-u^2/2}, \qquad \mathcal{J}_P(\mathbf{x},z_0;\omega)=\mathcal{H}_P(\mathbf{x},z_0;\omega)=\oint_{C_{u}}du\,\zeta_2(u)e^{-u^2/2}.
\end{equation}
 The equality between $\mathcal{J}_{P}$ and $\mathcal{H}_{P}$ follows because $\mathrm{Res}\left(F_2;u_{0 +}\right)=\mathrm{Res}\left(F_1;u_{0 +}\right)$. The remaining calculation of  $\mathcal{J}_{SD}$ follows the same steps as that of 
 $\mathcal{H}_{SD}$ (\ref{H SD explicit}) in the  Appendix \ref{Integral H Appendix} and leads finally to Eq. (\ref{J final}), which can be shown to be  convergent at $\theta=0,\pi/2,\pi$. 
 The expansion of ${\cal J}$ in the UH  near the interface yields
 \beq
\mathcal{J}(\xi\rightarrow0)=\frac{e^{i\tilde{k}_0r}}{i\tilde{k}_0r}\left(i\tilde{k}_0z_0-\frac{\xi}{8}\right)+\sqrt{\frac{2}{\pi i\tilde{k}_0r}}e^{i\tilde{k}_0r}\left(\frac{\pi}{2i}+\sqrt{\frac{\pi i\tilde{k}_0r}{2}}\xi\right) + O(\xi^2)\;.
\eeq
The calculation of  $\mathcal{I}$, defined in Eq.  (\ref{integral I}) follows similar steps. After introducing the Hankel function $H_{0}^{(1)}(k_{\perp}\rho)$ and performing the  chain of transformations from the $k_\perp$-plane to the $u$-plane previously described we obtain
\begin{equation}
\mathcal{I}(\mathbf{x},z_0;\omega)=\frac{e^{i\tilde{k}_0r}}{ir}\sqrt{\frac{1}{2\pi\sin\theta}}\oint_{C_{u}}duF_3(u)e^{-u^2/2}, \qquad 
F_3(u)=\frac{\sin^{1/2}\left[\theta+w(u)\right]e^{i\tilde{k}_0z_0\cos\left[\theta+w(u)\right]}}{\sqrt{1-\frac{u^2}{4i\tilde{k}_0r}}}\;,
\end{equation}
in the UH. Then, we realize that the only poles remaining in $F_3(u)$ are those arising from the square root in the denominator. Nevertheless we neglect them since, as previously mentioned in the Appendix  \ref{Integral H Appendix},  these poles will only matter when we seek for correction  terms of higher order than $r^{-1}$, which is not intended  in this work. Therefore, this integral does not need a pole extraction in contrast with the former integrals $\mathcal{H}$  and $\mathcal{J}$. Following the same steps  previously carried out   for $\mathcal{H}_{SD}$  in the  Appendix  \ref{Integral H Appendix} we 
obtain Eq. (\ref{I final}). The expansion of ${\cal I}$ in the UH  near the interface is  given by 
 \begin{equation}
\mathcal{I}(\xi\rightarrow0)=\frac{e^{i\tilde{k}_0r}}{ir} (1+i\tilde{k}_0z_0\xi + O(\xi^2)).
\end{equation}

 \section{Some properties of the function $Z(\Lambda)$ }

\label{FDF}

The function  $Z(\Lambda)$ is  known as the Faddeeva (plasma dispersion) function \cite{Michalski-Mosig,FriedConte}
and has been much studied in the literature \cite{Felsen-Marcuvitz, Brekhovskikh, Ishimaru, Baccarelli, Collin}.
Let us recall the definition
\begin{equation}
Z(\Lambda)\equiv \frac{1}{\sqrt{\pi}}\int_{- \infty} ^{+ \infty} dx \frac{e^{-x^2}}{x-\Lambda}=i \sqrt{\pi} \, e^{-\Lambda^{2}}\mathrm{erfc}(-i\Lambda),
\label{Z}
\end{equation}
where the last relation in Eq. (\ref{Z}) in terms of  the complementary error function  
  is taken from  Refs.  
 \cite{Abramowitz, Michalski-Mosig}, and yields
\begin{equation}
{\mathcal {\tilde W}}(u_{0 +})= \frac{1}{i \sqrt{\pi}} Z(\Lambda)=i\pi e^{-i\tilde{k}_0r\left(1-\sin\theta\right)}\mathrm{erfc}\left[-i\sqrt{i\tilde{k}_0r\left(1-\sin\theta\right)}\right].\label{erfc 1}
\end{equation}
The function ${\cal {\tilde W}}(u_0)$, already introduced in Eq. (\ref{HPFIN}), can be written in terms of    the alternative expressions for the plasma dispersion function:  $\omega(\Lambda)$ defined in Ref. \cite{Abramowitz} and $Z(\Lambda)$ defined in Ref. \cite{FriedConte}, as follows 
\begin{equation}
{\cal {\tilde W}}(u_{0 +})={\cal {\tilde  W}}(\sqrt{2}\Lambda)= \omega(\Lambda)=\frac{1}{i \sqrt{\pi}} Z(\Lambda), \quad \Lambda=\sqrt{i\tilde{k}_0r\left(1-\sin\theta\right)}, 
\label{RELATIONS}
\end{equation}
where $\Lambda=x+iy$ is a complex variable, with $x,y$ being real numbers. 

The function $Z(\Lambda)$ satisfies the useful expression
$
Z(\Lambda^{*})=-\Big[ Z(-\Lambda))\Big]^{*},
$
together with
\begin{eqnarray}
&& Z(\Lambda)= i\pi^{1/2}e^{-\Lambda^2}-\Lambda \sum_{n=0}^\infty \pi^{1/2} (-\Lambda^2)^n/(n+1/2)!, \qquad |\Lambda| \rightarrow 0,
\label{POWERSERIES} \\
&& Z(\Lambda)= i\pi^{1/2} \sigma(\Lambda)e^{-\Lambda^2}-\frac{1}{\Lambda}\sum_{n=0}^\infty {\Lambda}^{-2n}(n-1/2)!)/\pi^{1/2}, \qquad |\Lambda| \rightarrow \infty.
\label{ASYMPT} 
\end{eqnarray}
Here  $\sigma(\Lambda)= 0, 1, 2 $ when $y>0, \, y=0,\, y<0$,  respectively. From Eq. (\ref{RELATIONS}) we  write $\Lambda=\sqrt{i} s= e^{i\pi/4} s, \,  s=\sqrt{\tilde{k}_0r\left(1-\sin\theta\right)} \, $, and we require to calculate $Z(e^{i\pi/4} s)$ which we could read  from  Ref. \cite{FriedConte}. However we find a misprint in the expression for $Z(e^{-i\pi/4} s)$  given there. The correct result is 
\begin{equation}
Z(s e^{-i\pi/4})=i\pi ^{1/2}e^{is^{2}}\left[ 1+\sqrt{2} e^{-i 3\pi/4}\left[ C(
t) -iS(t) \right] \right], \quad t=\sqrt{{2}/{\pi}} \, s, 
\end{equation}   
where $C(t)$ and $S(t)$  are the Fresnel functions   and with the identification $s=\rho$. 
Finally we obtain
\begin{equation}
|Z(s e^{i\pi/4})|^2=2 \pi \left[ \frac{1}{2}+ C^2(t)+  S^2(t)-  C(t) - S(t) \right].
\label{MODSQUARE}
\end{equation}

\

\end{document}